\pdfoutput=1
\documentclass[floatfix,nofootinbib,twocolumn,english,pra,aps]{revtex4}
\usepackage{silence}
\usepackage{float}
\usepackage[cal=boondox]{mathalfa}
\usepackage{mathtools}
\usepackage{graphicx}
\usepackage{babel}
\makeatletter
\pdfpageheight\paperheight
\pdfpagewidth\paperwidth

\begin{document}
\title{\noindent Invisible gravitons and large-scale magnetism}
\author{\noindent Massimo Giovannini}
\email{massimo.giovannini@cern.ch}
\affiliation{\noindent Department of Physics, CERN, 1211 Geneva 23, Switzerland and INFN, Section of Milan-Bicocca, 20126 Milan, Italy}
\begin{abstract}
\noindent 
The large-scale limits  on the relic signals of gravitational radiation complement the bounds coming from the interferometric detectors (in the audio band) and from the pulsar timing arrays (in the nHz range). Within this inclusive perspective the spectral energy density of the gravitons is sharply suppressed in the aHz region even though the high frequency signal can be comparatively much larger both in the kHz and GHz domains. For there are no direct tests on the expansion rate prior to the formation of the light nuclei, a modified postinflationary timeline affects the total number of $e$-folds and additionally suppresses the tensor to scalar ratio by making the relic signals effectively invisible in the aHz range.  The expansion rate prior to nucleosynthesis is further bounded by the evolution of the hypercharge field and the large-scale magnetism also constrains the decelerated expansion rate. The magnetogenesis requirements are compatible with a potentially detectable spectral energy density of the relic gravitons between the MHz and the THz while the tensor to scalar ratio remains suppressed in the aHz region. A maximum of the spectral energy density of the gravitons in the audio domain leads instead to a larger magnetic field when the scale of the gravitational collapse of the protogalaxy (of the order of the Mpc) gets comparable with the Hubble radius before equality. Along a converse viewpoint the results obtained here imply that a long decelerated stage expanding faster than radiation does not affect the high frequency range but reduces the effective number of $e$-folds by so enhancing the tensor to scalar ratio, possibly beyond its observational limit.
\end{abstract}
\maketitle
\renewcommand{\theequation}{1.\arabic{equation}}
\setcounter{equation}{0}
\section{Introduction and Motivations}
\label{sec1}
According to the so-called adiabatic paradigm  \cite{CONF1} (see also \cite{CONF1p,CONF1a})
the dominant source of large-scale inhomogeneities should come from 
the Gaussian fluctuations of the spatial curvature. 
The soundness of this working hypothesis 
has been repeatedly confirmed by the observations 
of the last score year starting from the WMAP results  
\cite{CONF2,CONF3,CONF4,CONF5,CONF6} and ending 
with the current determinations of the cosmological parameters 
(see e.g. \cite{CONF7,CONF8,CONF9,CONF10,CONF11}).
If the curvature inhomogeneities arose during a 
stage of  conventional inflationary expansion \cite{CONF12}, 
the large-scale fluctuations should have a quantum mechanical origin 
as postulated almost sixty years ago \cite{CONF13} 
well before the formulation of the current theoretical framework. Thus in a given cosmological 
scenario the relic phonons \cite{REL1,REL2} (associated with the inhomogeneities of the scalar 
curvature) must be produced together with the relic gravitons \cite{REL3,REL4,REL5,REL6} (corresponding to the tensor modes of the geometry).  This perspective holds, a fortiori, in the case of single-field
inflationary scenarios where the quasi-flat spectrum of curvature 
inhomogeneities measured by the large-scale experiments 
\cite{CONF2,CONF3,CONF4,CONF5,CONF6,CONF7,CONF8,CONF9,CONF10,CONF11}
is complemented by an equally nearly scale invariant spectrum of relic gravitons 
\cite{RELINF1,RELINF2,RELINF3,RELINF4} that has not been 
observed so far neither in the ultra-low frequency domain\footnote{In the ultra-low frequency range $\nu= {\mathcal O}(\nu_{p})$ and  $\nu_{p} = k_{p}/(2\pi) = 3.092\,\,\mathrm{aHz}$ where $k_{p} = 0.002\, \mathrm{Mpc}^{-1}$ is the common pivot scale  at which the scalar and tensor power spectra are assigned \cite{CONF2,CONF3,CONF4,CONF5,CONF6,CONF7,CONF8,CONF9,CONF10,CONF11} prior to photon decoupling. In the audio region (between few Hz and $10$ kHz) the wide band interferometers are now operating.} (probed by the large-scale observations) nor in the audio band where the interferometers have been setting bounds on diffuse backgrounds of gravitational radiation in the last twenty 
years \cite{INT1,INT2,INT3,INT4,INT5}.  To avoid potential confusions we stress that, in this paper, the conventional prefixes of the International System of units will be consistently used (e.g. $1\, \mathrm{aHz} =10^{-18} \, \mathrm{Hz}$,
$1 \, \mathrm{fHz} = 10^{-15} \mathrm{Hz}$ and so on and so forth).

 Following the standard practice the constraints on the aHz gravitons are introduced as limits on the tensor-to-scalar-ratio $r_{T} = {\mathcal A}_{T}/{\mathcal A}_{{\mathcal R}}$ where ${\mathcal A}_{T}$ and ${\mathcal A}_{{\mathcal R}}$ denote, respectively,  the amplitudes of the tensor and of the scalar power spectra at a conventional reference wavenumber $k_{p} = 0.002\, \mathrm{Mpc}^{-1}$ that corresponds to comoving 
frequencies $\nu_{p} = {\mathcal O}(\mathrm{aHz})$. While the WMAP collaboration did set upper limits $r_{T} < {\mathcal O}(0.1)$ \cite{CONF2,CONF3,CONF4,CONF5,CONF6}, the recent determinations suggest $r_{T} < {\mathcal O}(0.06)$ or even $r_{T} < {\mathcal O}(0.03)$ \cite{CONF7,CONF8,CONF9,CONF10,CONF11}. In single-field 
inflationary models the spectral slope in the aHz range and the slow-roll parameter $\epsilon$  are all related to $r_{T}$ by the so-called consistency relations stipulating that $n_{T} \simeq - 2\epsilon \simeq - r_{T}/8$.  Although it is true that, in concordance scenario, the $B$-mode polarization is only induced by the relic gravitons (and not by the curvature inhomogeneities), it must be nonetheless stressed that the tensor modes 
democratically affect the $E$-mode polarization and the temperature autocorrelations\footnote{It is occasionally stated that the limits on $r_{T}$ chiefly come from the so-called $B$-mode polarization; however the value of $r_{T}$ controls the magnitude of the tensor contribution affecting {\em both} the temperature and the polarization anisotropies.}. The suppression of $r_{T}$ must therefore be associated with the early initial conditions of the long-wavelength fluctuations as argued long ago even before the formulation of the conventional inflationary scenarios \cite{REL3,REL4,REL5,REL6}.

The reduction of $r_{T}$ in the large-scale limit may occur either in a model-dependent perspective (because of the specific features of the potential) or thanks to the timeline of the decelerated evolution: between these two options the former is usually more emphasized than the latter even if, in our opinion, it should probably be the opposite \cite{MGINV}. Concerning 
the first possibility it is useful to recall that the slow-roll parameter $\epsilon(\tau)$ evaluated for\footnote{This occurs 
when the comoving frequency $\nu$ crosses the Hubble radius during inflation.} $\tau_{\nu} \simeq 1/(2\pi\nu)$ scales approximately as $\epsilon_{\nu} \propto 1/N_{\nu}$ in the case of monomial potentials \cite{CONF12} (see also \cite{POT1,POT2,POT3}) but the scaling is modified for plateau-like potentials (i.e. $\epsilon_{\nu} \propto 1/N_{\nu}^2$); more complicated scalings are expected for hill-top potentials \cite{POT1,POT2,POT3,POT4} (see also \cite{POT5,POT6,POT7}). In spite of the specific potential the value of $N_{\nu}$ measures the number of $e$-folds elapsed since the bunch of scale $\nu = {\mathcal O}(\nu_{p})$ crossed the Hubble radius. This means  that both $\epsilon_{\nu}$ and $r_{T}$ inherit a further suppression if the postinflationary evolution 
does not simply coincides with a radiation-dominated stage \cite{MGINV} (see also \cite{MGST1,MGST2,MGST3}), as usually assumed in the concordance paradigm \cite{CONF2,CONF3,CONF4,CONF5,CONF6,CONF7,CONF8,CONF9,CONF10,CONF11}.

In practice the value of $N_{\nu}$ depends on the decelerated evolution\footnote{If the decelerated timeline prior to BBN is faster than radiation $N_{\nu}$ 
may get smaller than ${\mathcal O}(60)$ but the opposite is true if the expansion rate is slower 
than radiation: in this case $N_{\nu} > {\mathcal O}(60)$ and $\epsilon_{\nu}$ gets 
more reduced than in the conventional situation (i.e. 
when radiation dominates right after inflation). } between the end of inflation and the onset of big-bang nucleosynthesis (BBN). Indeed, $N_{\nu}$ can be much smaller than ${\mathcal O}(60)$ (for a prolonged postinflationary stage expanding faster than radiation \cite{POT1,POT2,POT3})
and could even reach the typical values $r_{T} = {\mathcal O}(0.2)$ suggested by the Bicep2 observations \cite{POT8} that turned out to be affected by large foreground contaminations. Because the current data suggest
a much lower value of the tensor to scalar ratio it is interesting to explore, as suggested long ago, 
more general timelines where the expansion rate can be slower than radiation: in this case 
$N_{\nu}$ exceeds ${\mathcal O}(60)$ and $r_{T}$ undershoots ${\mathcal O}(0.06)$ \cite{MGINV}. For invisible gravitons  in the aHz region, the spectral energy density in the kHz and GHz domains can be much larger than in the case of the concordance paradigm since the same timeline that suppresses $r_{T}$ in the aHz range also increases the spectral energy density in critical units for much larger frequencies \cite{MGST1,MGST2,MGST3}.
The expansion histories that reduce the ultra-low frequency signals may also impact on the 
power spectra of other quantum modes eventually produced during the accelerated stage of expansion and a particularly interesting case is represented by the gauge fields whose amplification is physically related with the problem of large-scale magnetism.

Prior to the formulation of the adiabatic paradigm the existence of galactic magnetism has been 
often ascribed to the explicit breaking of spatial isotropy in the early stages of the hot big-bang scenario \cite{ANIS1,ANIS2,ANIS3}. This viewpoint is today untenable and we also know that the gauge fields can be parametrically amplified without breaking the spatial isotropy provided the Weyl invariance is broken  \cite{MMAA1,MMAA1a,MMAA1b} (see also \cite{REL1,REL2}). Besides the invariance under local gauge transformations the Weyl  and the duality symmetries \cite{MMAA2,MMAA3} determine the gauge power spectra whose late-time expressions  depend upon the decelerated expansion rate, exactly as in the case of the relic gravitons \cite{MGG1,MGG2}.  As firstly pointed out by Hoyle \cite{HOYLE} the existence of fields with huge correlation scales points towards a cosmological origin of large-scale magnetism. Since the early 1950s \cite{FERMI} it has been repeatedly argued that  magnetic fields with typical strengths of few $\mu$G should be widespread in spiral galaxies \cite{gal1,gal2,gal3,gal4,gal5,gal6,gal7}, extended 
radio sources, clusters of galaxies \cite{cluster1,cluster2,cluster3} and superclusters \cite{cluster4}.  In a nutshell, the problem of magnetogenesis rests on the hierarchies separating the diffusion distance of the intergalactic medium and the typical scale associated with the gravitational collapse of the protogalaxy \cite{gal6}. While the diffusivity scale in the interstellar medium is of the order of the AU ($1 \, \mathrm{AU} = 1.49 \times 10^{13} \,\, \mathrm{cm}$), magnetic fields are observed over much larger scales ranging between the $30\, \mathrm{kpc}$ and few $\mathrm{Mpc}$ ($1\, \mathrm{pc}= 3.08\times 10^{18} \, \mathrm{cm}$). 

The comoving scale associated with the 
gravitational collapse of the protogalaxy is of the order of the Mpc and it corresponds to 
(comoving) frequencies ${\mathcal O}(\nu_{g})$ where $\nu_{g} = 10\, \mathrm{fHz}$.  If the large-scale magnetic fields would have been produced at a topical moment during the decelerated stage of expansion, their maximal correlation scale would be bounded by the Hubble radius whose evolution is always {\em faster} than the correlation scale\footnote{For instance a magnetic field with typical correlation scale of the order 
of the Hubble radius at the electroweak epoch (i.e. approximately few cm)
corresponds to a cocoon of the order of the astronomical unit, at least for the 
conventional decelerated timeline of the concordance scenario where radiation dominates right after inflation. Although various ad hoc suggestions exist to increase this figure up to 100 AU, the final scales are anyway too small in comparison with the spatial region of the gravitational collapse of the protogalaxy.}.  By definition the problems related to cosmic magnetism involve then  distances that are (at least) of the order of the $\mathrm{Mpc}$ and 
the size of  the correlation scale makes it unlikely that the 
magnetic fields in clusters (or even superclusters) could be in any way the result 
of a specific mechanism operating inside the Hubble radius. The quantitative aspects of this 
conclusion  ultimately depend upon the decelerated timeline, as already argued long ago \cite{SYMM13,SYMM14}; when these suggestions have been originally formulated
the defining features of the concordance were much less clear than today. The purposes of the present investigation is thus to consider the interplay between invisible gravitons and large-scale magnetism since both problems involve frequencies between the aHz and and the fHz.
For the same reason the maxima of the spectral energy of the relic gravitons either in the ultra-high frequency domain or in the audio band pin down different postinflationary timelines 
that can be ultimately constrained by the magnetogenesis requirements. 
In the adiabatic paradigm (possibly complemented by an early stage of inflationary expansion) the large-scale gauge fields could be parametrically amplified and later behave as vector random fields.  One of the first concrete suggestions along this perspective has been the introduction of a pseudoscalar coupling \cite{SYMM4,SYMM5,SYMM6} not necessarily coinciding with the Peccei-Quinn axion \cite{SYMM7,SYMM8,SYMM9}. It has been later argued that the resulting action could be complemented by a direct coupling of the inflaton with the kinetic term of the gauge fields both in the case of inflationary and contracting Universes \cite{SYMM13,SYMM14} (see also \cite{SYMM10,SYMM11,SYMM12,SYMM15}). The origin of the scalar and of the pseudoscalar couplings may involve not only the inflaton but also some other  spectator field with specific physical properties \cite{SYMM15}. This class of problems together with their physical implications has been dubbed magnetogenesis in Ref. \cite{SYMM13} and we shall occasionally stick to this general terminology also in this paper. 

The viewpoint pursued in this investigation is that the relic gravitons and the gauge spectra can be mutually constrained when a decelerated stage of expansion precedes the conventional 
radiation dominated evolution. Along this perspective the layout of this paper is, in short, the following. Section \ref{sec2} is devoted to the low-frequency  effects of postinflationary stages expanding at rates that are either faster or slower than radiation. Section \ref{sec3} instead focuses on the evolution of the hyeprelectric and hypermagnetic fields when the gauge coupling is dynamical both during inflation and at later times. The impacts of the decelerated phases on the spectra of relic gravitons and of the hypermagnetic fields are considered in Sec. \ref{sec4} and in Sec. \ref{sec5} respectively. In Sec. \ref{sec4} we analyze the dependence of the signal upon the decelerated expansion rates. In the first part of Sec. \ref{sec5} the dependence of the hypermagnetic power spectra on  the different timelines is explicitly investigated with particular attention to the magnetogenesis requirements. In the second part of Sec. \ref{sec5} all the constraints deduced both from the relic gravitons and from the large-scale magnetism are combined together. Section \ref{sec6} contains a brief summation and the concluding considerations.
\renewcommand{\theequation}{2.\arabic{equation}}
\setcounter{equation}{0}
\section{Invisible gravitons}
\label{sec2} 
The uncertainties in the total number of $e$-folds are not a feature specifically associated with the dynamics of single-field inflationary models. In this sense the scrutiny of the decelerated 
timeline of the geometry is per se relevant. However, since the single field 
scenarios are compatible with the adiabaticity and with the Gaussianity of the large-scale inhomogeneities (and are directly constrained by observations) \cite{CONF2,CONF3,CONF4,CONF5,CONF6,CONF7,CONF8,CONF9,CONF10,CONF11}  for the present ends we are going to focus 
on the following tree-level effective action\footnote{The Greek and Latin (lowercase) indices run, respectively,  over the four space-time dimensions and over the three spatial dimensions. The signature of the four-dimensional metric $G_{\mu\nu}$ us mostly minus [i.e. 
$(+, \,, -\,, -\,, -)$]; the Ricci tensor follows from the contraction between the first and third indices of the Riemann tensor as $R_{\mu\nu} = R^{\alpha}_{\,\,\,\,\,\mu\alpha\nu}$.} 
\begin{equation}
S= \int d^{4}x \sqrt{- G} \biggl[ - \frac{R}{2 \ell_{P}^2} + \frac{1}{2} G^{\alpha\beta} \partial_{\alpha} \varphi \partial_{\beta} \varphi - V(\varphi)\biggr],
\label{LFP1}
\end{equation}
where $G_{\alpha\beta}$ indicates the four-dimensional metric tensor with determinant 
$G = \mathrm{det}\,G_{\alpha\beta}$; $\varphi$ is the inflaton field and 
$V(\varphi)$ denotes the related potential. The Planck length introduced in Eq. (\ref{LFP1}) is the 
inverse of the (reduced) Planck mass\footnote{Natural units $\hbar = c = k_{B} =1$ 
(where $k_{B}$ is the Boltzmann constant) are employed throughout; in these units $M_{P} = 1.22\times 10^{19} \mathrm{GeV}$.}
\begin{equation}
\ell_{P} = 1/\overline{M}_{P}, \quad \overline{M}_{P} = M_{P}/\sqrt{8\pi},
\label{PM}
\end{equation}
where $M_{P} =1.22 \times 10^{19} \, \mathrm{GeV}$. Equation  (\ref{LFP1}) is just the first term of a low energy description \cite{EFFAC1} and the higher derivatives potentially present in action are suppressed by the negative powers of a large mass $M_{eff}$ associated with the fundamental theory that underlies the effective Lagrangian. The first correction to Eq. (\ref{LFP1}) consists of all possible terms containing four derivatives involving the inflaton field, the Ricci scalar, the Riemann tensor and the scalar curvature. 

Following the analyses of Refs. \cite{EFFAC1,EFFAC2} the leading correction to Eq. (\ref{LFP1}) consists of $12$  terms [see also, in this respect, the section VI of Ref. \cite{EFFAC3} where slight differences in the counting appear in comparison with the logic of Refs. \cite{EFFAC1,EFFAC2}].  Among the $12$ aforementioned terms two break parity and may polarize the relic gravitons but their magnitude is anyway too small to be observable  \cite{EFFAC4}. The remaining terms control the corrections to the two-point functions and are conceptually relevant to establish the limitations 
of the effective description of Eq. (\ref{LFP1}); these aspects will not play a direct role in the forthcoming discussions but have a direct counterpart in the analysis of the gauge fields (see the initial part of Sec. \ref{sec3}). 

Depending upon the properties of $V(\varphi)$, the tensor to scalar ratio (denoted by
$r_{T}$ in what follows)  may exhibit different scaling properties as a function of the number 
of $e$-folds $N_{\nu}$ elapsed since the frequencies $\nu= {\mathcal O}(\nu_{p})$ 
were of the order of the Hubble rate during inflation. This stage will be referred to as 
the horizon crossing although this popular locution not completely inaccurate 
and has nothing to do with causality (see, for instance, Ref. \cite{CONF1a}). 
Although the previous observation does not fix the value of $N_{\nu}$,
it is customary to assume that $N_{\nu} = \overline{N}_{\nu} = {\mathcal O}(60)$ but this estimate is valid 
provided expansion timeline is dominated by radiation between the end of inflation and the equality 
time \cite{CONF2,CONF3,CONF4,CONF5,CONF6,CONF7,CONF8,CONF9,CONF10,CONF11} (see also \cite{CONF12}). As we are going to see both the shape of the  potential and the decelerated evolution contribute to the  suppression of $r_{T}(\nu)$ where $\nu$ denotes, as explained in Sec. \ref{sec1}, 
 the comoving frequency.

\begin{figure} 
\begin{center}
\includegraphics[width=8cm,height=6cm]{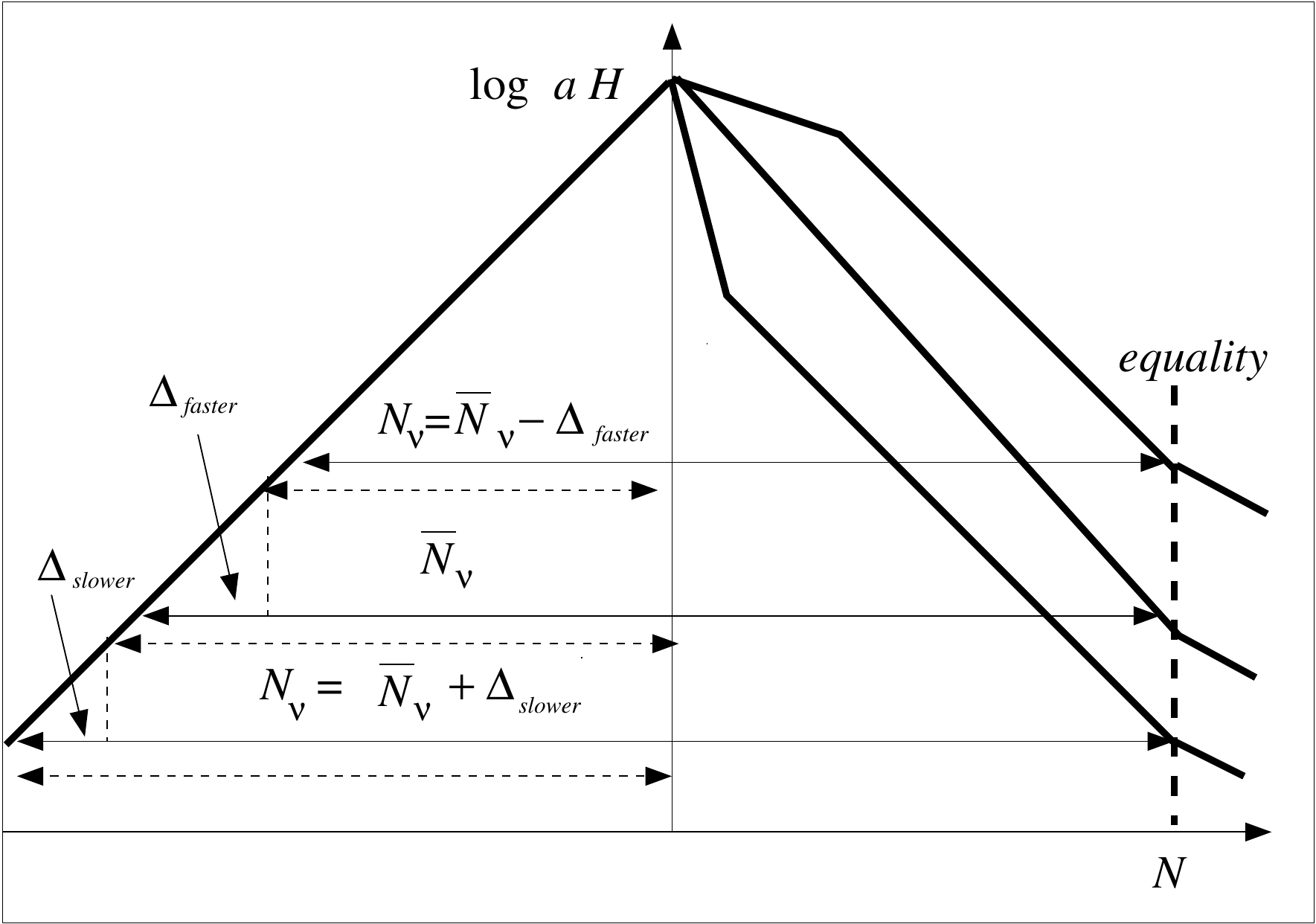}
\end{center}
\caption{\label{FIGURE1} The common logarithm of the (comoving) expansion rate 
$a \, H$ is illustrated as a function of the number of $e$-folds. During the inflationary stage 
in the leftmost region the plot $a\, H$ is proportional to the scale factor. In the 
radiation stage $a\,H$ scales instead as $a^{-1}$ and the number of $e$-folds elapsed since the crossing of $\nu= {\mathcal O}(\nu_{p})$ is ${\mathcal O}(60)$. If the postinflationary 
evolution is  slower than radiation $N_{\nu} < \overline{N}_{\nu}$; the opposite is true 
if the expansion rate after inflation is faster than radiation.  The crossing time 
$\tau_{\nu}$ (frequently mentioned in the discussion) occurs when a given frequency 
$\nu$ approximately equals the comoving Hubble rate during inflation.}
\end{figure}

In the cartoon of Fig. \ref{FIGURE1} the variation of the number of $e$-folds 
elapsed since the crossing of the frequencies $\nu = {\mathcal O}(\nu_{p})$ is schematically illustrated. On a physical ground three kinds of timelines must be distinguished: {\it (i)}
 when the postinflationary evolution only consists of a radiation stage the number of $e$-folds from the crossing time is conventionally indicated by $\overline{N}_{\nu}$ and it is approximately ${\mathcal O}(60)$; {\it (ii)} if the expansion rate {\em after} inflation is faster than radiation the upper curve at the right hand side of Fig. \ref{FIGURE1} demonstrates that the value of $N_{\nu}$ is comparatively smaller and it is given by $N_{\nu} = \overline{N}_{\nu} - \Delta_{faster} < 60$; {\it (iii)}  finally, when the postinflationary expansion rate is slower than radiation (see the 
lower line at the right hand side of Fig. \ref{FIGURE1}) $N_{\nu} = \overline{N}_{\nu} +  \Delta_{slower}$ and the number of $e$-folds elapsed since the crossing of the frequencies $\nu = {\mathcal O}(\nu_{p})$ gets larger than in the radiation-dominated evolution.

\subsection{The number of $e$-folds}
By definition the number of $e$-folds elapsed between the crossing time of a given comoving frequency 
 $\nu$ and the final stages of inflation is given by
\begin{equation}
N_{\nu} = \int_{\tau_{\nu}}^{\tau_{f}} {\mathcal H} \, d\tau, \quad {\mathcal H} = a\, H,
\label{INV1}
\end{equation}
where $\tau$ denotes the conformal time coordinate and 
$H$ is the expansion rate. In Eq. (\ref{INV1}) we also introduced the standard notation ${\mathcal H} = a^{\prime}/a$ where $a$ is the scale factor and the prime denotes a derivation with respect to $\tau$.
The value of the crossing time $\tau_{\nu}$ follows from the condition
$ 2 \pi \, \nu \, \tau_{\nu} = {\mathcal O}(1)$ while $\tau_{f}$ indicates the end of inflation. During the 
inflationary stage $\epsilon \ll 1$ where $\epsilon = - \dot{H}/H^2$ and the overdot denotes 
a derivation with respect to  $t$; we recall that the cosmic time coordinate $t$ and the conformal time $\tau$ are  
related as $ a(\tau) d\tau = d\, t$. 

Since inflation ends
when $\epsilon(\tau_{f}) \to 1$ (i.e.  $\dot{H} \to - H^2$) for single-field scenarios we have $V(\varphi_{f}) = \dot{\varphi}_{f}^2$ and the expression of Eq. (\ref{INV1}) can also be phrased in terms of the excursion of the inflaton field
\begin{equation}
N_{\nu} =(1/\overline{M}_{P}^{2})\,\int_{\varphi_{f}}^{\varphi_{\nu}} d\varphi\,\,(V/V_{,\varphi}),\quad V_{,\varphi} = \partial_{\varphi} \, V.
\label{INV2}
\end{equation}
Once the class of potentials governing the dynamics is specified,  
Eq. (\ref{INV2}) relates directly $\varphi_{\nu} = \varphi(\tau_{\nu})$ and $N_{\nu}$. The same is true also for 
the slow-roll parameters\footnote{Within the present notations the slow-roll parameters are $\epsilon(\tau) =(V_{,\varphi}/V)^2\,  \overline{M}_{P}^2 /2$, $\overline{\eta}(\tau) = (V_{,\varphi\varphi}/V)\overline{M}_{P}^2$
and so on. } at the crossing time $\tau_{\nu}$, i.e. 
 $\epsilon_{\nu} = \epsilon(\tau_{\nu})$ and $\overline{\eta}_{\nu}= \overline{\eta}(\tau_{\nu})$.
A further slow-roll parameter [i.e. $ \eta(\tau) = \ddot{\varphi}/(H\, \dot{\varphi})$] is often introduced but
it is expressed as a function of $\epsilon_{\nu}$ and $\overline{\eta}_{\nu}$ (i.e. $\eta_{\nu} = \epsilon_{\nu} - \overline{\eta}_{\nu}$). For different classes of potentials 
Eqs. (\ref{INV1})--(\ref{INV2}) imply different scalings for the slow-roll parameters 
with $N_{\nu}$; we then conclude that the effective suppression of $r_{T}$ 
is a combination of the shape of the potential {\em and} of the decelerated timeline.

\subsection{The quantum normalization}
An accelerated stage of expansion suppresses the spatial gradients eventually present during
the protoinflationary epoch \cite{INCON1,INCON2,INCON3} and after few $e$-folds the only source of the gauge-invariant curvature inhomogeneities is provided by the zero-point fluctuations of the corresponding quantum fields:
\begin{equation}
\widehat{{\mathcal R}}(\vec{x},\tau) = \int \frac{d^{3} k}{(2\pi)^{3/2} z_{\varphi}(\tau)}\, \bigl[ \widehat{a}_{\vec{k}}\,\,f_{k}^{(s)}e^{- i \vec{k}\cdot\vec{x}} + \mathrm{H.c.}\bigr],
\label{CURV1}
\end{equation}
where $[ \widehat{a}_{\vec{k}}, \, \widehat{a}_{\vec{p}}^{\dagger}] = \delta^{(3)}(\vec{k} - \vec{p})$ while 
$f_{k}^{(s)}= f_{k}^{(s)}(\tau)$ is the scalar mode function.  The stenographic notation ``H. c.'' of Eq. (\ref{CURV1}) indicates the Hermitian conjugate of the first term inside the square bracket; 
for the sake of conciseness we also write $z_{\varphi} = z_{\varphi}(\tau) = {\mathcal H} \varphi^{\prime}/a$. In full analogy with Eq. (\ref{CURV1}) the expansion valid for the transverse 
and solenoidal quantum fields describing the tensor modes of the geometry is given by\footnote{Consistently with the current 
observational determinations \cite{CONF1,CONF2,CONF3,CONF4,CONF5,CONF6,CONF7,CONF8,CONF9,CONF10,CONF11,CONF12}, we consider here a conformally flat background geometry; the conditions defining the solenoidal and traceless modes of the 
geometry read, in this case, $\partial_{i} \widehat{h}^{i}_{\,\,j} =0$ and $\widehat{h}_{i}^{\,\,\,i}=0$.}
\begin{eqnarray}
&&\widehat{h}_{i\,j}(\vec{x},\tau) = \frac{\sqrt{2} \, \ell_{P} }{(2\pi)^{3/2} a(\tau)} \sum_{\alpha=\oplus,\otimes}\int d^{3}\,k 
\nonumber\\
&&\times  e^{(\alpha)}_{i\,j}(\hat{k})\bigl[ \widehat{a}_{\vec{k},\,\alpha} \, f_{k,\alpha}^{(t)} e^{- i \vec{k}\cdot\vec{x}} + \mathrm{H.c.}\bigr],
\label{TENS1}
\end{eqnarray}
where the index $\alpha=\oplus,\,\otimes$ runs over the two tensor polarizations defined by $e^{(\oplus)}_{i\,j}(\hat{k}) = (\hat{m}_{i} \, \hat{m}_{j} 
- \hat{n}_{i}\, \hat{n}_{j})$ and by $e^{(\otimes)}_{i\,j}(\hat{k}) = (\hat{m}_{i} \, \hat{n}_{j} 
+ \hat{n}_{i}\, \hat{m}_{j})$ ($\hat{m}$, $\hat{n}$ and $\hat{k}$ are three unit vectors satisfying $\hat{m} \times \hat{n} = \hat{k}$). The tensor mode functions\footnote{Note that $f_{k,\oplus}^{(t)} =f_{k,\otimes}^{(t)}= f_{k}^{(t)}$ in the unpolarized case.} appearing in Eq. (\ref{TENS1}) are denoted by  $f_{k,\alpha}^{(t)} =  f_{k,\alpha}^{(t)}(\tau)$. The corresponding power spectra are obtained from the expectation values of two field operators for spatially separated points (but at the same conformal time) over the initial vacuum state i.e.\begin{eqnarray}
\hspace{-1cm}
&& \langle \widehat{{\mathcal R}}(\vec{x},\tau)\,\widehat{{\mathcal R}}(\vec{x}+ \vec{r},\tau) \rangle = \int_{0}^{\infty}  P_{{\mathcal R}}(k,\tau) j_{0}(k\, r) dk/k,
\label{PS1}\\
\hspace{-1cm}
&& \langle \widehat{h}_{ij}(\vec{x},\tau)\,\widehat{h}^{ij}(\vec{x}+ \vec{r},\tau) \rangle = \int_{0}^{\infty} P_{T}(k,\tau) j_{0}(k\, r)dk/k,
\label{PS2}
\end{eqnarray}
where $j_{0}(k\,r)$ is the spehrical Bessel function of zeroth order \cite{abr1,abr2} while $P_{{\mathcal R}}(k,\tau)$ and $P_{T}(k,\tau)$ denote, respectively, the scalar and tensor power spectra  $P_{{\mathcal R}}(k,\tau) = k^3 \bigl| f_{k}^{(s)}(\tau)\bigr|^2/(2\pi^2)$ and $P_{T}(k,\tau) = 4 \ell_{P}^2 k^3 \bigl| f_{k}^{(t)}(\tau)\bigr|^2/\pi^2$.  The tensor to scalar ratioI follows from the quotient of $P_{T}(k,\tau)$ and $P_{{\mathcal R}}(k,\tau)$, i.e. $r_{T}(\nu,\tau)= P_{T}(\nu,\tau)/P_{{\mathcal R}}(\nu,\tau)$. The frequency $\nu$ crosses the horizon 
when $\tau \to \tau_{\nu}$ so that the corresponding value of $r_{T}(\nu,\tau)$ becomes
\begin{equation}
r_{T}(\nu,\tau_{\nu})\to 8 \ell_{P}^2\, a(\tau_{\nu})/z_{\varphi}(\tau_{\nu}) = 16 \epsilon_{\nu}.
\label{TS1}
\end{equation}
To comply with standard practice (and to avoid the proliferation of arguments) the following notation will be adopted $r_{T} = r_{T}(\nu,\tau_{\nu})$. Up to numerical factors, $r_{T} = {\mathcal O}(1/N_{\nu})$ 
in the case of monomial potentials. Conversely $r_{T} = {\mathcal O}(1/N_{\nu}^2)$ for plateau-like potentials 
and even more complicated scalings may arise (see, for instance, the illustrative examples at the end of this section). Depending on  $N_{\nu}$ the suppression of $r_{T}$ can be substantially different.

\subsection{The actual values of $N_{\nu}$}
The qualitative viewpoint conveyed in Fig. \ref{FIGURE1} shall now be scrutinized quantitatively with the purpose of deriving the dependence of $N_{\nu}$ upon the rates and the durations of the 
postinflationary stages.  Thanks to Eq. (\ref{INV1}) the number of $e$-folds elapsed since the crossing of the frequencies $\nu = {\mathcal O}(\nu_{p})$ (where $\nu_{p} = k_{p}/(2\pi) = 3.09 \, \mathrm{aHz}$) can be explicitly computed and from the condition $2 \pi \nu \tau_{\nu} = 2\pi \nu/(a_{\nu} H_{\nu})=  {\mathcal O}(1)$ we have
\begin{equation}
\biggl(\frac{\nu}{\nu_{0}}\biggl) \biggl(\frac{a_{0} H_{0}}{a_{r} H_{r}}\biggr) \biggl(\frac{a_{r} H_{r}}{a_{f} H_{f}}\biggr) \biggl(\frac{a_{f} H_{f}}{a_{\nu} H_{\nu}}\biggr)= {\mathcal O}(1),
 \label{NN1}
 \end{equation}
 where the subscripts {\em r} and {\em f} denote, respectively, the onset of the radiation epoch and final stages of inflation 
 defined by the conditions established prior to Eq. (\ref{INV2}); the value of $\nu_{0}$ is related to the current 
 value of the Hubble rate, i.e. $\nu_{0} = H_{0}/(2\pi) = {\mathcal O}(\mathrm{aHz})$. Since 
  $N_{\nu} = \ln{(a_{end}/a_{\nu})}$, the following general expression can be obtained\footnote{The $\ln$ denotes throughout
the natural (or Neperian) logarithm; the $\log$ indicates instead the common logarithm (i.e. $\log = \log_{10}$).}  \cite{MGINV}:
 \begin{equation} 
N_{\nu} = \overline{N}_{\nu} + \frac{1}{2} \sum_{\ell=1}^{n -1} \biggl(\frac{\delta_{\ell} -1}{\delta_{\ell} +1}\biggr) \, \ln{(H_{\ell+1}/H_{\ell})}.
\label{NN2}
\end{equation}
The first contribution appearing in Eq. (\ref{NN2}) (denoted by $\overline{N}_{\nu}$) gives the 
number of $e$-folds computed during a radiation stage extending between the end of inflation and the equality time (see also Fig. \ref{FIGURE1}). The second contribution at the right hand side of Eq. (\ref{NN2})  follows from the modified decelerated evolution where, prior 
to matter-radiation equality, there are $n$ successive stages with a progressively decreasing rate (i.e. $H_{\ell+1}/H_{\ell} < 1$). 
When $n = 1$ a single radiation dominated stage extends between the end of inflation and the equality time; in this situation $N_{\nu} = \overline{N}_{\nu}$. However, if $n = 2$ the conventional radiation epoch is complemented at early time by a second intermediate stage of expansion taking place between the end of the inflationary phase and the BBN time. More complicated situations\footnote{As an example when $n = 3$ there will be two successive stages of expansion preceding 
the conventional radiation-dominated phase.} are equally described by Eq. (\ref{NN2}) where $\delta_{\ell}$ (with $\ell = 1,\,.\,.\,.\,n-1$) indicates the expansion rate in each of the different stages. 

The example illustrated in Fig. \ref{FIGURE1} 
corresponds to the case $n=2$: in the first decelerated stage of expansion the values of $\delta$ are both larger and smaller than $1$ while in the second phase (coinciding with radiation) $\delta \to 1$. When 
$\delta_{\ell} \to 1$ (for {\em all} the different $\ell$) the whole postinflationary evolution collapses to a single radiation phase since $N_{\nu}$ equals $\overline{N}_{\nu}$. The first contribution to Eq. (\ref{NN2}) follows from Eq. (\ref{NN1}) when $a_{f}$ and $a_{r}$ coincide and it is given by:
\begin{equation} 
e^{\overline{N}_{\nu}} = (2 \, \Omega_{R0})^{1/4} d(g_{s}, \, g_{\rho})\,\sqrt{H_{\nu}/H_{0}}\, (\nu_{0}/\nu),
\label{NN3}
\end{equation}
where $\Omega_{R0}$ is the current radiation fraction in critical units 
and  $d(g_{s}, \, g_{\rho}) = (g_{s,\, eq}/g_{s,\, r})^{1/3} \, (g_{\rho,\, r}/g_{\rho,\,eq})^{1/4}$
accounts for the evolution of the number of relativistic species between the onset 
of the radiation epoch and the equality time\footnote{This term follows from the radiation-dominated evolution between $a_{r}$ and $a_{eq}$ so that $(H_{r}/H_{eq})^{1/2}= (a_{eq}/a_{r}) d(g_{s}, g_{\rho})$. In a stage of local thermal equilibrium, the entropy density is conserved and the total energy density depends on $g_{\rho}$ (i.e. the number of relativistic degrees of freedom in the plasma) while $g_{s}$ denotes the effective number of relativistic degrees of freedom appearing in the entropy density. In the standard situation where $g_{s,\, r}= g_{\rho,\, r} = 106.75$ and $g_{s,\, eq}= g_{\rho,\, eq} = 3.94$ we have that $d(g_{s}, \, g_{\rho})= 0.75$.}. The contribution of $d(g_{s}, \, g_{\rho})$ to Eq. (\ref{NN3}) is conceptually relevant but numerically not essential for the determination of $\overline{N}_{\nu}$ whose explicit value depends instead upon $H_{\nu}/M_{P} = \sqrt{\pi \, \epsilon_{\nu} \, {\mathcal A}_{{\mathcal R}}}$ (i.e. the Hubble rate at the crossing time). We recall, in this respect,
that at $\tau_{\nu}$ the power spectrum of curvature inhomogeneities given in Eq. (\ref{PS1}) can be explicitly written as $P_{{\mathcal R}}(\tau_{\nu}) = (H_{\nu}^2/M_{P}^2)/(\pi \epsilon_{\nu})$. Moreover $P_{{\mathcal R}}(\tau_{\nu}) = {\mathcal A}_{\mathcal R} = 2.41\times 10^{-9}$ for $\nu = {\mathcal O}(\nu_{p})$ since the parametrization of the scalar power spectrum adopted here corresponds to 
\begin{equation}
P_{{\mathcal R}}(\nu) = {\mathcal A}_{{\mathcal R}} (\nu/\nu_{p})^{n_{s} -1}, \quad n_{s} = 1 - 6 \epsilon_{\nu} + 2 \overline{\eta}_{\nu},
\label{NN4}
\end{equation}
where $n_{s}$ is the scalar spectral index of curvature inhomogeneities expressed in terms of the slow-roll parameters at Hubble crossing. 

After keeping track of the actual numerical values of all the factors entering Eq. (\ref{NN3}), the value of $\overline{N}_{\nu}$ is
\begin{eqnarray}
\hspace{-0.8cm}
&&\overline{N}_{\nu} = 59.4  + \frac{1}{4} \ln{\biggl(\frac{\epsilon_{\nu}}{0.001}\biggr)} +\frac{1}{4} \ln{\biggl(\frac{{\mathcal A}_{{\mathcal R}}}{2.41\times 10^{-9}}\biggr)}  
\nonumber\\
\hspace{-0.8cm}
&&+  \ln{d(g_{s}, \, g_{\rho})} - \ln{\biggl(\frac{\nu}{\nu_{p}}\biggr)}
+   \frac{1}{4} \ln{\biggl(\frac{h_{0}^2 \, \Omega_{R0}}{4.15\times 10^{-5}}\biggr)}.
\label{NN5}
\end{eqnarray}
For $H_{\ell+1} < H_{\ell}$, when  all the $\delta_{\ell}$ are smaller than $1$  Eqs. (\ref{NN2})--(\ref{NN5}) suggest that $N_{\nu} > \overline{N}_{\nu} = {\mathcal O}(60)$; this is because the second contribution at the right hand side of Eq. (\ref{NN2}) is always positive. In the opposite situation (i.e. $\delta_{\ell} >1$ for all $\ell$) the supplementary contribution in Eq. (\ref{NN2}) is negative so that $N_{\nu} < \overline{N}_{\nu} = {\mathcal O}(60)$. In case 
the $\delta_{\ell}$ are both positive and negative what counts is the amount of time where the expansion rate is, in an averaged sense, either slower or faster than radiation.
 \begin{figure} 
\begin{center}
\includegraphics[width=8cm,height=6cm]{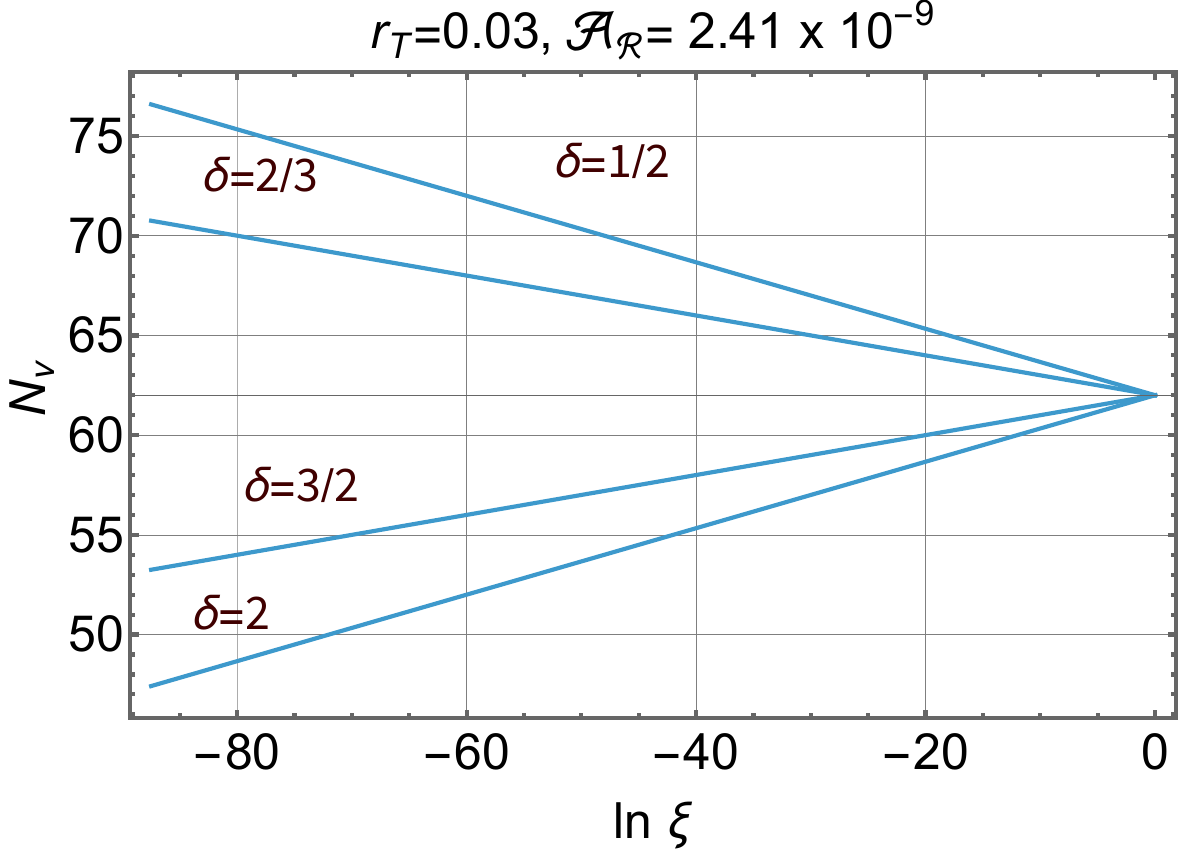}
\end{center}
\caption{\label{FIGURE2} The parameters of a single postinflationary stage preceding the radiation epoch (i.e. $n=2$ in Eq. (\ref{NN2})) are illustrated. On the vertical axis $N_{\nu}$ is plotted while on the horizontal axis the natural logarithm of $\xi = H_{r}/H_{\nu}$ is reported. As expected when $\delta > 1$ we have $N_{\nu} < \overline{N}_{\nu}$ while for $\delta < 1$ we have instead $N_{\nu} > \overline{N}_{\nu}$. To comply with the late-time constraints we must require that $H_{r}$ always exceeds the 
expansion rate at BBN and this implies that  $H_{r} \geq 10^{-44} M_{P}$ where $H_{\nu}$ is estimated from
$H_{\nu}/M_{P} = \sqrt{\pi {\mathcal A}_{{\mathcal R}} r_{T}}/4$. By looking at the maximal excursions on $N_{\nu}$ on the vertical axis it follows that, in practice, $N_{\nu} = \overline{N}_{\nu} \pm {\mathcal O}(15)$. We stress that, on the horizontal axis, we illustrate the {\em natural} logarithm of $\xi$: since $\xi \geq  {\mathcal O}(10^{-38})$ we also have that $\ln{\xi} \geq - {\mathcal O}(87)$, and this fixes the lower limit of the horizontal axis.}
\end{figure}

 The allowed values of $N_{\nu}$ are illustrated in Fig. \ref{FIGURE2} for a single 
 postinflationary stage. When more than one stage is present what counts is the maximal excursion of $N_{\nu}$; this 
quantity can be estimated  when all the $\delta_{\ell}$ collapse to a single value (i.e. $\delta$) and, in this way, Eq. (\ref{NN2}) reduces to
\begin{equation} 
N_{\nu} = \overline{N}_{\nu} + \,\alpha(\delta) \ln{(H_{r}/H_{\nu})}, 
\label{NN6}
\end{equation}
where the variable $\alpha(\delta) = (\delta -1)/[2 (\delta +1)]$ (repeatedly mentioned in the forthcoming considerations) has been introduced. The maximal and the minimal values of $N_{\nu}$ depend both on $(H_{r}/H_{\nu})$ and on $\alpha(\delta)$. Since $H_{r}$ indicates the expansion rate at radiation dominance,  its minimal 
value is provided by $H_{r}^{(min)}=  {\mathcal O}(10^{-44}) \, M_{P}$ where  
 $H_{r}^{(min)}$ is obtained from the typical expansion 
rate at the epoch of big-bang nucleosynthesis (BBN). The maximal 
value of $H_{\nu}$ can be instead estimated as $ H^{(max)}_{\nu} \simeq \sqrt{ \pi {\mathcal A}_{{\mathcal R}} \, \epsilon_{\nu}}\, M_{P}$ with $\epsilon_{\nu} = r_{T}/16$. This means 
that an upper limit on $\ln{(H_{\nu}/H_{r})}$ is about $88$ so that, broadly speaking,
$\ln{(H_{r}/H_{\nu})} = - {\mathcal O}(90)$. From Eq. (\ref{NN6}) we 
can then estimate $N_{\nu} = \overline{N}_{\nu} - {\mathcal O}(90) \alpha(\delta)$.
In the case of perfect barotropic fluids the value of $\alpha(\delta)$ eventually 
depends on the barotropic index $w$ as $\alpha(w) = (1- 3 w)/[6 (1+ w)]$. 

Barring for more exotic requirements,  ordinary matter must obey all the energy conditions, so that $w$ eventually ranges between $0$ and $1$ and this 
consideration implies that $-1/\leq \alpha\leq 1/6$. The maximal and minimal values of $N_{\nu}$ are therefore 
given by $N_{\nu}^{(min)} = \overline{N}_{\nu} + {\mathcal O}(15) = {\mathcal O}(75)$ 
and by $N_{\nu}^{(min)} = \overline{N}_{\nu} - {\mathcal O}(15) = {\mathcal O}(55)$.
These results also clarify the cartoon of Fig. \ref{FIGURE1}: a stage expanding
faster than radiation (i.e. $\delta > 1$) reduces the number of $e$-folds 
elapsed since the crossing of the frequencies $\nu = {\mathcal O}(\nu_{p})$;
the opposite is true when the expansion is slower than radiation since, in this 
case, ${\mathcal O}(60) < N_{\nu} < {\mathcal O}(75)$. The values of 
$\epsilon_{\nu}$ are comparatively more suppressed if the 
decelerated timeline expands, for a certain period at a rate slower than radiation. 
For the sake of illustration in Fig. \ref{FIGURE2} we also plot $N_{\nu}$ as a function of $\ln{\xi}$. In Fig. \ref{FIGURE2} the values of $N_{\nu}^{(max)}$ and $N_{\nu}^{(min)}$ correspond to the two straight lines with $\delta = 1/2$ and $\delta =2$.  

\subsection{Timeline and potentials}
According to the previous discussion the explicit value of $N_{\nu}$  (and the consequent suppression or enhancement of $r_{T}$) is determined from a timeline that spans $38$ orders of magnitude between ${\mathcal O}(10^{-6}) M_{P}$ and ${\mathcal O}(10^{-44})\, M_{P}$. Different decelerated stages leave specific signatures both in the spectrum of relic gravitons and in other phenomena like the ones associated with large-scale magnetism. This model-independent perspective can be complemented by particular classes of potentials that may effectively lead to a modified postinflationary history. For instance if the reheating stage is delayed by a long phase dominated by the coherent oscillations of the inflaton (as suggested, with various motivations, in Refs. \cite{POT1,POT2,POT3}) the radiation dominance is preceded by an epoch expanding faster than radiation\footnote{In this case, however,  the total number of $e$-folds gets smaller than in the radiation case (i.e. $N_{\nu} < {\mathcal O}(60)$) and $r_{T} > 0.03$; it is also possible to get to $r_{T} = {\mathcal O}(0.2)$ as suggested by the 
 Bicep2 collaboration \cite{POT8} in an attempt to interpret what turned to be, after a more careful analysis, a foreground contamination.}.  Another possibility is a stage dominated by the kinetic energy of the scalar field; in this situation the intermediate phase expands at a rate slower than radiation, as it happens in the case of quintessential inflationary scenarios \cite{quint1,quint2} (see also \cite{MGST1,MGST2,MGST3} and \cite{quint3,quint4}); see also \cite{quint5} for an extended review. Since the monomial potential do not suppress enough $r_{T}$, plateau-like potentials are more promising: in this second case the inflationary limit of the potential corresponds to $V(\varphi) \to M^4$ for $\Phi\gg 1$ where $\Phi= \varphi/\overline{M}_{P}$; the mass $M$ fixes the scale of the potential (see \cite{MGINV} and references therein). Overall it is always possible to parametrize $V(\varphi)$ as 
 $V(\varphi) =  M^4 \,\, v(\Phi)$ where
 \begin{equation}
\lim_{\Phi\gg 1} v(\Phi) =1, \qquad \lim_{\Phi\ll 1} v(\Phi) \propto \Phi^{2 q}.
\label{PP1}
\end{equation}
While different forms of $v(\Phi)$ can be envisaged a rather general parametrization 
involves the ratio of two functions  approximately scaling with the same power of $\Phi$ for $\Phi \gg 1$. Given a specific form of $v(\Phi)$ the property spelled out in Eq. (\ref{PP1}) guarantees that for $\Phi \ll 1$ the coherent oscillations of the inflaton could trigger an extended stage of expansion where the energy density $\rho_{\Phi}$ of the scalar field 
is approximately constant \cite{OSC1,OSC2,OSC3,OSC4}
\begin{equation}
\rho_{\Phi} = \overline{M}_{P}^2 \dot{\Phi}^2/2 + M^4 v(\phi), \quad 3 H  \overline{M}_{P}^2 \dot{\Phi}^2 \ll \dot{\rho}_{\Phi},
\label{PP2}
\end{equation}
where, as usual the overdot denotes a derivation with respect to the conformal 
time coordinate.  In the case of Eq. (\ref{PP2}) we also have   $\overline{M}_{P}^2\,\dot{\Phi}^2 = 2 M^4 (v_{max} - v)$ where $v_{max} = v(\Phi_{max})$. If we then average over the period of oscillation we also deduce
\begin{equation} 
- \dot{H}/H^2 = \frac{3 \int_{0}^{1} \sqrt{ 1 - y^{2q}} dy}{ \int_{0}^{1} dy/\sqrt{ 1 - y^{2q}}} ,
\label{PP3a}
\end{equation}
where $y = \Phi/\Phi_{max}$. After performing explicitly the integrals in Eq. (\ref{PP3})
we obtain $- \dot{H}/H^2=  3 \,q/(q +1)$ and this also means that  $\delta = (q+1)/(2 q -1)$.
\begin{figure} 
\begin{center}
\includegraphics[width=8cm,height=6cm]{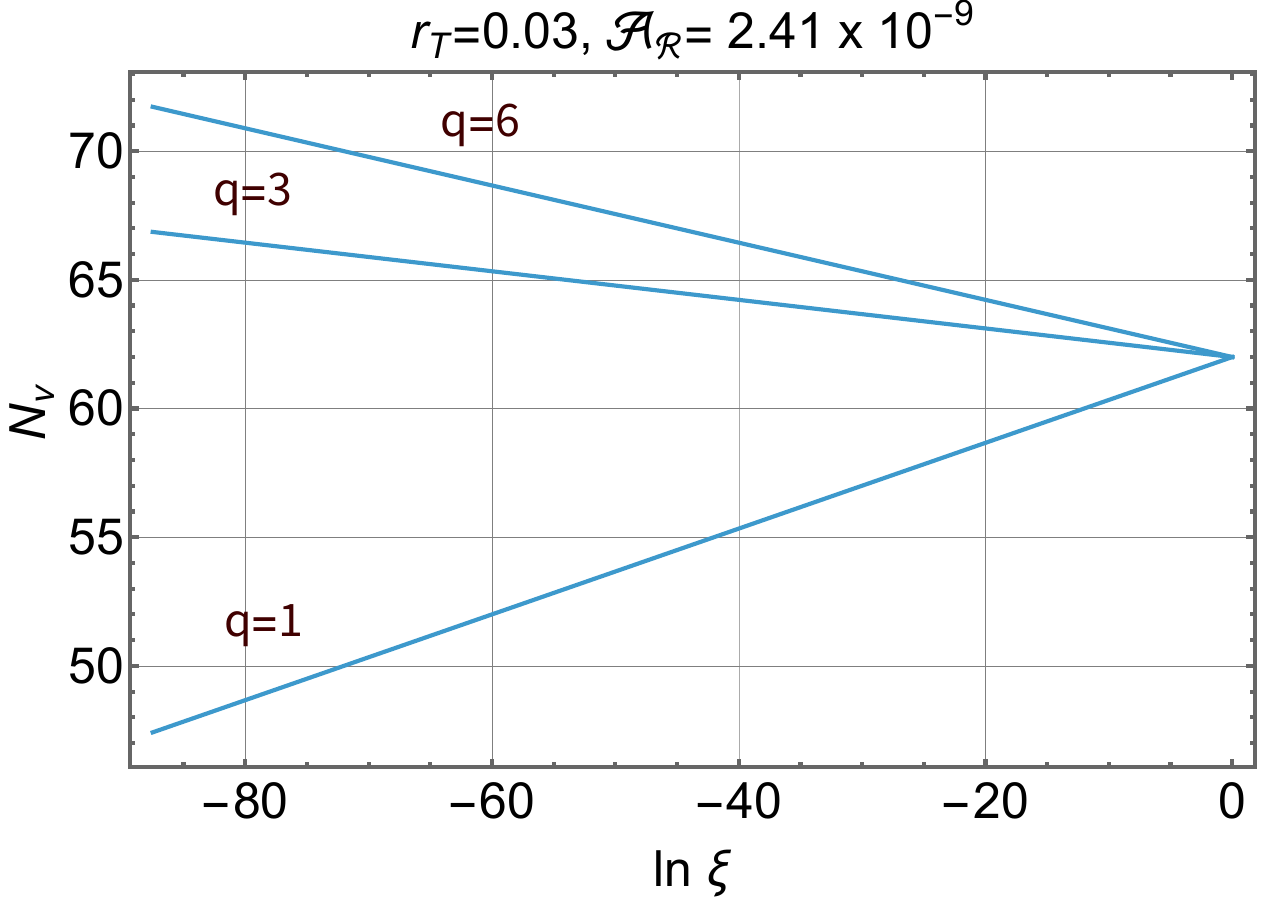}
\end{center}
\caption{\label{FIGURE3} As in Fig. \ref{FIGURE2} $N_{\nu}$ is illustrated as a function 
of $\ln{\xi}$ for different values of $q$ (see Eq. (\ref{PP3a}) and discussion thereafter).}
\end{figure}
The condition $\delta \geq 1$ implies that $0< q \leq 2$; furthermore for $q \gg 1$ the asymptote is $\delta \to 1/2$ exactly as in the case of a stiff background dominated by the kinetic energy of the inflaton. In  we illustrate again The maximal excursion of $N_{\nu}$ (already discussed in Fig. \ref{FIGURE2})
is further illustrated in Fig. \ref{FIGURE3} where the $q$-dependence, 
possibly arising as a consequence of the coherent oscillations of the inflaton (see Eqs. (\ref{PP2})--(\ref{PP3a})),
is specifically analyzed. Overall the obtained results suggest that different potentials may lead to decelerated timelines eventually modifying the total number of $e$-folds as illustrated in Fig. \ref{FIGURE1}. This phenomenon is however more general and not necessarily related to the shapes and properties of the inflationary potentials in the small field limit.

\subsection{A class of illustrative examples}
A concrete class of potentials satisfying the conditions (\ref{PP1})--(\ref{PP2})
can be constructed by combining monomial potentials
\begin{equation}
v(\Phi) =\beta^{p} \Phi^{2 q}/[ 1 + \beta^2\Phi^{4 q/p}]^{p/2},
\label{PP3}
\end{equation}
where, for the sake of simplicity, we require that $4 q > p$ and $\beta>0$. From the expression (\ref{PP3}) it follows that a $q$-dependent  oscillating stage occurs for $\Phi \ll 1$; in this limit the potential can be written as $v(\Phi) = \beta^{p} \Phi^{2 \, q}$. From Eq. (\ref{PP3}) the explicit expressions of the tensor-to-scalar ratio and the scalar spectral index become:
\begin{eqnarray}
\hspace{-0.67cm}
&& r_{T}(\Phi)= \frac{ 32 \, q^2}{\Phi^2 (1 + \beta^2 \Phi^{4q/p})^2}, 
\nonumber\\
\hspace{-0.67cm}
&& n_{s}(\Phi) = 1 - \frac{4\, p \, q( 1+ q) + 4 q ( q + 4 p) \beta^2 \Phi^{4q/p}}{p \Phi^2 ( 1 + \beta^2 \Phi^{4q/p})^2}.
\label{PP7}
\end{eqnarray}
Recalling Eqs. (\ref{INV1})--(\ref{INV2}), $N_{\nu}$ becomes:
\begin{equation}
N_{\nu} =\int_{\Phi_{f}}^{\Phi_{\nu}} \Phi\biggl( 1 + \beta^2 \Phi^{4 q/p}\biggr)/(2 q) \, d \Phi,
\label{PP8}
\end{equation}
where $\Phi_{\nu}= \Phi(\tau_{\nu})$ denotes the value of the field when the 
frequency  $\nu$ crosses the comoving Hubble radius while  $\Phi_{f} \to 1$ coincides with the 
end of inflation\footnote{In terms of $\Phi$ the conditions discussed after Eq. (\ref{INV1}) imply $\epsilon(\Phi_{f}) \to 1$ and $H^2 = - \dot{H}$. The condition $\epsilon(\Phi_{f})=1$ also demands 
$\Phi_{f}^2 \bigl( 1 + \beta^2 \Phi_{f}^{4 q/p}\bigr)^2 = 2 q^2$.
When $\beta < 1$ then  $\Phi_{f} \simeq 1/(\sqrt{2}\, q)$ and this is a quantity ${\mathcal O}(1)$. For $\beta>1$ we get instead $\Phi_{f} \simeq (\sqrt{2}\, q\, \beta^2)^{q/(4 p +q)}$ which is however 
also of order $1$. }. The value of $N_{\nu}$ for $\nu = {\mathcal O}(\nu_{p})$ is then given by: 
\begin{equation}
N_{\nu} = p \, \beta^2 \, \bigl( \Phi_{\nu}^{2 + 4 q/p} -1\bigr)/[4 q ( p + 2 q)],
\label{PP10}
\end{equation}
Since  $\Phi_{\nu}$ corresponds to the crossing of the bunch of frequencies $\nu = {\mathcal O}(\nu_{p})$
during inflation we may evaluate Eq. (\ref{PP10}) for  $\Phi_{\nu} \gg 1$ and obtain 
$N_{\nu} = p \, \beta^2/[4 q\, ( p + 2 q)]  \Phi_{\nu}^{2 + 4q/p}$. Thus, thanks to the previous results  we may 
finally obtain the suppression of  $n_{s}$, $r_{T}$ and $n_{T}$ after trading $\Phi_{\nu}$ for $N_{\nu}$ 
in Eq. (\ref{PP7}):
\begin{eqnarray}
 n_{s}(N_{\nu}) &=& 1 - \frac{12 q^2 \, \beta^{ - 2/(1+ 2q/p)}}{[ 4\, q\, ( p + 2 q)\, N_{\nu}/p]^{(p+ 4 q)/(p+ 2 q)}}
 \nonumber\\
&-& \frac{p +4 q}{(p + 2\, q) \, N_{\nu}},
\label{PPS1}\\
 r_{T}(N_{\nu}) &=& \frac{32 \, q^2 \,  \beta^{ - 2/(1+ 2q/p)}}{[ 4\, q\, ( p + 2 q)\, N_{\nu}/p]^{(p+ 4 q)/(p+ 2 q)}}, 
\label{PPS2}\\
n_{T}(N_{\nu})
&=& - \frac{4 \, q^2 \, \beta^{ - 2/(1+ 2q/p)}}{[ 4\, q\, ( p + 2 q)\, N_{\nu}/p]^{(p+ 4 q)/(p+ 2 q)}}.
\label{PPS3}
\end{eqnarray}
For different values of  $q$, $p$ and $\beta$ the suppression 
of $r_{T}$ is larger than in the case of monomial potentials. If we require that 
 $n_{s}$ falls within the $1\,\sigma$ observational limits set by the large scale observations supplemented by the lensing observations (e.g. $\overline{n}_{s} = 0.9649\pm 0.0042$ or $\overline{n}_{s} = 0.9665\pm0.0038$ with the addition of the baryon acoustic oscillations) we can better constrain the various parameters 
 (see, in this respect, Ref. \cite{MGINV} and discussions therein). What matters for the present considerations is that the combination of the shapes of the potential and of the decelerated evolution can easily make the relic gravitons invisible in the aHz range\footnote{ While the concrete realization of this general possibility is per se relevant, in what follows we are going to pursue a complementary 
 approach with the aim of constraining, in a model-independent perspective the duration 
 and the expansion rate of the decelerated timeline prior to radiation dominance.}.
 
\renewcommand{\theequation}{3.\arabic{equation}}
\setcounter{equation}{0}
\section{Hypermagnetic fields}
\label{sec3}
\subsection{Action and symmetries}
The same class of timelines suppressing $r_{T}$ also impact on the gauge spectra. To avoid the constraints 
imposed by Weyl invariance the gauge fields are amplified because of the evolution of the gauge coupling $g_{y}$ and of its pseudoscalar analog $\overline{g}_{y}$ (see, in this respect, \cite{EFFAC3}). The result of this 
process is a background of  solenoidal random fields that  do not  break the spatial isotropy (as it happens instead in the case of the fossil remnants discussed in Refs. \cite{REM1,REM2}). With this idea in mind we therefore focus on the following general action\footnote{As usual $Y^{\mu\nu}$ and $\widetilde{\,Y\,}^{\mu\nu}= E^{\mu\nu\alpha\beta} Y_{\alpha\beta}/2$ are, respectively, the gauge field strength and its dual in curved space; note that $E^{\mu\nu\alpha\beta} = \epsilon^{\mu\nu\alpha\beta}/\sqrt{- G}$.}:
\begin{equation}
 S_{Y} = -\int d^{4} x\sqrt{-G} \bigl[ Y_{\alpha\beta}Y^{\alpha\beta}/g_{y}^2 
+ Y_{\alpha\beta}\widetilde{Y}^{\alpha\beta}/\overline{g}_{y}^2\bigr]/4,
\label{AG}
\end{equation}
where $g_{y} = (4\pi/\lambda)^{1/2}$ and $\overline{g}_{y}=(4\pi/\overline{\lambda})^{1/2}$ denote the  gauge couplings that can be always expressed in terms of the corresponding susceptibilities conventionally denoted by $\lambda$ and $\overline{\lambda}$. Although both $g_{y}$ and $\overline{g}_{y}$ may not only depend on the inflaton, for the present ends, what matters is the overall evolution during inflation and in the decelerated stage of expansion of $g_{y}$ \cite{MGB1,MGB2}. The coupling $\overline{g}_{y}$ has been included for the sake of completeness; it can be shown, on a general ground, that $\overline{g}_{y}$ does not affect the shape of the large-scale gauge spectra but it mildly modifies  their amplitude \cite{EFFAC3}. More specifically, from Eq. (\ref{AG}) the corresponding equations can be written as: 
\begin{equation}
\nabla_{\mu} \bigl[Y^{\mu\nu}/g_{y}^2 + \widetilde{Y}^{\mu\nu}/\overline{g}_{y}^2] =0, \quad \nabla_{\mu} \widetilde{Y}^{\mu\nu}=0,
\label{BG}
\end{equation}
where $\nabla_{\mu}$ denotes the covariant derivative defined with respect to the curved metric $G_{\mu\nu}$.  Equation (\ref{BG}) can be directly expressed in terms of the corresponding susceptibilities as:
\begin{equation}
\nabla_{\mu} \bigl[\lambda Y^{\mu\nu} + \overline{\lambda} \widetilde{Y}^{\mu\nu}] =0, \quad \nabla_{\mu} \widetilde{Y}^{\mu\nu}=0.
\label{CG}
\end{equation}
If $\overline{\lambda} \to 0$, the second term inside the square bracket of  Eq. (\ref{CG}) disappears. 
The gauge spectra following from Eqs. (\ref{BG}) or (\ref{CG}) can be related by using the duality symmetry \cite{MGB1,MGB2} that connects the first dynamical equation to the Bianchi identity and vice-versa\footnote{After inserting Eq. (\ref{CG2})  into the second equation of (\ref{CG}) implies $\nabla_{\mu} \bigl[\lambda Z^{\mu\nu} + \overline{\lambda} \widetilde{Z}^{\mu\nu}] =0$; conversely if we use Eqs. (\ref{CG1})--(\ref{CG2}) into the first equation of (\ref{CG}) we simply get $\nabla_{\mu}  \widetilde{Z}^{\mu\nu}=0$.} \cite{MMAA2,MMAA3}:
\begin{eqnarray}
Y^{\mu\nu} &=& - \overline{\lambda} \,\,Z^{\mu\nu} -  (1 + \overline{\lambda}^2)/\lambda\,\,\widetilde{Z}^{\mu\nu},
\label{CG1}\\
\widetilde{Y}^{\mu\nu} &=& \lambda \,\,Z^{\mu\nu} + \overline{\lambda} \,\,\widetilde{Z}^{\mu\nu}.
\label{CG2}
\end{eqnarray} 
As in the case of Eq. (\ref{LFP1}), Eq. (\ref{AG}) is just the first term of an effective theory  whose higher derivatives are suppressed by the negative powers of a large mass $M_{eff}$ that appears in the fundamental theory underlying the effective description.  The first correction to Eq. (\ref{AG}) consists of all possible terms that contain $4$ space-time derivatives and involve the gauge fields, the inflaton and the metric tensor.
This analysis can be found in Ref. \cite{EFFAC3} (see also \cite{EFFAC5} for a shorter account of the basic idea) and the corrections to Eq. (\ref{AG}) consist of $14$ terms that can be schematically written as:
\begin{eqnarray}
\hspace{-0.6cm}
&& \Delta {\mathcal L}_{gauge} = \frac{\sqrt{-G}}{16 \, \pi\, M_{eff}^2} \biggl[ \lambda_{1}(\phi) \, R\, Y_{\alpha\beta}\, Y^{\alpha\beta} + \,.\,.\,.\,.\,
\nonumber\\
\hspace{-0.6cm}
&& + \lambda_{7}(\phi) \nabla_{\mu}\nabla^{\nu} \phi\,  Y_{\nu\alpha} Y^{\mu\alpha}
+ \overline{\lambda}_{1}(\phi) \, R\, Y_{\alpha\beta}\, \widetilde{\, Y\,}^{\alpha\beta}  
\nonumber\\
\hspace{-0.6cm}
&&+  \,.\,.\,.\,.\,+ \overline{\lambda}_{7}(\phi) \nabla_{\mu}\nabla^{\nu} \phi\,  Y_{\nu\alpha} \, \widetilde{\,Y\,}^{\mu\alpha}
 \biggr],
\label{DG1}
\end{eqnarray}
where, by definition, $\phi = \varphi/M_{eff}$ is the inflaton field rescaled through the effective mass scale and the terms appearing in the complete expression of Eq. (\ref{DG1}) have been analyzed, one by one, in different contexts (see e.g. Refs. \cite{EFFAC6,EFFAC7,EFFAC8,EFFAC9,EFFAC10}).  The first $7$ contributions of  Eq. (\ref{DG1})  are parity-even while the remaining  $7$ are parity-odd; the contributions that do not break parity are associated with $\lambda_{i}(\phi)$  while the ones that break parity are multiplied by  $\overline{\,\lambda\,}_{i}(\phi)$ where, in both cases, $i =1,\,\, ...\,,\,\,7$. The various $\lambda_{i}$ leads to a mismatch between electric and magnetic gauge couplings. In the case of inflationary backgrounds these differences can be explicitly estimated and they depend on different dimensionless combinations involving the rate of inflationary expansion, $M_{eff}$, $M_{P}$ and the slow-roll parameters \cite{EFFAC3}. The corrections of Eq. (\ref{DG1})
would imply that the electric and magnetic gauge coupling differ by factors smaller than ${\mathcal O}(10^{-10})$ \cite{EFFAC3}; for the present ends, the electric and the magnetic gauge coupling coincide.
The magnetogenesis scenarios based on Eqs. (\ref{AG}) and (\ref{CG}) are, overall, as generic as the conventional models of inflation of Eq. (\ref{LFP1}) where the  dependence of the Lagrangian on the inflaton field is practically unconstrained by symmetry. This means that there are classes of models where this conclusion does not immediately follow, at least in principle\footnote{Some of the couplings $\lambda_{i}(\phi)$ and $\overline{\lambda}_{i}(\phi)$ could be (artificially) tuned to be very large. It could also happen that the inflaton has some particular symmetry (like a shift symmetry $\varphi \to \varphi + \mathrm{const}$); this possibility reminds of the relativistic theory of Van der Waals (or Casimir-Polder) interactions  \cite{EFFAC8,EFFAC9,EFFAC10}. Another non-generic possibility implies that the rate of inflaton roll defined by $\eta$ remains constant (and possibly much larger than $1$), as it happens in certain fast-roll scenarios \cite{SYM4,SYM5,SYM6}. In all these cases $\lambda$ and $\overline{\lambda}$ may have asymmetric evolutions and the general results reported here are not immediately applicable. }. For the sake of simplicity we shall focus, in what follows, on the case $\overline{\lambda} \to 0$; however the presence of $\overline{\lambda}$ does not affect the shape of the large-scale gauge spectra but it slightly modifies their amplitude (see Ref. \cite{EFFAC3} and discussion therein). 
 
\subsection{Evolutions of the gauge coupling}
The evolution of the gauge fields  during the conventional stage of accelerated expansion outlined in Sec. \ref{sec2} demands that the gauge couplings are always perturbative  throughout all their evolution. As already stressed in \cite{MGB1,MGB2} it is imperative to consider a  complete scenario where the gauge coupling first increases and then flattens out at late times; if the gauge coupling is {\em not} continuous across the inflationary boundary incorrect conclusions can be drawn on the asymptotic behaviour of the gauge fields. This strategy naturally follows from the continuity of the mode functions and of the extrinsic curvature throughout all the stages of the dynamical evolution. During the accelerated stage of expansion (i.e. for $\tau \leq - \tau_{1}$) $\gamma$ indicates the rate of increase of the gauge coupling in the conformal time parametrization
\begin{equation}
g_{y}(\tau) = g_{1}( -\tau/\tau_{1})^{-\gamma}, \,\,\, \tau \leq - \tau_{1}. 
\label{GC1}
\end{equation}
We shall be considering values of $g_{1}$ always smaller than ${\mathcal O}(0.01)$ so that the gauge coupling remains always perturbative both during inflation and even later on. Indeed, for a reliable estimate the gauge power spectra the value of $g_{y}(\tau)$ must be continuous and differentiable across $-\tau_{1}$:
\begin{equation}
g_{y}(\tau) = g_{1} [(\gamma/\zeta) ( \tau/\tau_{1} + 1) +1 ]^{\zeta}, \qquad \tau \geq - \tau_{1},
\label{GC2}
\end{equation}
where $\zeta$ controls the evolution in the postinflationary stage. The explicit form of Eqs. (\ref{GC1})--(\ref{GC2}) is dictated by the continuity of $g_{y}(\tau)$ and of $g_{y}^{\,\prime}$: absent this essential requirement the evolution of the mode functions would be singular in $-\tau_{1}$; this means that the transient regime (where the gauge coupling relaxes 
and it does it in a computable manner) must be carefully taken into account. 

Since the gauge coupling increases during inflation (i.e. $\gamma>0$) and 
flattens out in the decelerated stage, the growth rate of $g_{y}$ must eventually get much smaller than its inflationary value so that the physical situation corresponds to $0 \leq  \zeta \ll  \gamma$. If the gauge field strengths are expressed in terms of the (physical) hyperelectric and hypermagnetic components (i.e.  $Y^{i\,j} = - a^2(\tau)\epsilon^{i\,j\,k}\, B_{k}^{(ph)}$ and  $Y_{i\,0} = a^2(\tau) E_{i}^{(ph)}$) the (comoving)
normal modes of the system are given by $E_{i}$ and $B_{i}$
and their relation to the physical fields is given by:
\begin{equation}
E^{(ph)}_{i} = g_{y}(\tau)E_{i}/a^2(\tau),\,\, B^{(ph)}_{i}= g_{y}(\tau) B_{i}/a^2(\tau).
\label{GC3}
\end{equation}
The quantum mechanical operators corresponding to the comoving hyperelectric and hypermagnetic fields are therefore expressed as:
\begin{eqnarray}
\hspace{-1.cm}
&&\hat{B}_{i}(\vec{x},\tau) = - \frac{i\, \epsilon_{m n i}}{(2\pi)^{3/2}}  \sum_{\alpha} \int d^{3} k \,k_{m} \,e^{(\alpha)}_{n}(\hat{k}) \times
\nonumber\\
\hspace{-1.cm}
&&\biggl[ f_{k,\,\alpha}(\tau)\, \hat{a}_{\vec{k}, \alpha} e^{- i \vec{k} \cdot \vec{x}}  - \mathrm{H.\,c.}\biggr], 
\label{ONE6}\\
\hspace{-1cm}
&&\hat{E}_{i}(\vec{x},\tau) = - \frac{1}{(2\pi)^{3/2}}  \sum_{\alpha} \int d^{3} k \,e^{(\alpha)}_{i}(\hat{k}) \,\times 
\nonumber\\
\hspace{-1.cm}
&&\biggl[ g_{k\,\alpha}(\tau)  \hat{a}_{\vec{k}, \alpha} e^{- i \vec{k} \cdot \vec{x}}   + \mathrm{H.\,c.} \biggr],
\label{ONE7}
\end{eqnarray}
and the sum over $\alpha=1,\,2$ is performed over the vector polarizations that  are directed along the (orthogonal) unit vectors $\hat{e}_{1}$ and $\hat{e}_{2}$ (with $\hat{k} \cdot\hat{e}_{\alpha} =0$ and $\hat{e}_{1} \times \hat{e}_{2} = \hat{k}$). In Eq. (\ref{ONE7}) $ \hat{a}_{\vec{k}, \alpha} $ and  $ \hat{a}_{\vec{k}, \alpha}^{\dagger} $ are the 
creation and annihilation operators obeying $[\hat{a}_{\vec{q}, \alpha}, \, \hat{a}_{\vec{p}, \beta}^{\dagger}] = \delta^{(3)}(\vec{q} - \vec{p})\, \delta_{\alpha\beta}$. In Eqs. (\ref{ONE6})--(\ref{ONE7}) $f_{k,\alpha}$ and $g_{k,\,\alpha}$ obey the following pair of  equations:
\begin{equation}
f_{k,\,\alpha}^{\,\prime} = g_{k,\,\alpha} + {\mathcal F} f_{k,\,\alpha},\,\,\, g_{k,\,\alpha}^{\, \prime}= -  k^2 f_{k,\,\alpha} - {\mathcal F} g_{k,\,\alpha},  
\label{ONE8}
\end{equation}
where $ {\mathcal F} = (1/g_{y})^{\,\,\prime}\,g_{y}$ and the prime denotes a derivation with respect to the conformal time coordinate $\tau$; because of the relation between 
$g_{y}$ and $\lambda$ we also have that $ {\mathcal F} = \sqrt{\lambda}^{\prime}/\sqrt{\lambda}$. The mode functions must  be correctly normalized and their Wronskian 
 must satisfy, for each polarization, the condition $f_{k,\,\alpha}(\tau) \, g_{k, \alpha}^{*}(\tau) - f_{k,\,\alpha}^{*}(\tau) \, g_{k, \alpha}(\tau) = i$.
 
The field operators of Eqs. (\ref{ONE6})--(\ref{ONE7}) can be finally represented in Fourier space and the corresponding two-point functions become: 
\begin{eqnarray}
 &&\langle \hat{B}_{i}(\vec{k}, \tau)\, \hat{B}_{j}(\vec{p},\tau) \rangle = \frac{2\pi^2}{k^3}\, P_{B}(k,\tau)\, p_{ij} \,\delta^{(3)}(\vec{k} + \vec{p}),
\nonumber\\
&&\langle \hat{E}_{i}(\vec{k},\tau)\, \hat{E}_{j}(\vec{p},\tau) \rangle = \frac{2\pi^2}{k^3}\, P_{E}(k,\tau)\, p_{ij} \,\delta^{(3)}(\vec{k} + \vec{p}),
\nonumber
\end{eqnarray}
where the expectation values are evaluated with respect to the state annihilated by $ \hat{a}_{\vec{k}, \alpha}$;
note also that $p_{ij}\equiv p_{ij}(\hat{k}) = (\delta_{i j} - \hat{k}_{i} \, \hat{k}_{j} )$. The (comoving) hypermagnetic power spectrum
$P_{B}(k,\tau)$ appearing in the previous equation is:
\begin{equation}
P_{B}(k,\tau) = k^{5}\,|f_{k}(\tau)|^2/(2\pi^2).
\label{ONE14b1}
\end{equation}
If the mode functions for the two polarizations coincide the sums appearing in 
Eq. (\ref{ONE6})--(\ref{ONE7}) are trivial since $f_{k,\, \oplus} = 
f_{k,\, \otimes} =f_{k}$  and $g_{k,\, \oplus} = 
g_{k,\, \otimes} =g_{k}$. In full analogy with Eq. (\ref{ONE14b1}) the (comoving) hyperelectric power spectrum is given by 
\begin{equation}
P_{E}(k,\tau) = k^3\,|g_{k}(\tau)|^2/(2\pi^2).
\label{ONE14b}
\end{equation}
Thanks to Eq. (\ref{GC3}) the relation between the physical and the comoving power 
spectra can be written as
\begin{equation}
{\mathcal P}_{X}(k,\tau) =\frac{g_{y}^2(\tau)}{a^{4}(\tau)} P_{X}(k,\tau), 
\label{ONE14c}
\end{equation}
where $X= B,\, E$. Equations (\ref{ONE14b})--(\ref{ONE14c})
the expressions of the comoving and physical power spectra
is obviously different throughout the various stages of the dynamical evolution. 

\subsection{Evolution of the gauge fields}
The evolution of the gauge fields across the inflationary phase is encoded in the 
explicit expressions of the mode functions and when $g_{y}(\tau)$ evolves as in Eq. (\ref{GC1}),  the solution of Eq. (\ref{ONE8}) compatible with the Wronskian normalization is \cite{MGB1}:
\begin{equation}
f_{k}(\tau) = N_{\mu}\, \sqrt{- k\tau} \, H_{\mu}^{(1)}(-k\tau)/\sqrt{2 k}.
\label{TWO2}
\end{equation}
Equation (\ref{TWO2}) is valid during the accelerated stage of expansion (i.e. for $\tau \leq - \tau_{1}$); $\mu = |\gamma -1/2|$ and 
$N_{\mu}= \sqrt{\pi/2}\, e^{i \pi ( 2 \mu + 1)/4}$ while $H_{\mu}^{(1)}(- k \tau)$ are the Hankel functions of the first kind; the index $\mu$ shall always be real and positive semi-definite follows from Eq. (\ref{TWO2}) since, from Eq. (\ref{ONE8}), $g_{k} = f_{k}^{\prime} - {\mathcal F} f_{k}$. Because of the properties of the Hankel functions for $\gamma > 1/2$
\begin{equation}
g_{k}(\tau) = N_{\mu} \,\sqrt{k/2} \,  \sqrt{- k\tau} \, H_{\mu+1}^{(1)}(-k\tau),\,\, \tau \leq - \tau_{1},
\label{TWO3}
\end{equation}
whereas in the case $0< \gamma < 1/2$ we would have instead $g_{k}(\tau)  = - N_{\mu} \sqrt{k/2} \,  \sqrt{- k\tau} \, H_{\mu-1}^{(1)}(-k\tau)$. 

When $\tau\geq - \tau_{1}$ the mode functions differ substantially from Eq. (\ref{TWO2})--(\ref{TWO3}). It is therefore misleading (as sometimes propounded) to derive the properties of the gauge power spectra at late times (and for large scales)  by only taking into account the inflationary expressions of the mode functions. 
The key point, in this respect, is that the amplified gauge fields at the
end of inflation {\em do not} coincide with the gauge fields at late time.
To clarify this point, the continuous parametrization of Eqs. (\ref{GC1})--(\ref{GC2}) implies that the late-time values values of $f_{k}(\tau)$ and $g_{k}(\tau)$ for $\tau \geq - \tau_{1}$ can be written as\footnote{Since the Wronskians of ($f_{k}$, $g_{k}$)  and of ($\overline{f}_{k}$, $\overline{g}_{k}$) are both equal to the imaginary unit,
the determinant  of the matrix formed by the coefficients entering Eq. (\ref{LT1}) must be
$A_{f\, f} A_{g\, g}- A_{f\, g} A_{g\, f}=1$. From the continuity of the mode functions it also follows that $A_{f\,g}(k, - \tau_{1}, \tau_{1}) =A_{g\,f}(k, - \tau_{1}, \tau_{1})=0$ and that $A_{f\,f}(k, - \tau_{1}, \tau_{1}) =A_{g\,g}(k, - \tau_{1}, \tau_{1})=1$.}
\begin{eqnarray}
f_{k}(\tau) &=& A_{f f}\,\overline{f}_{k} + A_{f g}\,\overline{g}_{k}/k,
\nonumber\\
g_{k}(\tau) &=& k\, A_{g f} \,\overline{f}_{k} + A_{g g} \,\overline{g}_{k} ,
\label{LT1}
\end{eqnarray}
In Eq. (\ref{LT1}), by definition, 
$\overline{f}_{k}= f_{k}(-\tau_{1})$ and $\overline{g}_{k} = g_{k}(-\tau_1)$ 
denote the values of the mode functions at end of the inflationary phase while the 
the other terms all depend upon $k$, $\tau$ and $\tau_{1}$ [i.e. $A_{f\, f}\equiv  A_{f\, f}(k,\tau, \tau_{1})$ and similarly for the other coefficients].  From the explicit expression of Eq. (\ref{GC2}) the matrix elements are\footnote{To avoid confusions we remind that the Bessel indices $\mu$ and $\nu$ should not be confused with the comoving frequencies.} 
\begin{eqnarray}
\hspace{- 1cm}
&&A_{f\, f}= \ell(x_{1}, y) \biggl[ Y_{\beta}(w x_{1}) J_{\nu}(k y) - J_{\beta}(w x_{1}) Y_{\nu}(k y) \biggr],
\nonumber\\
\hspace{- 1cm}
&& A_{f\, g}= \ell(x_{1},y) \biggl[ J_{\nu}(w x_{1}) Y_
{\nu}(k y) - Y_{\nu}(w x_{1}) J_{\nu}(k y) \biggr],
\nonumber\\
\hspace{- 1cm}
&& A_{g\, f} = \ell(x_{1}, y) \biggl[ Y_{\beta}(w x_{1}) J_{\beta}(k y) - J_{\beta}(w x_{1}) Y_{\beta}(k y) \biggr],
\nonumber\\
\hspace{- 1cm}
&&A_{g\, g}= \ell(x_{1}, y)\biggl[ J_{\nu}(w x_{1}) Y_{\beta}(k y) - Y_{\nu}(w x_{1}) J_{\beta}(k y) \biggr],
\label{LT2}
\end{eqnarray}
where $\ell(x_{1}, y) = \pi \sqrt{w x_{1}} \sqrt{k y} /2$. For the sake of conciseness, in Eq. (\ref{LT2}) the following 
shorthand notations have been introduced:  $\beta = (\nu-1)$, $w = \zeta/\gamma$, $y =  \tau + \tau_{1}[1 + w]$ and $\nu= \zeta +1/2$. Within these notations $y(-\tau_{1}) = w \tau_{1}$ which also 
implies that  $k y(-\tau_{1}) = w \, k\, \tau_{1} = w x_{1}$. 

The matrix elements of Eq. (\ref{LT2})  depend on
the dimensionless variables $x = k \tau$, $x_{1} = k \tau_{1}$ and $\nu$. In practice $x_{1} = k \tau_{1} \leq 1$ measures $k$ in units of the maximal wavenumber of the spectrum (i.e. $1/\tau_{1} = a_{1} H_{1}$). This is why, for a more explicit form of the gauge power spectra in the decelerated stage of expansion, the matrix elements of Eq. (\ref{LT2}) can be systematically expanded in powers of  $x_{1} < 1$ for fixed $k y$ with the subsidiary conditions $ 0\leq \zeta \ll 1/2$. The leading terms of the expansion are therefore given by:
\begin{eqnarray}
\hspace{-0.4cm}
&& A_{f\, f} = (w x_{1}/2)^{\zeta} \sqrt{x/2} \,\,\Gamma(1/2 -\zeta) J_{- \zeta -1/2}(x),
\nonumber\\
\hspace{-0.4cm}
&& A_{f\, g} =(w x_{1}/2)^{-\zeta} \sqrt{x/2} \,\,\Gamma(1/2 +\zeta) J_{\zeta +1/2}(x),
\nonumber\\
\hspace{-0.4cm}
&& A_{g\, f} = - (w x_{1}/2)^{\zeta}  \sqrt{x/2}  \,\, \Gamma(1/2 -\zeta) J_{1/2- \zeta}(x),
\nonumber\\
\hspace{-0.4cm}
&& A_{g\, g}= (w x_{1}/2)^{-\zeta} \sqrt{x/2}  \,\, \Gamma(1/2 +\zeta) J_{\zeta -1/2}(x),
\label{LT8}
\end{eqnarray}
where we stress that the condition $x_{1} < 1$ also implies that $k \, y \simeq x = k\tau$.

\subsection{Comoving gauge spectra}
After inserting the correctly normalized mode functions of Eqs. (\ref{TWO2})--(\ref{TWO3}) into Eq. (\ref{ONE14b}) the comoving power spectra during inflation turn out to be\footnote{We recall that $|k \tau| = (-k\tau)$ since, during a stage of accelerated expansion, the conformal time coordinate is always negative.}:
\begin{eqnarray}
\hspace{-1.1cm}
&& P_{B}(k,\tau) = \frac{a^4 H^4}{8\pi} |k\tau|^5 \bigl|H_{\mu}^{(1)}(|k\tau|)\bigr|^2 
\label{GS1}\\
\hspace{-1.1cm}
&& P_{E}(k,\tau) = \frac{a^4 H^4}{8\pi} |k\tau|^5 \bigl|H_{\mu+1}^{(1)}(|k\tau|)\bigr|^2, \,\, \gamma > 1/2,
\label{GS2}\\
\hspace{-1.1cm}
&& P_{E}(k,\tau) = \frac{a^4 H^4}{8\pi} |k\tau|^5 \bigl|H_{\mu-1}^{(1)}(|k\tau|)\bigr|^2,\,\,  \gamma < 1/2.
\label{GS3}
\end{eqnarray}
 The spectra of Eqs. (\ref{GS1}) and (\ref{GS2})--(\ref{GS3}) hold during the inflationary stage (i.e. for $\tau < - \tau_{1}$) and can be explicitly estimated when the relevant scales 
are larger than the effective horizon (i.e. when $|k\tau| < 1$)
 \begin{eqnarray} 
\hspace{-0.67cm}
P_{B}(k,a) &=&  a^4 \, H^4\, D(|\gamma -1/2|) \bigl|k/(a\, H)\bigr|^{ 5 - |2 \gamma-1|}, 
\label{GS4}\\
\hspace{-0.67cm}
P_{E}(k,a) &=&  a^4 \, H^4\, D(\gamma +1/2) \bigl|k/(a\, H)\bigr|^{4 - 2 \gamma}.
\label{GS5}
\end{eqnarray}
The function $D(x) = 2^{2 x- 3} \Gamma^2(x)/\pi^3$ has been introduced in Eqs. 
(\ref{GS4})--(\ref{GS5}) for the sake of conciseness and it is consistently employed throughout to simplify the obtained expressions. Note that the two different intervals of $\gamma$ mentioned in Eqs. (\ref{GS2})--(\ref{GS3}) 
lead eventually to the same limit for $|k \tau| \ll 1$ since the corresponding Hankel 
functions are estimated using their limit for small arguments \cite{abr1,abr2}. 

The spectral energy density follows from the energy-momentum tensor of the gauge fields and it is directly expressed in terms of Eqs. (\ref{GS4})--(\ref{GS5}) as 
\begin{equation}
\Omega_{Y}(k,a) = [ P_{E}(k,a) + P_{B}(k,a)]/(3 \, H^2\, a^4\,\overline{M}_{P}^2).
\label{GS6}
\end{equation}
Since $\Omega_{Y}(k,a)$ must always be subcritical, we have from Eqs. (\ref{GS4})--(\ref{GS5}) and (\ref{GS6}) that $\gamma \leq 2$: when $\gamma>2$ the hypermagnetic power spectra get progressively steeper while their hyperelectric counterpart  diverge in the large scale limit (i.e. $k \ll a\, H$). In summary when the gauge coupling increases during a quasi-de Sitter stage of expansion the spectral energy density is subcritical for $0 < \gamma \leq 2$ and overcritical for $\gamma >2$; thus the latter range is excluded while the former is still viable. Since the hypermagnetic spectrum is steep (i.e. violet) when $\gamma=2$ the conventional wisdom is that it will also be minute at the galactic scale after the gauge coupling flattens out. 

\subsection{Late-time spectra}
The conclusion contained in the previous paragraph is only sound if the hypermagnetic power spectra at the end of inflation remain unaltered for $\tau \geq -\tau_{1}$.  To compute the late-time power spectra it is therefore mandatory to extend the analysis of the gauge power spectra in the regime where the gauge coupling flattens out (i.e. for $\tau \geq - \tau_{1}$). The late-time hypermagnetic spectrum {\em does not} coincide with the hypermagnetic spectrum at the end of inflation and, after the gauge coupling flattens out (i.e. $\zeta \ll \gamma$), the late-time hypermagnetic power spectra outside the horizon are determined by the hyperelectric fields at the end of inflation. On a general ground, the (comoving) power spectra at late times follow from Eqs. (\ref{ONE14b})  and (\ref{LT1})--(\ref{LT2}):
\begin{eqnarray}
&& P_{B}(k,\tau) = \frac{k^5}{2\pi^2} \bigl| A_{f\, f} \, \overline{f}_{k} + A_{f\, g} \,\overline{g}_{k}/k\bigr|^2,
\label{LT3}\\
&&P_{E}(k,\tau) = \frac{k^3}{2\pi^2} \bigl| k\, A_{g\, f} \, \overline{f}_{k} + A_{g\, g} \,\overline{g}_{k}\bigr|^2.
\label{LT4}
\end{eqnarray}
Since $x_{1}$ is always strictly smaller than $1$, at late times (i.e. $\tau \gg \tau_{1}$) we also have that $x_{1} \ll x\ll 1$ and, in this limit, $|A_{f\, g}\,\overline{g}_{k}| \gg | A_{f\, f}\, k\,  \overline{f}_{k} |$ for all ranges of $\gamma \leq 2$, as required by the constraints imposed by the spectral energy density. When $x_{1} \ll 1$ and $x\gg 1$  the functions whose argument  coincides with $k y \simeq x \gg 1$ can be always represented as $J_{\nu}(k y) = M_{\nu} \cos{\theta_{\nu}}$ and $Y_{\nu}(k y) = M_{\nu} \sin{\theta_{\nu}}$. When $x\gg 1$, $\theta_{\nu}(x) \to  x $ while $M_{\nu}(x) \to \sqrt{2/\pi} x^{-1/2}[ 1 + {\mathcal O}(x^{-2})]$; this is the so-called modulus-phase approximation for the Bessel functions \cite{abr1,abr2}. Thanks to this observation the comoving spectrum of Eq. (\ref{LT3}) becomes:
\begin{equation}
P_{B}(k,\tau) = \frac{k^5}{2\pi^2} \bigl| A_{f\, g}(\zeta,\, x_{1}, \,x)\, \overline{g}_{k}/k\bigr|^2.
\label{LT6}
\end{equation}
The results of Eqs. (\ref{LT1})--(\ref{LT6}) show that the 
hyperelectric field at the end of inflation determines the late-time hypermagnetic field 
for $\tau \gg - \tau_{1}$. This 
happens provided the gauge coupling first increases during inflation and then flattens out in the radiation-dominated epoch\footnote{If the gauge coupling would instead {\em decrease} during inflation and then flatten out the late time hypermagnetic fields are fixed by the hypermagnetic fields at the end of inflation. This case is however unphysical for many reasons related to the presence of a strongly coupled stage at the beginning of inflation \cite{MGB1,MGB2}. The power 
spectra can be however determined by using the duality symmetry discussed in Eqs. (\ref{CG1})--(\ref{CG2}). The duality symmetry exchanges 
electric and magnetic power spectra as explicitly discussed in Ref. \cite{MGB1} (see also \cite{MMAA1,MMAA2}).}. Even if the value of $x$ can be either smaller or larger 
than $1$, as soon as $ x = k \tau = {\mathcal O}(1)$ the conductivity cannot be neglected and this situation will be more specifically discussed below; in this section we just consider the case $x\gg 1$ without taking into account further suppressions. 

For the hyperelectric spectrum the inequality of Eq. (\ref{LT1}) is in fact replaced by 
the  condition $| A_{g\, g}\,  \overline{g}_{k}| \gg | A_{g\, f}\,k\, \overline{f}_{k} |$
which can be verified explicitly by using the same strategy illustrated in the case of Eq. (\ref{LT1}); for the sake of conciseness these details will not be explicitly discussed.
Therefore, thanks to Eq. (\ref{LT2}), the late-time expression of the comoving hyperelectric spectrum  is:
\begin{eqnarray}
P_{E}(k,\tau) = \frac{k^3}{2\pi^2} \bigl|  A_{g\, g}(\zeta, \, x_{1},\, x) \, \overline{g}_{k} \bigr|^2.
\label{LT7}
\end{eqnarray}
Equation (\ref{LT7}) mirrors the result of Eq. (\ref{LT6}) and it shows that the hyperelectric power spectrum for $\tau \gg - \tau_{1}$ is determined by the hyperelectric power spectrum at $\tau = - \tau_{1}$. As we shall see
in a moment when the gauge coupling decreases the dual result will hold.
Inserting Eq. (\ref{LT8}) into Eq. (\ref{LT6}) and recalling the expressions for $\overline{f}_{k}$ and $\overline{g}_{k}$ the hypermagnetic power spectrum becomes:
\begin{equation}
P_{B}(k,\tau) = a_{1}^{4} H_{1}^4 \, D(\gamma + 1/2) \, x_{1}^{\alpha(\gamma,\zeta)} F_{B}(k \tau),
\label{LT9}
\end{equation}
where $x_{1} =k/(a_{1} H_{1})$ and
$\alpha(\gamma, \zeta) = 4 - 2 \gamma - 2 \zeta$;  moreover $F_{B}(x)=(w/2)^{-\,2 \zeta} \, (x/2) \, \Gamma^2(\zeta+1/2) \, J_{\zeta+1/2}^2(x)$. Similarly, from Eqs. (\ref{LT7}) and (\ref{LT8}) the hyperelectric spectrum is
\begin{equation}
P_{E}(k,\tau) = a_{1}^{4} \, H_{1}^4 \, D(\gamma + 1/2) \,x_{1}^{\alpha(\gamma,\zeta)} F_{E}(k \tau),
\label{LT10}
\end{equation}
where $F_{E}(x) = (w/2)^{-\,2 \zeta} \, (x/2) \, \Gamma^2(\zeta+1/2) \, J_{\zeta-1/2}^2(x)$.  The results of Eqs. (\ref{LT9})--(\ref{LT10}) only assume $x_{1} <1 $ and $0\leq \zeta \ll \gamma$ and can be evaluated either for $k\tau \ll 1$ or for $k\tau \gg 1$. As long as  $ k \tau \ll 1$  it is enough to recall that $J_{\alpha}(z) \simeq (z/2)^{\alpha}/\Gamma(\alpha +1)$ \cite{abr1,abr2}.  Equations (\ref{LT9}) and (\ref{LT10}) hold for any value of $k \tau$; however, as we shall argue hereunder, for $\tau > \tau_{k} \sim 1/k$ the power spectra will be modified  by the finite value of the conductivity.

Another interesting limit is the sudden approximation which is not well defined a priori but only as the   $\zeta \to 0$ limit; in this case $x$ and $x_{1}$ are kept fixed and the matrix elements of Eq. (\ref{LT2}) assume a rather simple form implying:
\begin{eqnarray}
\hspace{-0.67cm}
&& P_{B}(x,x_{1})=\frac{k^5}{2\pi^2} \bigl| \cos{(x+ x_{1})} \overline{f}_{k} + \sin{(x+ x_{1})} \overline{g}_{k}/k\bigr|^2,
\nonumber\\
\hspace{-0.67cm}
&& P_{E}(x,x_{1}) = \frac{k^5}{2\pi^2} \bigl| -\sin{(x+ x_{1})} \, \overline{f}_{k} + \cos{(x+ x_{1})}\overline{g}_{k}/k\bigr|^2.
\nonumber
\end{eqnarray}
The previous expressions also imply that the gauge power spectra become
\begin{eqnarray}
\hspace{-0.6cm}
&&P_{B}(k,\tau) = a_{1}^{4} \, H_{1}^4 \, D(\gamma + 1/2) \, x_{1}^{4 - 2 \gamma} \, \sin^2{k\tau},
\label{LT11}\\
\hspace{-0.6cm}
&&P_{E}(k,\tau) = a_{1}^{4} \, H_{1}^4 \, D(\gamma + 1/2) \, x_{1}^{4 - 2 \gamma} \, \cos^2{k\tau}.
\label{LT12}
\end{eqnarray}
The same results of Eqs. (\ref{LT11})--(\ref{LT12}) follow immediately from  Eqs. (\ref{LT9})--(\ref{LT10})
 by recalling that $w^{-\zeta} = (\zeta/\gamma)^{-\zeta} \to 1$  in the limit $\zeta \to 0$. All in all, in the sudden approximation $x_{1}$ and $x$ are kept fixed while $\zeta\to 0$; in the smooth limit  
$\zeta$ may be very small (i.e.  $\zeta \ll 1$) but it is always different 
from zero. 
\renewcommand{\theequation}{4.\arabic{equation}}
\setcounter{equation}{0}
\section{Ultra high-frequency gravitons}
\label{sec4} 
Before analyzing the impact of different decelerated timelines on the 
gauge spectra deduced in Sec. \ref{sec3} it is appropriate to deduce 
the corresponding spectra of relic gravitons. In Sec. \ref{sec5} the 
concurrent constraints will be explicitly deduced.
The suppression potential 
of $r_{T}$ in the aHz domain and the increase of $N_{\nu}$ 
are associated with the presence of high-frequency spikes in the spectral energy density \cite{MGST1,MGST2,MGST3}. 
Since after the inflationary stage the background expands (at least 
for some time) at a rate that is slower than radiation, $N_{\nu}$ increases 
and $r_{T}$ gets suppressed. This situation has been illustrated in Sec. \ref{sec2} (see Figs. \ref{FIGURE1}, \ref{FIGURE2} and \ref{FIGURE3} and discussions therein). 
We intend to present here the estimates of $\Omega_{gw}(\nu, \tau_{0})$ (i.e. the spectral energy density of the relic gravitons in critical units) for two relevant situations that will be 
analyzed in conjunction with the constraints coming from large-scale magnetism. The first class of scenarios involves to a maximum in the ultra-high frequency region while the second case leads to a maximum in the audio band\footnote{The spectra of relic gravitons at high-frequencies can be 
computed within different approximation schemes and, for the present purposes, we shall make use  of some recent analyses \cite{UFS1,UFS2,UFS3} by focussing on the dependence upon the postinflationary timeline.}. 

\subsection{The maximal frequency}
For the present ends the first important observation is that the maximal frequency of the relic gravitons never exceeds the THz domain \cite{UFS1}. Indeed, in the high-frequency region the spectral energy density can be always 
written in terms of the averaged multiplicity of the produced pairs of gravitons with 
opposite three-momenta (i.e. $\overline{n}_{\nu}$) \cite{UFS2,UFS3}:
\begin{equation}
\Omega_{gw}(\nu, \tau_{0}) = \frac{128 \pi^3}{3} \biggl(\frac{\nu}{\sqrt{H_{0} \, M_{P}}}\biggr)^{4} \, \overline{n}_{\nu}.
\label{UF1}
\end{equation} 
Equation (\ref{UF1}) suggests that the maximal frequency of the spectrum corresponds to the production of a single pair of gravitons (i.e.  $\overline{n}_{\nu_{max}} \to 1$). The unitarity of the process of graviton production implies that the averaged multiplicity is exponentially suppressed for $\nu> \nu_{max}$ \cite{UFS2} (see also \cite{REL1,REL2,REL3,REL4}). 
The quantum mechanical perspective leading to Eq. (\ref{UF1}) \cite{UFS1,UFS2}  can also be appreciated 
by noting that the spectral energy density of the relic gravitons vanishes in the limit $\hbar\to 0$  \cite{hbar}.
Although in this paper the natural system of units is consistently employed, 
$\hbar$  dependence can be restored by recalling that the energy of a single graviton is given by $\hbar\, \omega$ where $\omega = k c$ (and $c$ is the speed of light); another $\hbar$ comes from the definition of Planck mass. This means that $\Omega_{gw}(\nu,\tau_{0}) \propto \hbar^2$ \cite{hbar} which is consistent with the quantum mechanical origin\footnote{In spite of this observation, as mentioned at the beginning of section \ref{sec2} the natural units $\hbar = c = k_{B} =1$ will be used throughout.} of the diffuse backgrounds of relic gravitons. The same conclusion can also be reached along a classical perspective where the maximal frequencies correspond to the bunch of wavenumbers that experience the minimal amplification and that reenter the comoving Hubble radius right after inflation.

All the wavelengths reentering the Hubble radius between the end of inflation and the big-bang nucleosynthesis epoch (BBN) must  comply with the bound\footnote{In Eq. (\ref{UF4}) $h_{0}$ is the Hubble rate expressed in units of $100\,\mathrm{Hz}\, \mathrm{km}/\mathrm{Mpc}$ 
and its presence introduces a further indetermination that is eliminated
provided $\Omega_{gw}(\nu,\tau_{0})$ is multiplied by $h_{0}^2$. If is often convenient to study directly $h_{0}^2 \Omega_{gw}(\nu, \tau_{0})$ rather than $ \Omega_{gw}(\nu,\tau_{0})$. Indeed, $ \Omega_{gw}(\nu,\tau_{0})$ contains $\rho_{crit} $ in its denominator and $h_{0}^2/\rho_{crit}$ is eventually independent of $h_{0}$.} \cite{bbn1,bbn2,bbn3,bbn4,bbn5}
\begin{eqnarray}
&&h_{0}^2 \, \int_{\nu_{bbn}}^{\nu_{\mathrm{max}}} \,\Omega_{gw}(\nu,\tau_{0}) \,\,d\ln{\nu} < 5.61\times 10^{-6} 
\nonumber\\
&& \times \biggl(\frac{h_{0}^2 \,\Omega_{\gamma0}}{2.47 \times 10^{-5}}\biggr) \, \Delta N_{\nu},
\label{UF4}
\end{eqnarray}
where $\Omega_{\gamma\,0}$ is the (present) critical fraction of CMB photons and $\nu_{bbn}$ is the typical frequency associated with BBN\footnote{Note that $g_{\rho,\, bbn}$ denotes the effective number of relativistic species at the nucleosynthesis epoch
and $T_{bbn}$ is the corresponding themperature. For typical values of the parameters (i.e. $T_{bbn} = {\mathcal O}(\mathrm{MeV})$, $g_{\rho,\, bbn}=10.75$) we have that $\nu_{bbn} = {\mathcal O}(10^{-2})\,\, \mathrm{nHz}$.}
\begin{eqnarray}
\nu_{bbn} = 8.17 \times 10^{-33} g_{\rho,\, bbn}^{1/4} \, T_{bbn} \biggl(\frac{h_{0}^2 \Omega_{R0}}{4.15\times 10^{-5}} \biggr)^{1/4}.
\label{UF4a}
\end{eqnarray}
Equation (\ref{UF4}) sets a constraint  on the extra-relativistic species possibly present at the BBN time and since the limit is often expressed via $\Delta N_{\nu}$ (i.e. the contribution of supplementary neutrino species), the actual bounds on $\Delta N_{\nu}$ range from $\Delta N_{\nu} \leq 0.2$ to $\Delta N_{\nu} \leq 1$ so that the integrated spectral density in Eq. (\ref{UF4}) must vary, at most, between  $10^{-6}$ and $10^{-5}$. For all practical purposes Eq. (\ref{UF1}) can be always referred to a putative $\nu_{max}$ beyond which the 
averaged multiplicity of the gravitons is exponentially suppressed:
\begin{equation}
\Omega_{gw}(\nu, \tau_{0}) =   \frac{128 \pi^3}{3} \frac{\nu_{max}^{4}}{H_{0}^2 \, M_{P}^2}
 (\nu/\nu_{max})^{4}  \, \overline{n}_{\nu_{max}},
 \label{UF2}
 \end{equation} 
 where, by definition, $\overline{n}_{\nu_{\mathrm{max}}} = {\mathcal O}(1)$. Thanks to Eq. (\ref{UF2}) 
from Eqs. (\ref{UF4})--(\ref{UF4a}) we can deduce the absolute upper bound on the maximal frequency of the cosmic gravitons \cite{UFS1}
\begin{equation}
\nu_{max} < {\mathcal O}(10^{-2}) \, \sqrt{H_{0} \, M_{P}} < {\mathcal O}(\mathrm{THz}).
\label{UF3}
\end{equation}
More detailed estimates of the averaged multiplicity above $\nu_{max}$ can be performed within different approximation schemes and we mention here the results obtained in Refs. \cite{UFS2,UFS3}:
\begin{equation}
\overline{n}_{\nu} = 3\eta \, {\mathcal Q}(\delta, r_{T}) (\nu/\nu_{max})^{m_{T} - 3}/\bigl[e^{\eta\,(\nu/\nu_{\mathrm{max}}) } -1\bigr],
\label{UF5}
\end{equation}
where $\eta$ is a numerical factor determined from the direct numerical integration 
of the evolution of the tensor mode functions; ${\mathcal Q}(\delta, r_{T})$
and $m_{T}$ are given by:
\begin{eqnarray}
&&{\mathcal Q}(\delta, r_{T}) = \frac{2^{2(p + \delta) -3}}{3 \, \pi^2\, q^{2 \delta}}\, \Gamma^2(p)\, \Gamma^2(\delta +1/2), 
\nonumber\\
&&m_{T} = \frac{2 - 4 \epsilon}{1 - \epsilon} - 2 \delta = \frac{32 - 4 r_{T}}{16 - r_{T}} - 2 \delta.
\label{UF6}
\end{eqnarray}
where $p = (48 - r_{T})/(32 -2\,r_{T})$. In Eq. (\ref{UF5}) the second equality follows by enforcing the consistency relations. Furthermore, in the limit $r_{T} \ll 1$ we can expand the first term in Eq. (\ref{UF6}) and obtain $m_{T} = 2 ( 1 - \delta) - r_{T}/8 + {\mathcal O}(r_{T}^2)$. We note that in the limit  $\delta \to 1$ we have instead $m_{T} \to - r_{T}/8 + {\mathcal O}(r_{T}^2)$, as expected in the case of the standard quasi-flat spectrum \cite{RELINF1,RELINF2,RELINF3,RELINF4}.
 
\subsection{Single postinflationary stage of expansion}
Broadly speaking the case of a single postinflationary stage 
of expansion preceding the radiation epoch corresponds to the timeline illustrated in the cartoon of 
Fig. \ref{FIGURE1} where a single stage of decelerated expansion takes place between the end of inflation 
and the onset of the radiation-dominated phase.  From Eqs. (\ref{UF1}) and (\ref{UF6}) the spectral energy density for $\nu_{r} < \nu \leq \nu_{max}$ can be approximated as 
\begin{equation}
\Omega_{gw}(\nu,\tau_{0}) = \overline{\Omega}_{gw}\,\,(H_{r}/H_{1})^{4\alpha(\delta)}\, (\nu/\nu_{max})^{m_{T}}, 
\label{UF7a}
\end{equation}
where the overall amplitude $\overline{\Omega}_{gw}$ now depends on $r_{T}$ and $\delta$ (i.e. the postinflationary 
expansion rate):
\begin{equation}
\overline{\Omega}_{gw} =  r_{T} \,{\mathcal Q}(\delta, r_{T})\, {\mathcal A}_{{\mathcal R}} \,\Omega_{R0} \, d^{4}(g_{s}, g_{\rho}).
\label{UF7}
\end{equation}
We remind that $d(g_{s},g_{\rho})$ and $\alpha(\delta)$ have been already introduced in Eqs. (\ref{NN3}) and (\ref{NN6}) respectively. By definition $\nu_{max}$ and $\nu_{r}$ are given by
\begin{equation}
\nu_{max} = \xi^{\alpha(\delta)}\,\, \overline{\nu}_{max}, 
\quad \nu_{r} = \sqrt{\xi}\,\, \overline{\nu}_{max},
\label{UF8}
\end{equation}
where, as in Figs. \ref{FIGURE2} and \ref{FIGURE3}, $\xi= H_{r}/H_{\nu}$ measures 
the duration of the postinflationary stage preceding the conventional radiation epoch.
We relate $\nu_{max}$ to $\overline{\nu}_{max}$ 
which corresponds to the maximal frequency is the case $\delta\to 1$; indeed 
when $\delta \to 1$ (as it happens for a postinflationary evolution dominated by radiation)
$\alpha(\delta) \to 0$ (see Eq. (\ref{NN6})), $\xi \to 1$ (because $H_{r} = H_{1} \simeq H_{\nu}$)
 and $\nu_{max} = \overline{\nu}_{max}$:
\begin{equation} 
\overline{\nu}_{max}= ( 2 \Omega_{R0})^{1/4}\,d(g_{s},g_{\rho})\, \sqrt{H_{0} \, H_{1}}/(2\pi).
\label{UF8a}
\end{equation}
For typical values of the parameters (e.g. $r_{T} \to 0.06$, $h_{0}^2 \Omega_{R0} \to 4.15 \times 
10^{-5}$, ${\mathcal A}_{{\mathcal R}} = 2.41\times 10^{-9}$) Eq. (\ref{UF8a}) gives 
$\overline{\nu}_{max} = 271.93 \, d(g_{s}, g_{\rho})\, \mathrm{MHz}$.
The same approximation scheme leading to Eqs. (\ref{UF7a}) and (\ref{UF8})--(\ref{UF8a}) can also be employed in the range $\nu_{eq} < \nu < \nu_{r}$ where the spectral energy density in critical units becomes: 
\begin{equation}
\Omega_{gw}(\nu, \tau_{0}) = \overline{\Omega}_{gw} \, (H_{r}/H_{1})^{\frac{ m_{T}}{\delta +1}} \,(\nu/\nu_{r})^{n_{T}}.
\label{UF10}
\end{equation}
For $\nu < \nu_{r}$ we have that $\Omega_{gw}(\nu, \tau_{0})$ is quasi-flat since $n_{T} = - r_{T}/8$. 
We remind that 
 $\nu_{r}$ cannot be arbitrarily small since it must always exceed 
 $\nu_{bbn}$; given the specific expressions of $\nu_{r}$ and $\nu_{bbn}$ this 
condition follows because $H_{r} \geq H_{bbn}$. For a single stage 
preceding the radiation epoch the limits on $\nu_{r}$ must be 
combined with the BBN bound (\ref{UF4}) that constrains $\Omega_{gw}(\nu, \tau_{0})$ for 
$\nu \leq \nu_{max}$; this discussion will be specifically presented in Sec. \ref{sec5}.
We finally mention that, although there are numerical ways of setting the low-frequency normalization (see e.g. \cite{MGINV}), we prefer here to employ directly the results of Eqs. (\ref{UF7}) and (\ref{UF10}) 
since their accuracy is sufficient for the purposes of the present analysis.
 
\subsection{Double postinflationary stage of expansion}
\label{subsec4D}
Before getting into the phenomenological aspects of the problem 
it is useful to relax the timeline discussed in the previous subsection and consider 
the case where the postinflationary stage consists of two 
separate expanding phases (with rates $\delta_{1}$ and $\delta_{2}$) both preceding 
the standard radiation-dominated evolution. The frequency 
$\nu_{max}$ introduced in Eq. (\ref{UF8}) now becomes
\begin{equation}
\nu_{max} = \xi_{1}^{\alpha(\delta_{1})}\, \xi_{2}^{\alpha(\delta_{2})}\, \overline{\nu}_{max},
\label{AU1}
\end{equation}
where $\xi_{1} = H_{2}/H_{1}<1$ and $\xi_{2} = H_{r}/H_{2}<1$; as before, $\overline{\nu}_{max}$ 
is given by Eq. (\ref{UF8a}) and the three rates $H_{1} > H_{2} > H_{r}$ mark, respectively, the end 
of the inflationary stage, the end of the first intermediate stage characterized by the rate $\delta_{1}$ and the 
end of the second intermediate stage with rate $\delta_{2}$. When $\delta_{1} \to \delta_{2} = \delta $  the 
result of Eq. (\ref{UF8}) is recovered since $\nu_{max} \to \xi^{\alpha(\delta)} \, \overline{\nu_{max}}$ and  
$ \xi_{1} \xi_{2} = (H_{2}/H_{1})(H_{r}/H_{2})= \xi$ where, as before, $\xi= H_{r}/H_{1}$. 
Because there are now two phases taking place prior to radiation 
dominance, between $\nu_{max}$ and $\nu_{r}$ a further typical frequency appears, namely 
\begin{equation}
\nu_{2} = \sqrt{\xi_{1}}\, \xi_{2}^{\alpha(\delta_{2})} \overline{\nu}_{max}.
\label{AU2}
\end{equation}
In this situation we have that $\nu_{r} = \sqrt{\xi_{1}}\, \sqrt{\xi_{2}} \, \overline{\nu}_{max}$
but since $\xi_{1}\, \xi_{2} = \xi$ this result coincides exactly with the expression of Eq. (\ref{UF8}). 
The most interesting physical situation coincides, for the present ends, with the one which is also
more constrained from the observational data at intermediate frequencies.
For this purpose we now recall that, according to Eq. (\ref{UF7}), in the range $\nu_{2} < \nu \leq \nu_{max}$ the spectral energy density in critical units  becomes:
\begin{equation}
\Omega_{gw}(\nu,\tau_{0}) = \overline{\Omega}_{gw}\,\xi_{1}^{4 \, \alpha(\delta_{1})}\,  \xi_{2}^{4 \alpha(\delta_{2})} \,(\nu/\nu_{max})^{m_{T}^{(1)}}, 
\label{AU3}
\end{equation}
where $m_{T}^{(1)} = (1 - 3 r_{T}/16)/(1 - r_{T}/16) - | 2 \delta_{1} -1|$. In the 
range $\nu_{r} < \nu< \nu_{2}$  the spectral energy density is given by:
\begin{eqnarray}
\Omega_{gw}(\nu, \tau_{0}) = \overline{\Omega}_{gw} \xi_{1}^{\frac{2(\delta_{1} -1) + m_{T}^{(1)}}{(\delta_{1} +1)}}\, \xi_{2}^{4 \alpha(\delta_{2})}\, (\nu/\nu_{2})^{m_{T}^{(2)}},
\label{AU4}
\end{eqnarray}
with $m_{T}^{(2)} = (1 - 3 r_{T}/16)/(1 - r_{T}/16) - | 2 \delta_{2} -1|$. Finally when  $\nu_{eq} < \nu< \nu_{r}$ we have
\begin{eqnarray}
\Omega_{gw}(\nu, \tau_{0}) &=& \overline{\Omega}_{gw} \,\,\xi_{1}^{\frac{2(\delta_{1} -1) + m_{T}^{(1)}}{(\delta_{1} +1)}}\,  
\nonumber\\
&\times& \xi_{2}^{\frac{2(\delta_{2} -1) + m_{T}^{(2)}}{(\delta_{2} +1)}}\,\,(\nu/\nu_{r})^{\overline{m}_{T}}.
\label{AU5}
\end{eqnarray}
Three different dynamical situations can be envisaged. When  $\delta_{1}$ and $\delta_{2}$ are both smaller than $1$ the situation is, in practice, very similar to the one of a single stage expanding slower than radiation; in this case 
the two spectral indices $m_{T}^{(1)}$ and $m_{T}^{(2)}$ will both be positive and lead to a spike for $\nu = {\mathcal O}(\nu_{max})$. In spite of some irrelevant numerical differences, this is exactly the physical case already treated in the previous subsection. In the second case  $\delta_{1}$ and and $\delta_{2}$ are both larger than $1$ and this means that the spectral slopes of 
$\Omega_{gw}(\nu,\tau_{0})$ the high-frequency spectral indices are both negative ( i.e. $m_{T}^{(2)} <0$ and 
$m_{T}^{(1)} <0$); this means that at high-frequencies $h_{0}^2 \Omega_{gw}(\nu,\tau_{0})$ is always smaller than in the conventional case where $\delta_{1}= \delta_{2} \to 1$ and $h_{0}^2 \Omega_{gw}(\nu, \tau_{0}) = {\mathcal O}(10^{-17})$ for $\nu > \nu_{r}$. Furthermore, since $\delta_{1} >1$ and $\delta_{2} > 1$ the maximal frequency will be smaller than ${\mathcal O}(300)$ MHz [see, in this 
respect, Eq. (\ref{AU1}) and recall that $\overline{\nu}_{max} = {\mathcal O}(300) \mathrm{MHz}$]. 
This second case is, in practice, the situation of Refs. \cite{POT1,POT2} where 
$N_{\nu} < {\mathcal O}(60)$ and $r_{T}$ is enhanced instead of being further suppressed.
For the present purposes the relevant case is the third one where $\delta_{1} >1$ and $\delta_{2} < 1$: in this case the spectral energy density exhibits a maximum for $\nu = {\mathcal O}(\nu_{2})$. This happens because when $r_{T}\ll 1$ the spectral index $m_{T}^{(1)} \simeq 2 - 2 \delta_{1} <0$ while $m_{T}^{(2)} \simeq 2 - 2 \delta_{2} >0$: therefore $h_{0}^2 \Omega_{gw}(\nu, \tau_{0})$ 
increases between $\nu_{r}$ and $\nu_{2}$ and it decreases between $\nu_{2}$ and $\nu_{max}$. The presence of an intermediate maximum in $\nu=\nu_{2}$ 
represents the most constrained situation especially if $\nu_{2}$ is located in the audio band where direct constraints are now available from wide-band interferometers \cite{INT1,INT2,INT3,INT5} (see also \cite{LIGO3} and the discussion of 
Sec. \ref{sec5}). 

\renewcommand{\theequation}{5.\arabic{equation}}
\setcounter{equation}{0}
\section{The decelerated timeline}
\label{sec5} 
The spectra of the quantum fields deduced in Secs. \ref{sec3} and \ref{sec4}   
contain an explicit dependence upon the decelerated  
expansion rate. The evolution of the relic gravitons 
and of the gauge spectra lead then to complementary constraints on the 
postinflationary timeline.  It is therefore instructive to combine the 
two classes of physical considerations with the purpose of 
analyzing the simultaneous limits on the general ideas
illustrated in Figs. \ref{FIGURE1}, \ref{FIGURE2} and \ref{FIGURE3}.  
With this logic in mind the subsection \ref{sec5A} focuses on the large-scale magnetism 
while the subsection \ref{sec5B} is devoted to  the graviton 
 spectra. Finally, in subsection \ref{sec5C} the concurrent constraints are finally 
 scrutinized in an extended phenomenological study. 

\subsection{Constraints from large-scale magnetism}
\label{sec5A} 
Since the bunch of wavenumbers associated with the protogalactic collapse 
are of the order of the inverse Mpc, the corresponding 
(comoving) frequencies must be $\nu = {\mathcal O}(\nu_{g})$ 
where $\nu_{g} = k_{g}/(2 \pi) = 1.546 \times 10^{-15}\, \mathrm{Hz}$.
The crossing time $\tau_{\nu}$ associated with this bunch of wavelengths is always smaller than the equality time and $\tau_{\nu}/\tau_{eq}$ is given by\footnote{As before, $\Omega_{R0}$ and  $\Omega_{M0}$ denote the present critical fractions in radiation and matter.}
\begin{equation}
 1.01\times10^{-2} \biggl(\frac{\nu_{g}}{\nu}\biggr)
\biggl(\frac{h_{0}^2 \Omega_{M0}}{0.1386}\biggr) \sqrt{\frac{4.15\times 10^{-5}}{h_{0}^2 \Omega_{R0}}}.
\label{FIVE1}
\end{equation}
While the maximal frequency of the gauge spectrum depends on the postinflationary expansion rate, the crossing time (\ref{FIVE1}) is fixed. More specifically, in the case of a single postinflationary phase preceding the radiation epoch (see Fig. \ref{FIGURE1} and discussion therein) $(\nu/\nu_{max})$ can be related to $(\nu/\nu_{g})$ in the following manner:
\begin{eqnarray}
&& \biggl(\frac{\nu}{\nu_{max}}\biggr) = 5.75\times 10^{-24}\, \biggl(\frac{\nu}{\nu_{g}}\biggr)\, \biggl(\frac{0.06}{r_{T}}\biggr)^{1/4}\, \biggl(\frac{H_{r}}{H_{1}}\biggr)^{\alpha(\delta)}
\nonumber\\
&& \times \biggl(\frac{4.15\times 10^{-5}}{h_{0}^2 \Omega_{R0}}\biggr)^{1/4} \,\biggl(\frac{2.41\times10^{-9}}{{\mathcal A}_{{\mathcal R}}}\biggr)^{1/4}.
\label{FIVE2}
\end{eqnarray}
For a postinflationary expansion 
rate dominated by radiation (i.e. $\delta \to 1$ and $\alpha(\delta) \to 0$)
there are approximately $24$ orders of magnitude 
 between $\nu_{g} = {\mathcal O}( \mathrm{fHz})$ and $\nu_{max}$. 
 When the decelerated rate is slower than radiation 
(i.e. $\alpha(\delta)< 0$ in Eq. (\ref{FIVE2})) the ratio $(\nu_{g}/\nu_{max})$ can even become 
${\mathcal O}(10^{-18})$ since\footnote{For instance, when  $\delta \to 1/2$ (i.e. $\alpha(\delta) \to 1/6$) we have $H_{r}/H_{1}= {\mathcal O}(10^{-38})$.  Because of BBN considerations we must always require $H_{r} \geq 10^{-44}\,\, M_{P}$. Since $H_{\nu} \simeq H_{1} = {\mathcal O}(10^{-6})$ we have that $H_{r}/H_{1} \geq {\mathcal O}(10^{-38})$.}
$H_{r} < H_{1}$. In case a double postinflationary stage precedes the radiation Eq. (\ref{FIVE2}) gets slightly 
different since the term containing the ratio $(H_{r}/H_{1})$ is modified as:
\begin{equation} 
(H_{r}/H_{1})^{\alpha(\delta)} \to (H_{r}/H_{2})^{\alpha(\delta_{2})} \, (H_{2}/H_{1})^{\alpha(\delta_{1})}.
\label{FIVE3}
\end{equation}
Equation (\ref{FIVE3}) can be generalized to multiple phases following the same 
strategy leading to Eq. (\ref{NN2}). For the present ends, however, what matters are only the single and double expanding stages that precede the radiation epoch; this is why, for the 
sake of conciseness, we shall avoid more general expressions.

There are two separate physical regimes where the gauge power spectra of Eqs. (\ref{LT9})--(\ref{LT10}) and (\ref{LT11})--(\ref{LT12}) should be evaluated. The first regime corresponds to typical times $\tau< \tau_{\nu}$ where, as in Eq. (\ref{FIVE1}), $\tau_{\nu}$ denotes the crossing time of the bunch of wavelengths $\nu = {\mathcal O}(\nu_{g})$: in this range the gauge power spectra  do not oscillate but the amplitude of the (physical) power spectra is suppressed both by the expansion of the background and by the dynamics of the gauge coupling. The second range involves typical time scales comparable and larger than the crossing time, i.e. $\tau \geq {\mathcal O}(\tau_{\nu})$.

\subsubsection{Prior to reentry ($\tau \leq \tau_{\nu}$)}
The results of Eqs. (\ref{LT9})--(\ref{LT10}) imply that 
the physical power spectrum of the hypermagnetic fields follows from Eq. (\ref{ONE14c})  
\begin{eqnarray}
{\mathcal P}_{B}(\nu,\tau) &=& g_{y}^2 \,  H_{1}^4 \, D(\gamma+1/2) \, (\nu/\nu_{max})^{n_{B}}\,  
\nonumber\\
&\times& (a_{1}/a)^4\, F_{B}(\tau/\tau_{\nu}). 
\label{FIVE4a}
\end{eqnarray}
where $n_{B} = 5 - |2 \gamma -1| - 2 \zeta$. In the limit $\tau< \tau_{\nu}$ the function $F_{B}(\tau/\tau_{\nu})$ does not oscillate;
furthermore, as mentioned after Eqs. (\ref{LT9})--(\ref{LT10}), we must have that $\zeta \ll \gamma$ since the gauge coupling must flatten out after the end of inflation. In this limit  the Bessel functions entering  $F_{B}(\tau/\tau_{\nu})$ have a simple trigonometric form (i.e. $F_{B}(\tau/\tau_{\nu})\to \sin^2{(\tau/\tau_{\nu})}$) 
so that, after the end of inflation,  
$g_{y}\to g_{1}$ and the gauge coupling flattens out\footnote{The limit $\zeta \to 0$ 
is only well defined after the power spectra have been computed in the case of continuous variation of the gauge coupling. If we would roughly set 
$\zeta=0$ in the evolution of the gauge coupling $g_{y}(\tau)$ would {\em not} be continuous across $\tau = -\tau_{1}$ 
(see Eqs. (\ref{GC1})--(\ref{GC2}) and discussion thereafter).}. Since the non-screened vector modes of the hypercharge field project on the electromagnetic fields through the cosine of the Weinberg angle, an effective coupling 
${\mathcal G}(g_{1}, \cos{\theta_{W}}) = g_{1}^2 \, \cos^2{\theta_{W}}$ can be 
explicitly introduced. While $\cos{\theta_{W}}$ has a well defined value 
$g_{1}$ is undetermined and we shall keep it as free parameter subjected to the requirement $g_{1} \leq 0.01$. Equation (\ref{FIVE4a}) is valid down to the crossing time $\tau = {\mathcal O}(\tau_{\nu})$.
If we now recall Eq. (\ref{FIVE1}) we can see that $\tau_{\nu}$ falls necessarily in the radiation-dominated stage. This means 
that Eq. (\ref{FIVE4a}) can be directly evaluated after $\tau_{r}$, i.e. in the radiation-dominated stage: 
\begin{eqnarray}
 {\mathcal P}_{B}(\nu,\tau) &=& H_{1}^4 D(\gamma+1/2) {\mathcal G}(g_{1}, \cos{\theta_{W}}) (H_{r}/H_{1})^{4\alpha(\delta)}
\nonumber\\
&\times&  (\nu/\nu_{max})^{n_{B}}\, (a_{r}/a)^4\, F_{B}(\tau/\tau_{\nu}).
\label{FIVE4b}
\end{eqnarray}
The amplitude of the physical power 
spectrum appearing in Eq. (\ref{FIVE4b}) is controlled by 
$H_{1}^4$ and since $H_{1}/M_{P}$ can be estimated as $\sqrt{\pi r_{T} {\mathcal A}_{{\mathcal R}}}/4$ we have that the overall normalization of Eq. (\ref{FIVE4b}) can be estimated as\footnote{The Bohr magneton in natural units (i.e. $e/(2 m_{e})$) must equal $5.788\times 10^{-11} 
{\rm MeV}/{\rm T}$.  But since the relation between $\mathrm{T}$ and $\mathrm{G}$ is 
obviously given by $1 \, {\rm T} = 10^{4} \, {\rm G}$ we also have that $\mathrm{G} = 6.9241 \times 10^{-20}\, \mathrm{GeV}^2$. The normalization (\ref{FIVE4c}) follows immediately if we note that $M_{P}^2 = 2.137 \times 10^{48}\,\mathrm{nG}$. }:
\begin{equation}
\frac{\pi^2 r_{T}^2 {\mathcal A}_{{\mathcal R}}^2}{256} \,\,M_{P}^4= 1.76 \times 10^{95} \,\, r_{T}^2 \,\,{\mathcal A}_{{\mathcal R}}^2 \,\,\, \mathrm{nG}^2.
\label{FIVE4c}
\end{equation}
Thanks to Eq. (\ref{FIVE3}),  Eq. (\ref{FIVE4b}) can also be generalized to the case of a double expanding stage with rates $\delta_{1}$ and $\delta_{2}$. 

\subsubsection{After reentry ($\tau \geq \tau_{\nu}$)}
For $\tau \geq \tau_{\nu}$ the evolution equations of the mode functions are modified by the presence of the conductivity and as soon as $\tau = {\mathcal O}(\tau_{\nu})$ their evolution must incorporate the finite value of the conductivity $\sigma_{c}$. While there 
are different ways of accounting of this effect, probably the simplest 
approximation is given by \cite{MGB2}
\begin{equation}
g_{k}^{\prime} = - k^2 f_{k} - \sigma_{c} \, g_{k}, \qquad\qquad f_{k}^{\prime} = g_{k}.
\label{FIVE5}
\end{equation}
To solve Eq. (\ref{FIVE5})  we can use an expansion in $(k/\sigma_{c})$ and directly insert, as initial data at $\tau =\tau_{\nu}$, the values of the mode functions for $\tau \leq \tau_{\nu}$. Since the time  $\sigma_{c} \gg {\mathcal H} = {\mathcal O}(\tau^{-1})$
 the physical conductivity greatly exceeds the Hubble rate i.e. $\sigma_{ph} \gg H$ (where $\sigma_{ph}(\tau) = \sigma_{c}(\tau)/a(\tau)$). At the reentry epoch $\tau = {\mathcal O}(\tau_{\nu})$ and $\tau_{\nu} \sigma_{c} \gg 1$.  
 
 According to Eq. (\ref{FIVE1}) the reentry of the wavelengths corresponding 
 to the frequencies ${\mathcal O}(\nu_{g})$ occurs prior to equality when the 
 evolution is already dominated by radiation; at this stage we can safely estimate the physical conductivity  and get $\sigma_{ph}(t_{eq}) =\, \sqrt{T_{eq}/m_{e}} (T_{eq}/\alpha_{em})$ which is the standard result valid in the case of a cold plasma of electrons and ions \cite{PL1,PL2,PL3}. This means, once more, that $\sigma_{ph}(t_{eq})\gg H_{eq}$ and the hierarchy between these two scales also implies that, 
 out of the two solutions of Eq. (\ref{FIVE5}), only one is physically meaningful 
\begin{equation}
f_{k}(\tau) = f_{k}(\tau_{k}) \,\, e^{- k^2/k_{\sigma}^2}, \quad
g_{k}(\tau) = (k/\sigma) g_{k}(\tau_{k}) \,\,e^{- k^2/k_{\sigma}^2},
\label{FIVE6}
\end{equation}
where $k_{\sigma}(\tau)$ is the magnetic diffusivity scale 
 \begin{equation}
 k^2/k_{\sigma}^{2}  = k^2 \int_{\tau_{k}}^{\tau} \, \frac{d z}{\sigma_{c}(z)}\to\frac{{\mathcal O}(10^{-26}) (k/\mathrm{Mpc}^{-1})^2}{ \sqrt{2 \, h_{0}^2 \Omega_{M0} (z_{\mathrm{eq}}+1)}}.
 \label{FIVE7}
 \end{equation}
 The estimate of Eq. (\ref{FIVE7}) follows by assuming that $\tau = {\mathcal O}(\tau_{eq})$ and it can be refined by computing the transport coefficients of the plasma in different regimes (see, for instance, \cite{MAC}). For the present purposes, however, what matters is that the ratio $(k/k_{\sigma})^2$  is negligibly small  for $\nu = \nu_{g}$ so that the negative exponentials of Eq. (\ref{FIVE6}) evaluate to $1$ and the physical power spectra for  $\tau \gg \tau_{\nu}$ are therefore given by:
\begin{eqnarray}
{\mathcal P}_{B}(\nu,\tau) &=& {\mathcal P}_{B}(\nu, \tau_{\nu}) \,\, [a_{\nu}/a(\tau)]^4 \,\, e^{-2 \nu^2/\nu_{\sigma}^2} 
\nonumber\\
&\to& {\mathcal P}_{B}(\nu, \tau_{\nu}) \,\, [a_{k}/a(\tau)]^4,
\label{cond5}\\
{\mathcal P}_{E}(\nu,\tau) &=& (\nu/\sigma)^2\,\,{\mathcal P}_{E}(\nu, \tau_{\nu}) \,e^{-2 \nu^2/\nu_{\sigma}^2} 
\nonumber\\
&\to& (\nu/\sigma)^2\,\,\,\,{\mathcal P}_{E}(\nu, \tau_{\nu}) \,\,[a_{\nu}/a(\tau)]^4.
\label{cond6}
\end{eqnarray}
The limits appearing in Eqs. (\ref{cond5})--(\ref{cond6}) take into account 
the smallness of $(\nu/\nu_{\sigma})$ and the suppression of the electric 
power spectrum that is a consequence of the standard hydromagnetic evolution;  when the conductivity is large the Ohmic electric field is given by $\vec{E} = (\vec{\nabla} \times \vec{B})/\sigma_{c}$. Within the present notations the suppression of the electric power spectra can therefore be estimated as:
\begin{equation}
(k/\sigma_{c})^2 = {\mathcal O}(10^{-48}) (T/T_{eq})^{-3} \, \, (k/\mathrm{Mpc}^{-1})^2.
\label{cond7}
\end{equation}
The approximate estimate of Eq. (\ref{cond7}) implies that 
the standing oscillations of the gauge power spectra are overdamped by the finite value of the conductivity so that the electric fields get suppressed in comparison with their magnetic counterpart, as it is expected in a good conductor. Bearing in mind Eq. (\ref{cond7}), the results of Eq. (\ref{cond6}) will then be evaluated at the time of the gravitational collapse of the protogalaxy. 

If the protogalactic matter collapsed by gravitational instability over a typical scale ${\mathcal O}(\mathrm{Mpc})$ the mean matter density before collapse was of the order of $\rho_{crit}$. Compressional amplification increases the initial values of the magnetic fields by $4$ or even $5$ orders of magnitude since, after collapse, the mean matter density got larger while the magnetic flux itself is  conserved \cite{MAC,seven,mg04,MAC2,MAC3}. After the collapse, the protogalaxy starts 
rotating with a typical rotation period of ${\mathcal O}(3) \times 10^{8}$ yrs: in this process the kinetic energy associated with the bulk velocity 
of the plasma can turn into magnetic energy \cite{MAC3}. Although the efficiency of this conversion 
can be estimated in different ways the simplest argument is, in short, the following\footnote{If we compare the rotation period with the age of the galaxy (i.e. $\mathcal{O}(10^{10} {\rm yrs})$), the galaxy performed about $30$ rotations since the time 
of the protogalactic collapse. The achievable amplification produced by the 
dynamo instability will be, at most, of ${\mathcal O}(10^{13})$, i.e. about $30$ $e$-folds \cite{seven,mg04}.}.
 By putting together the compressional amplification and the dynamo 
conversion the typical requirements on the physical power spectra imply
\begin{equation}
{\mathcal P}_{B}(\nu, \tau_{0}) \geq {\mathcal O}(10^{-22}) \, \, \mathrm{nG}^2.
\label{REQ1}
\end{equation}
In the most optimistic cases we could even relax the requirement 
of Eq. (\ref{REQ1}) and demand ${\mathcal P}_{B}(k, \tau_{0}) \geq {\mathcal O}(10^{-32}) \, \, \mathrm{nG}^2$. 
This second estimate assumes perfect dynamo efficiency.
In what follows Eq. (\ref{REQ1}) will just be considered as a 
conventional reference value since, generally speaking, we would aim at larger values of the magnetic power 
spectra. 

Recalling Eqs. (\ref{FIVE4b})--(\ref{FIVE4c}) and (\ref{cond5})--(\ref{cond6})
the physical power spectrum associated with a single postinflationary 
stage preceding the radiation epoch be expressed as:
\begin{eqnarray}
\hspace{-0.3cm}
&& {\mathcal P}_{B}(\nu, \tau_{0}) = 2 \,H_{1}^2 \, H_{0}^2\, \Omega_{R0} {\mathcal G}(g_{1}, \cos{\theta_{W}})
D(\gamma+1/2) 
\nonumber\\
\hspace{-0.3cm}
&&\times (H_{r}/H_{1})^{4 \alpha(\delta)} \, d^4(g_{s}, g_{\rho})\, (\nu/\nu_{max})^{n_{B}}.
\label{FFPS1}
\end{eqnarray}
Equation (\ref{FFPS1}) can also be written in an even more explicit form by employing the physical units; in this 
way ${\mathcal P}_{B}(\nu, \tau_{0})/\mathrm{nG}^2$ becomes 
\begin{eqnarray}
\hspace{-0.8cm}
 {\mathcal P}_{B}(\nu, \tau_{0})/\mathrm{nG}^2= 5. 61 \times 10^{10} {\mathcal A}_{{\mathcal R}} \, r_{T}
h_{0}^2 \Omega_{R0} \, \xi^{4 \alpha(\delta)}
\nonumber\\
\hspace{-0.8cm}
\times{\mathcal G}(g_{1}, \cos{\theta_{W}})  \, d^4(g_{s}, g_{\rho})\, (\nu/\nu_{max})^{n_{B}},
\label{FFPS2}
\end{eqnarray}
where, following the previous notations, $\xi = H_{r}/H_{1}$.
The power spectra have been given in the case of a single postinflationary phase preceding 
radiation but they can be easily generalized to the situation of a double phase. In particular, recalling 
Eq. (\ref{FIVE3}) we have that the expression of Eq. (\ref{FFPS2}) can be first 
modified in the amplitude since $\xi^{4\alpha(\delta)} \to \xi_{1}^{4\alpha(\delta_{1})}  \xi_{2}^{4\alpha(\delta_{2})}$. This change in Eq. (\ref{FFPS2}) together with the modified maximal frequency implies that the magnetic power spectrum for a double phase can be expressed as:
\begin{eqnarray}
 {\mathcal P}_{B}(\nu, \tau_{0})/\mathrm{nG}^2= 5. 61 \times 10^{10} {\mathcal A}_{{\mathcal R}} \, r_{T}
h_{0}^2 \Omega_{R0} \, 
\nonumber\\
\times \xi_{1}^{(4 - n_{B}) \alpha(\delta_{1})}\xi_{2}^{(4 - n_{B}) \alpha(\delta_{2})}
\nonumber\\
\times{\mathcal G}(g_{1}, \cos{\theta_{W}})  \, d^4(g_{s}, g_{\rho})\, (\nu/\overline{\nu}_{max})^{n_{B}},
\label{FFPS3}
\end{eqnarray}
It can be directly verified that for $\delta_{1} = \delta_{2} = \delta$
the results of Eq. (\ref{FFPS2}) are recovered since, in this case, 
$\xi_{1} \xi_{2} = \xi = H_{r}/H_{1}$. The dependence on the decelerated timeline 
appearing in Eqs. (\ref{FFPS2})-(\ref{FFPS3}) 
can be constrained by Eq. (\ref{REQ1}) either 
in its conservative or in its relaxed form\footnote{In practice these requirements set a limit on the duration and on the rate of the postinflationary evolution; before addressing this relevant issue  the explicit formulae valid for the spectrum of the relic gravitons must be explicitly deduced. Then in subsection \ref{sec5C} the relevant constraints coming from the relic gravitons and from large-scale magnetogenesis are jointly analyzed.}. 

\subsection{Constraints from graviton spectra}
\label{sec5B}
The direct limits on diffuse backgrounds of gravitational radiation coming from operating interferometers lead to important constraints on the postinflationary timeline. 
These bounds are especially important for a succession 
of two expanding stages with different rates. As already mentioned  
in Sec. \ref{sec1} the wide-band detectors reported a series 
of direct limits implying \cite{INT1,INT2,INT3,INT4,INT5} (see also \cite{LIGO3}):
\begin{equation}
\Omega_{gw}(\nu, \tau_{0}) < 5.8 \times 10^{-9}, 
\label{CONS2}
\end{equation}
for  $20\,\, \mathrm{Hz} < \nu_{au} < 76.6 \,\, \mathrm{Hz}$; throughout the present discussion $\nu_{au}$ 
denotes  frequency of the audio band that we shall broadly consider between few Hz and $10$ kHz with 
a likely value\footnote{An upper limit on $\nu_{au}$  can be 
estimated from the first zero of the so-called overlap reduction function which is determined by the
relative locations and orientations of the two detectors. If the two detectors are colocated 
the overlap reduction function is equal to $1$. If the two detectors 
are not colocated (as it is usually the case) the overlap reduction function is given as a combination of spherical Bessel functions; the first zero of this combination occurs for $\nu_{au} = 1/(2\, d)$ where $d$ denotes the distance between 
the two detectors. For $\nu < \nu_{au}$ we have the most sensitive window for the detection of a relic graviton background.} $\nu_{au} = {\mathcal O}(100) \mathrm{Hz}$.
The result of Eq. (\ref{CONS2}) holds for an exactly scale-invariant spectrum and it improves on a series of bounds previously deduced by the same class of detectors (see Ref. \cite{LIGO3} for a review of the older results).  Within the present notations the parametrization of $\Omega_{gw}(\nu, \tau_{0})$ adopted by Ref.  \cite{INT5} can be written as:
\begin{equation}
\Omega_{gw}(\nu, \tau_{0}) = \overline{\Omega}_{\alpha} (\nu/\nu_{au})^{\alpha},  
\label{NOT1}
\end{equation}
and the three specific cases constrained in Refs. \cite{INT3,INT5} are summarized in Tab. \ref{TABLE1}. 
As the value of $\alpha$ increases from $0$ to $3$ the limits become apparently more restrictive for a fixed reference frequency; the results of Tab. \ref{TABLE1} can be summarized by the following interpolating formula $\log{\overline{\Omega}_{\alpha}} < ( -\,8.236 -\, 0.335\, \alpha- 0.018\, \alpha^2)$.
Since the limits coming from the audio band play a relevant role in the case 
of a double decelerated phase after inflation (but before radiation dominance), the considerations of subsection \ref{subsec4D} apply and the most constraining possibility arises when $\nu_{2} = {\mathcal O}(\nu_{au})$. In this case the dependence upon $\xi_{2}$ can be eliminated since
\begin{equation}
\xi_{2}^{4 \alpha(\delta_{2})} = \xi_{1}^{-1/2} (\nu_{au}/\overline{\nu}_{max}).
\label{SIMP1}
\end{equation}
If we now consider the spectral energy density in the ultra-high frequency branch 
(and evaluate $h_{0}^2 \Omega_{gw}(\nu,\tau_{0})$ in the limit $\nu\to \nu_{max}$)  Eq. (\ref{SIMP1}) implies:
\begin{equation}
h_{0}^2 \Omega_{gw}(\nu_{max}, \tau_{0}) = h_{0}^2 \overline{\Omega}_{gw} \xi_{1}^{4 \alpha(\delta_{1})}
\, \biggl(\frac{\nu_{au}}{\overline{\nu}_{max} \sqrt{\xi_{1}}}\biggr)^4.
\label{SIMP2}
\end{equation}
\begin{table}
\center
\caption{Selected limits on diffuse backgrounds of gravitational radiation from wide-band interferometers.}
\vskip 0.4 cm
\begin{tabular}{||c|c|c||}
\hline
\rule{0pt}{4ex}  $\alpha$ & $\nu_{au}$ [Hz] & Constraints \\
\hline
\hline
$0$ &  $20-81.9$ & $\overline{\Omega}_{0} < 6 \times 10^{-8}$ Ref. \cite{INT3}\\
$2/3$ & $20-95.2$ & $\overline{\Omega}_{2/3} < 4.8 \times 10^{-8}$ Ref. \cite{INT3}\\
$3$ & $20-301$ & $\overline{\Omega}_{3} < 7.9 \times 10^{-9}$ Ref. \cite{INT3}\\
$0$ & $20-76.6$ & $ \overline{\Omega}_{0} <    5.8\times10^{-9}$  Ref. \cite{INT5}\\
$2/3$ & $20-90.6$ & $  \overline{\Omega}_{2/3} < 3.4\times 10^{-9}$  Ref. \cite{INT5}\\
$3$ & $20-291.6$ & $\overline{\Omega}_{3} < 3.9\times 10^{-10}$ Ref. \cite{INT5}\\
\hline
\hline
\end{tabular}
\label{TABLE1}
\end{table}
Equation (\ref{SIMP2}) demonstrates that  $h_{0}^2 \Omega_{gw}(\nu_{max}, \tau_{0})$
is only determined by $\xi_{1}$ and $\alpha(\delta_{1})$; note that if we would require $h_{0}^2 \Omega_{gw}(\nu_{max},\tau_{0}) < 10^{-6}$, the values of $\delta_{1}$ and $\xi_{1}$ would be directly constrained. Similar considerations hold for $h_{0}^2 \Omega_{gw}(\nu_{2}, \tau_{0})$ 
that can be written as:
\begin{equation}
h_{0}^2 \Omega_{gw}(\nu_{2},\tau_{0}) = h_{0}^2 \overline{\Omega}_{gw} (\nu_{au}/\overline{\nu}_{max})^4 \xi_{1}^{\theta(\delta_{1}, m_{T}^{(1)})},
\label{SIMP3}
\end{equation}
where $\theta(\delta_{1}, m_{T}^{(1)}) = (m_{T}^{(1)} -4)/(\delta_{1} +1)$.
Thanks to the previous analytic parametrizations (see Eq. (\ref{NOT1}) and discussion thereafter) from the limits of Tab. \ref{TABLE1} and it makes sense 
to require $h_{0}^2 \Omega_{gw}(\nu_{2}, \tau_{0})< 10^{-9}$. Furthermore, in an optimistic perspective we may also 
impose a lower bound on $h_{0}^2 \Omega_{gw}(\nu_{2}, \tau_{0})$ and hope that $h_{0}^2 \Omega_{gw}(\nu_{2}, \tau_{0})> 10^{-16}$. Overall if $\nu_{2} = {\mathcal O}(\nu_{au})$ it makes sense to demand
\begin{equation}
10^{-16} < h_{0}^2 \Omega_{gw}(\nu_{2}, \tau_{0})< 10^{-9}, \,\,\, \nu_{2} = \nu_{au}.
\label{SIMP4}
\end{equation}
The condition (\ref{SIMP4}) plays some relevant role in the forthcoming phenomenological discussion (see below in this section).

We finally recall that between the pHz and the $100\, \mathrm{nHz}$ the pulsar timing arrays (PTA) might in principle set relevant constraints also for our problem. It turns out, however, that 
the observational limits provided so far are not directly relevant to constrain the postinflationary expansion history. Indeed the relic graviton spectra obtained from a modified postinflationary timeline are smaller than the experimental limits for frequencies ranging\footnote{The operating observational arrays are associated with the NANOgrav collaboration \cite{NANO1,NANO2}, with  the Parkes Pulsar Timing array (PPTA) \cite{PPTA1,PPTA2} and with the European Pulsar Timing array (EPTA) \cite{EPTA1,EPTA2}. There exist a consortium named International Pulsar Timing array (IPTA) \cite{IPTA1}.  The last data of the PTA collaborations have been released \cite{NANO2,PPTA2,EPTA2} together with the results of the Chinese Pulsar Timing array (CPTA) \cite{CPTA}. } between few pHz and the $100\, \mathrm{nHz}$.  The millisecond pulsars can be employed as effective detectors of random gravitational waves for a typical domain that corresponds to the inverse of the observation time during which the pulsar timing has been monitored  \cite{PP1a,PP1b,PP1c}. The signal coming from diffuse backgrounds of gravitational radiation, unlike other noises, should be correlated across the baselines so that the correlation signature of an isotropic and random gravitational wave background should follow the so-called Hellings-Downs curve \cite{PP1c}.  Various upper limits on the spectral energy density of the relic gravitons in the nHz range have been obtained  in the past \cite{PP2a,PP2b,PP2c,PP2d} and during the last six years the PTA reported an evidence that could be attributed to isotropic backgrounds of gravitational radiation. The observational collaborations customarily assign the chirp amplitude at a reference frequency $\nu_{P} = 1/\mathrm{yr} = 31.68\,\, \mathrm{nHz}$, i.e. $h_{c}(\nu,\tau_{0}) = \,Q \,\bigl(\nu/\nu_{P}\bigr)^{\beta}$; note that this exponent $\beta$ is not related to the $\beta$ introduced in section \ref{sec2} (see Eq. (\ref{PP3}) and discussion thereafter). Recalling now the relation between the spectral energy density and the chirp amplitude 
we have $\Omega_{gw}(\nu,\tau_{0})= 2 \pi^2 \nu^2  \,h_{c}^2(\nu, \tau_{0})/(3 H_{0}^2)$. After some 
algebra, recalling the experimental parametrization of the chirp amplitude, we obtain  \cite{LIGO3}:
\begin{equation}
h_{0}^2 \, \Omega_{gw}(\nu,\tau_{0}) = 6.287 \times 10^{-10} \, \, q_{0}^2\,\,\bigl(\nu/\nu_{P}\bigr)^{2 + 2 \beta},
\label{PTAR3}
\end{equation}
where $Q$ has been parametrized as $Q= q_{0} \times 10^{-15}$ 
(and $q_{0}$ is a number of order $1$). For $\nu \to \nu_{P}$  we have 
$h_{0}^2 \, \Omega_{gw}(\nu_{ref},\tau_{0}) =  6.287 \times 10^{-10} \, \, q_{0}^2$,
implying $h_{0}^2 \, \Omega_{gw}(\nu_{ref},\tau_{0}) = {\mathcal O}(2.57) \times 10^{-8}$ in the case of Ref. \cite{NANO2} (for $q_{0} =6.4$) and $h_{0}^2 \, \Omega_{gw}(\nu_{ref},\tau_{0}) = {\mathcal O}(6.04)\times 10^{-9}$ for Ref. \cite{PPTA2} (for $q_{0} =3.1$).  With the same logic we can also deduce 
the explicit relation between the spectral and the chirp amplitudes:
\begin{equation}
S_{h}(\nu, \tau_{0}) = 3.15 \, \times 10^{-23} \, \,  q_{0}^2\,\,\bigl(\nu/\nu_{P}\bigr)^{2 \beta-1} \, \, \mathrm{Hz}^{-1}.
\label{PTAR4}
\end{equation}
For a direct comparison with the spectral amplitude of the noise\footnote{ The existence of a spectral amplitude implicitly suggests that the signal comes from a stationary stochastic process. However relic gravitons lead to stochastic processes that are not stationary (see \cite{UFS2,UFS3} and references therein).}, it is also customary to employ $\sqrt{S_{h}(\nu, \tau_{0})} =5.61\times10^{-12} \, \, q_{0} (\nu/\nu_{P})^{\beta -1/2} \,\, \mathrm{Hz}^{-1/2}$. Because the largest contribution to $\Omega_{gw}(\nu, \tau_{0})$ from a modified decelerated timeline is obtained for maximally stiff stage of expansion lasting between $\nu_{bbn}$ and $\nu_{max}$. In this case, however, $\Omega_{gw}(\nu_{P}, \tau_{0}) = {\mathcal O}(10^{-13})$ which is always smaller than the potential constraint  provided by Eq. (\ref{PTAR3}). 

\subsection{Concurrent constraints}
\label{sec5C}
\begin{figure} 
\begin{center}
\includegraphics[width=7cm,height=7cm]{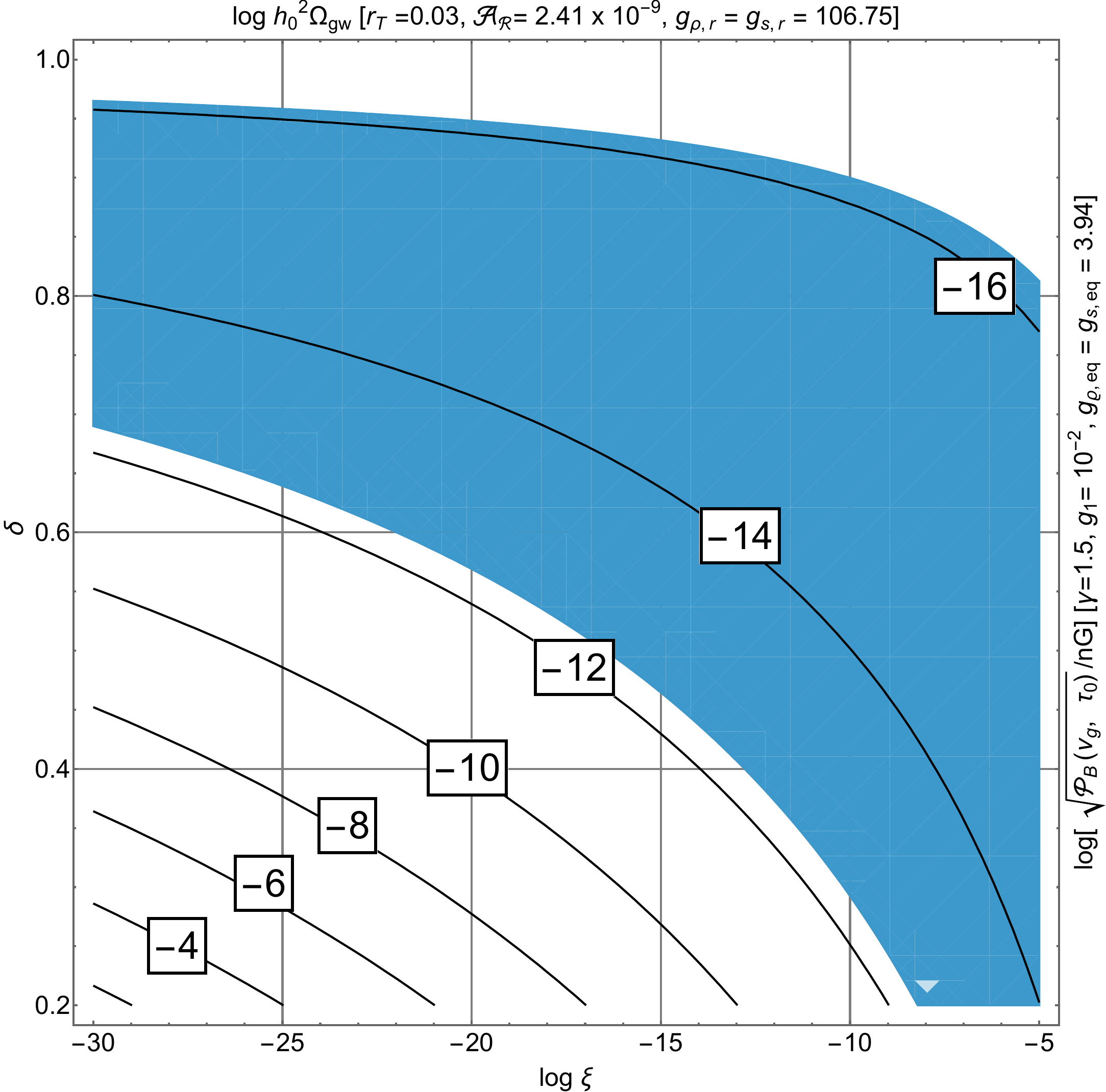}
\end{center}
\caption{\label{FIGURE4} The constraints on the duration and on the
rate of the decelerated stage are illustrated for a single phase preceding the radiation epoch. 
 The values of $\log{\xi}$ exceed $-38$ (as implied by the constraints 
associated with BBN) and the corresponding values of $\delta$ fall in the range of expansion rates 
slower than radiation. In the shaded region $h_{0}^2 \Omega_{gw}(\nu_{max},\tau_{0})$ 
ranges between $10^{-15}$ and $10^{-5}$. The labels appearing on the various contours 
indicate the common logarithm of $\sqrt{{\mathcal P}_{B}(\nu_{g},\tau_{0})}$ expressed in units of nG.}
\end{figure}
\begin{figure}
\begin{center}
\includegraphics[width=7cm,height=7cm]{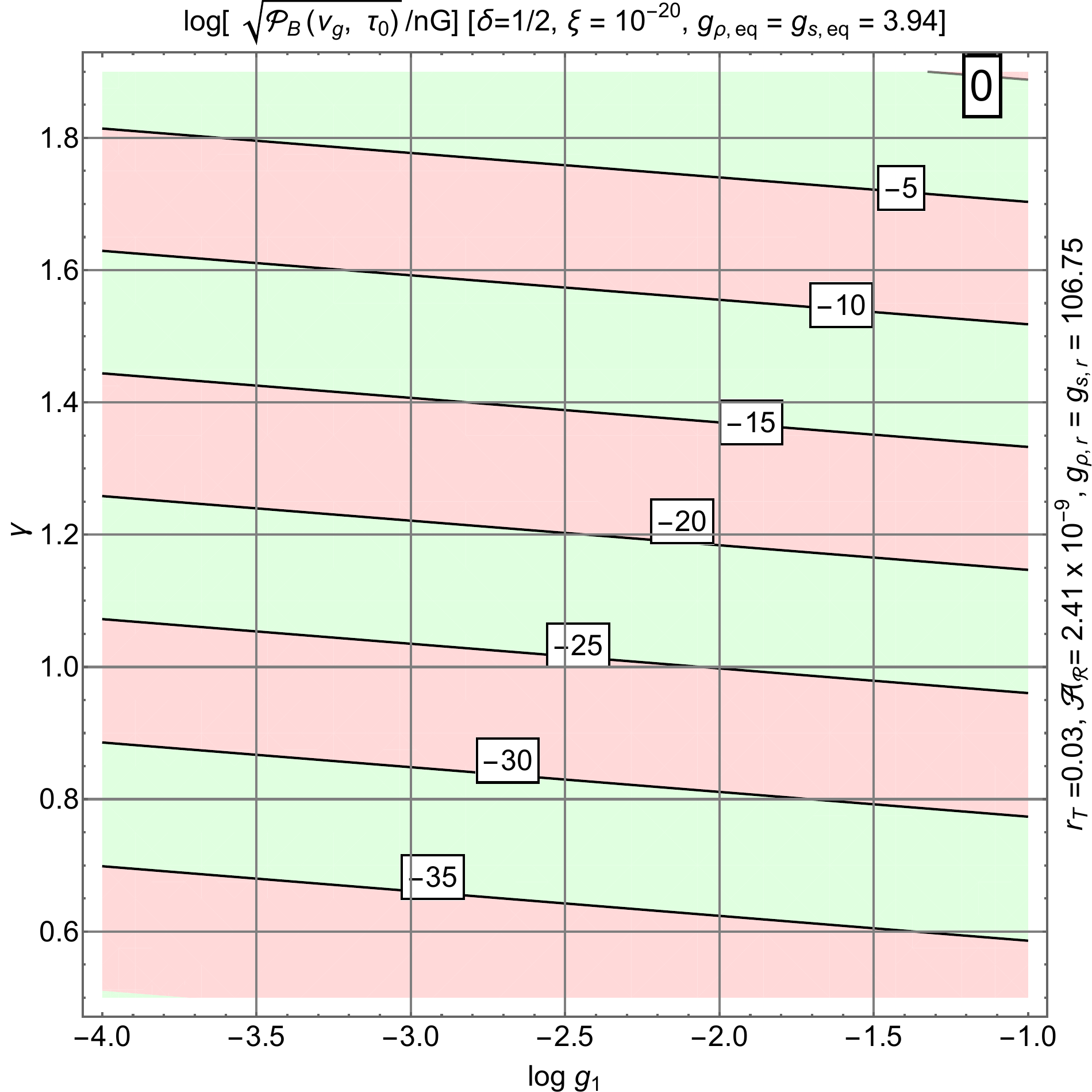}
\end{center}
\caption{\label{FIGURE5}
The common logarithm of $g_{1}$ is reported on the horizontal axis while, on the vertical axis, the value  $\gamma$ is illustrated. The magnetogenesis constraints are 
satisfied when  $\sqrt{{\mathcal P}_{B}(\nu_{g},\tau_{0})}> 10^{-11} \mathrm{nG}$. 
This demand can also be relaxed to  $\sqrt{{\mathcal P}_{B}(\nu_{g},\tau_{0})}> 10^{-16} \mathrm{nG}$ in the presence of an efficient 
dynamo action. As in Fig. \ref{FIGURE4} (and in all subsequent plots) the labels appearing on the contours 
indicate the common logarithm of $\sqrt{{\mathcal P}_{B}(\nu_{g},\tau_{0})}$ expressed in units of nG.}
\end{figure}
The constraints on the decelerated rate of expansion coming from the dynamics of the gauge fields and from the graviton spectra are now considered in a consistent perspective. In the first part of the discussion the attention is focused on a single postinflationary stage preceding the radiation epoch while the second part of the analysis is instead devoted to the presence of two successive decelerated phases taking place prior to radiation dominance. 

\subsubsection{Single postinflationary stage of expansion}
Since radiation becomes dominant at $H_{r}$, the condition $H_{r} \geq H_{bbn}$ must always be enforced so that the plasma will be dominated by radiation 
prior to BBN. This requirement complies with the limits coming from CMB physics \cite{CONF2,CONF3,CONF4,CONF5,CONF6,CONF7,CONF8,CONF9,CONF10,CONF11}
since the initial conditions of the temperature and polarization anisotropies are sensitive to the expansion rate and are set right after neutrino decoupling (i.e. approximately for temperatures smaller than the MeV) when the Universe is already dominated by radiation. The scale associated with gravitational collapse of the protogalaxy reenters 
prior to matter-radiation equality (i.e. for $H< H_{r}$) and, according to the present analysis, the parameters to be constrained are: {\it (i)} the tensor to scalar ratio $r_{T}$; {\it (ii)} the duration 
of the postinflationary stage prior to the onset of radiation (i.e. $\xi= H_{r}/H_{1}$); {\it (iii)} the rate of the postinflationary evolution $\delta$; {\it (iv)} the rate of the evolution of the gauge coupling. The duration of the postinflationary phase can be parametrized in terms of $\xi = H_{r}/H_{1}$ but since $H_{1}$ also contains a dependence upon $r_{T}$ it is possible to trade $\xi$ for $H_{r}/M_{P}$. For similar reasons, even if the expansion rate is parametrized by $\delta$,  when the stiff phase is associated with the coherent oscillations 
of an appropriate potential we employ $q$ as pivotal parameter (see Eq. (\ref{PP1}) and discussions thereatfer).  

In Fig. \ref{FIGURE4} the constraints are illustrated in the plane defined 
by the common logarithm of $\xi$ and by the expansion rate $\delta$. We have selected, for simplicity, $r_{T} =0.03$ and two fiducial values for $\gamma$ and $g_{1}$; the values of the other quantities have been listed in each of the plots of Fig. \ref{FIGURE4} and of all the subsequent figures. The labels appearing on the contours correspond to the common logarithm of $\sqrt{{\mathcal P}_{B}(\nu_{g},\tau_{0})}$ (expressed in nG) while the shaded area pins down the region of the parameter space where $h_{0}^2 \Omega_{gw}(\nu_{max}, \tau_{0})$ 
ranges between $10^{-15}$ and $10^{-5}$. Recalling that Eq. (\ref{REQ1}) would imply $\sqrt{{\mathcal P}_{B}(\nu_{g},\tau_{0})}/\mathrm{nG} \geq 10^{-11}$ (or $\sqrt{{\mathcal P}_{B}(\nu_{g},\tau_{0})}/\mathrm{nG} \geq 10^{-16}$ in the case of an efficient dynamo action) from the shaded area of Fig. \ref{FIGURE4} the power spectra at the galactic frequency 
always exceed $10^{-16} \mathrm{nG}$ but do not comply with the magnetogenesis requirement in its stricter form. While this result suggests the need of a complementary dynamo action (as already discussed in connection with Eq. (\ref{REQ1})), a large signal of relic gravitons near the maximal frequency is only marginally 
compatible with a phenomenologically relevant magnetic field coherent over the scale 
of the protogalactic collapse. This conclusion may slightly change depending on the growth rate of the gauge coupling and in Fig. \ref{FIGURE4} we have chosen $\gamma = 1.5$; as $\gamma\to 2$  the magnetic power spectra at late times experience a further increase. We also recall, in this respect, that $\gamma \leq 2$ and $g_{1} \leq 0.01$ since these conditions ensure that the hypermagnetic and hyperelectric fields are subcritical during inflation. 

In Fig.  \ref{FIGURE5} the variation of the physical power spectra is explored in the $(g_{1},\, \gamma)$ plane
while $\xi$ and $\delta$ have been fixed\footnote{As already stressed in this discussion 
the labels in the plot correspond to the common logarithm $\sqrt{{\mathcal P}_{B}(\nu_{g},\tau_{0})}$ expressed in nG [i.e.
$\log{(\sqrt{{\mathcal P}_{B}(\nu_{g},\tau_{0})}/\mathrm{nG})}$].}. 
This means that the magnetogenesis requirements for $\xi = {\mathcal O}(10^{-20})$ and 
$\delta \to 1/2$ are satisfied when $\gamma$ falls between $1$ and $2$.
Since $\gamma$ also determines the slopes of the gauge power spectra,
in the quasi-flat case (i.e. $1.5 < \gamma \leq 2$) we can safely assume that there are regions where $\sqrt{{\mathcal P}_{B}(\nu_{g},\tau_{0})}/\mathrm{nG} \geq 10^{-11}$.
\begin{figure}
\begin{center}
\includegraphics[width=7cm,height=7cm]{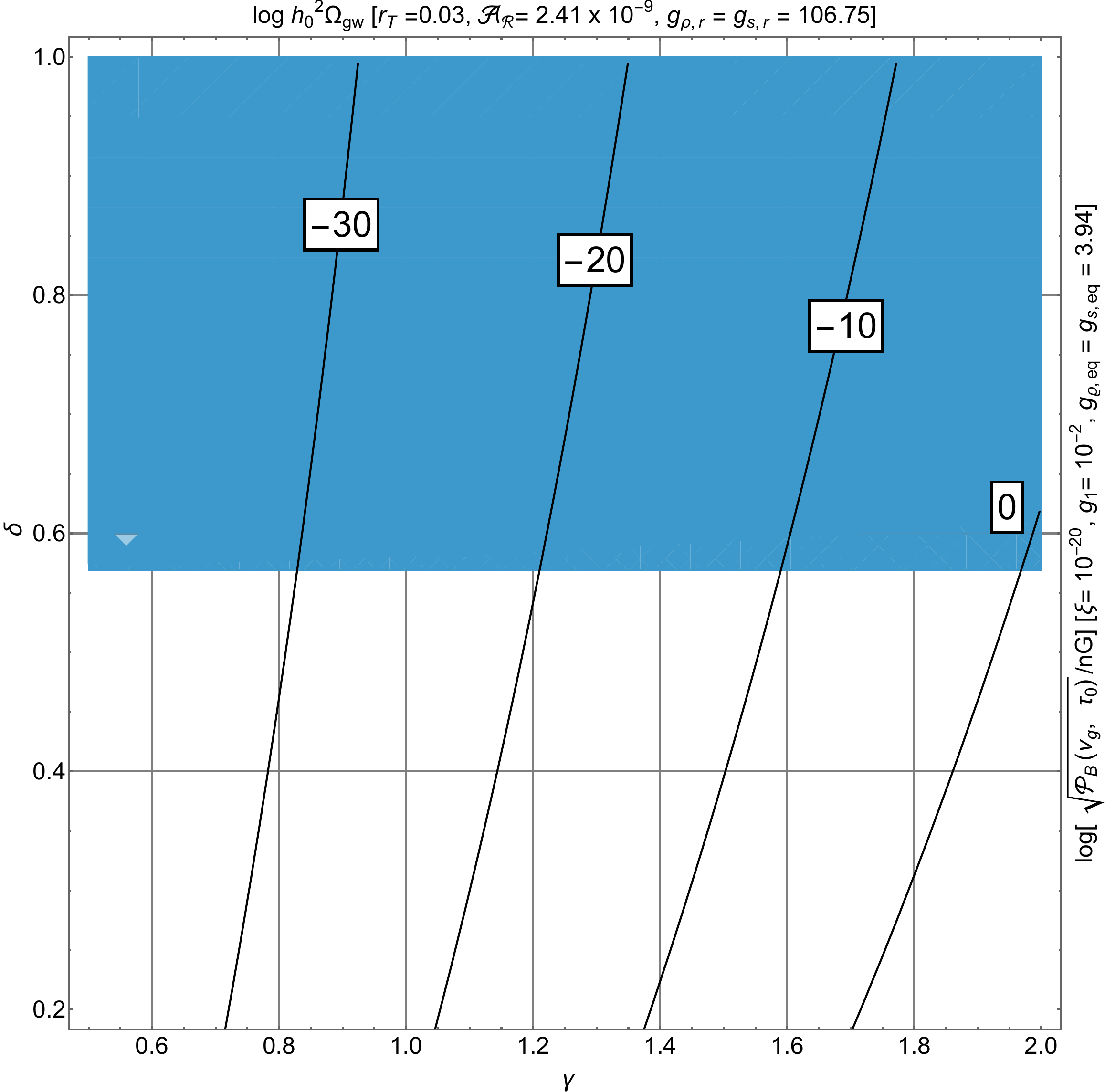}
\end{center}
\caption{\label{FIGURE6} The plane $(\gamma,\, \delta)$ is illustrated for a fixed 
duration of the decelerated stage of expansion prior to radiation (i.e. $\xi \to 10^{-20}$, 
the same value already assumed in Fig. \ref{FIGURE5}). The doubly shaded region corresponds to the critical density bounds applied to the spectral energy densities 
of the relic gravitons {\em and} of gauge fields. Once more the labels appearing on the various contours correspond to the common logarithm of $\sqrt{{\mathcal P}_{B}(\nu_{g},\tau_{0})}$ expressed in nG.}
\end{figure}
\begin{figure}
\begin{center}
\includegraphics[width=7cm,height=7cm]{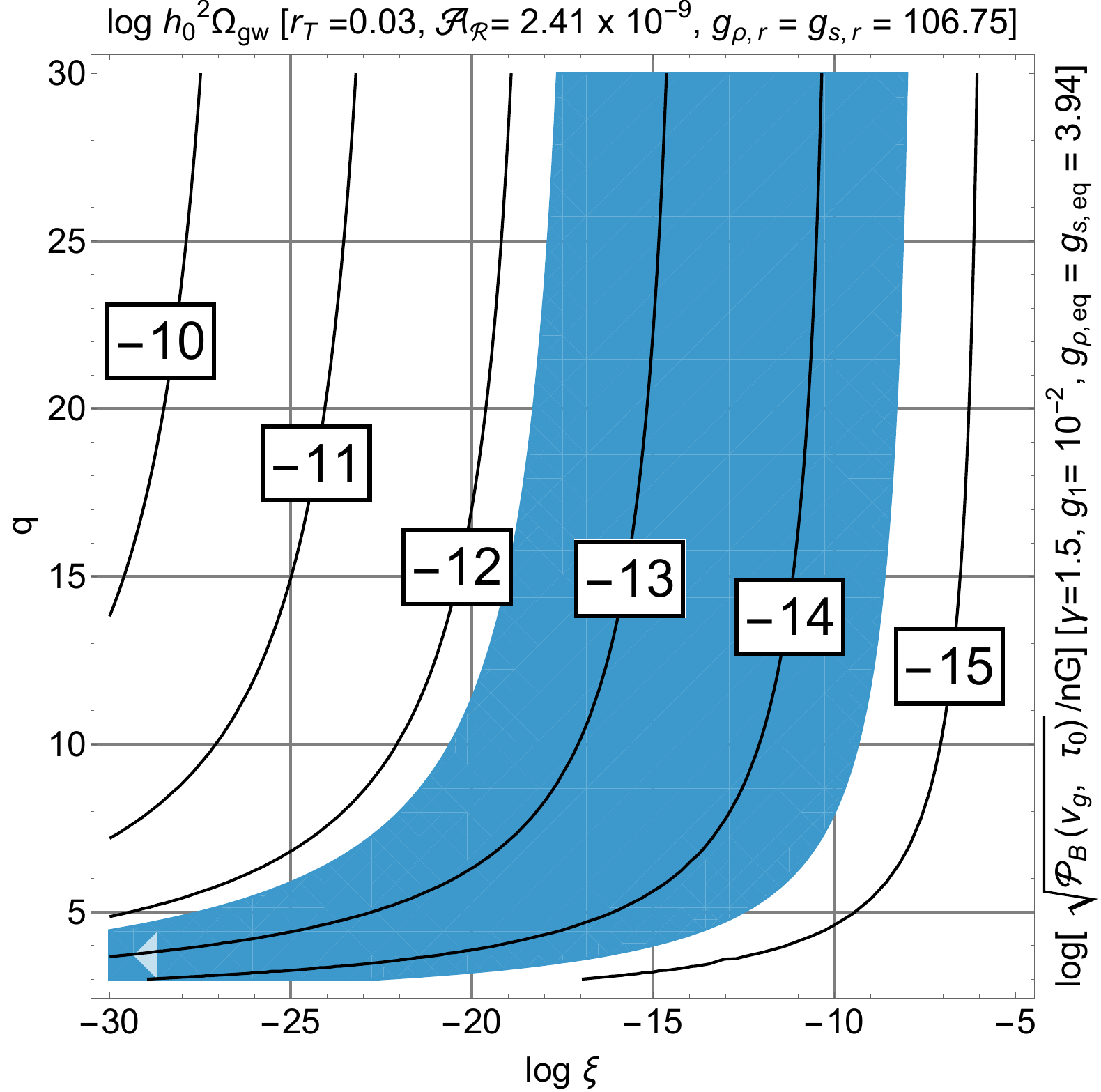}
\end{center}
\caption{\label{FIGURE7} The parameter space is illustrated in the plane defined by $\log{\xi}$ and $q$.  As in Fig. \ref{FIGURE4} the 
shaded area corresponds to the range where  $10^{-15} \leq h_{0}^2 \Omega_{gw}(\nu_{max},\tau_{0})\leq 10^{-6}$. As in the previous 
plots the labels on the different curves illustrate the common logarithm of $\sqrt{{\mathcal P}_{B}(\nu_{g},\tau_{0})}/\mathrm{nG}$.
The condition $\sqrt{{\mathcal P}_{B}(\nu_{g},\tau_{0})}/\mathrm{nG} \geq 10^{-11}$ (see Eq. (\ref{REQ1}))  is only partially
 satisfied. For an efficient dynamo action the condition (\ref{REQ1}) can be relaxed (e.g. $\sqrt{{\mathcal P}_{B}(\nu_{g},\tau_{0})}/\mathrm{nG} \geq 10^{-16}$); this second condition is compatible with the shaded region. }
\end{figure}

\begin{figure}
\begin{center}
\includegraphics[width=7cm,height=7cm]{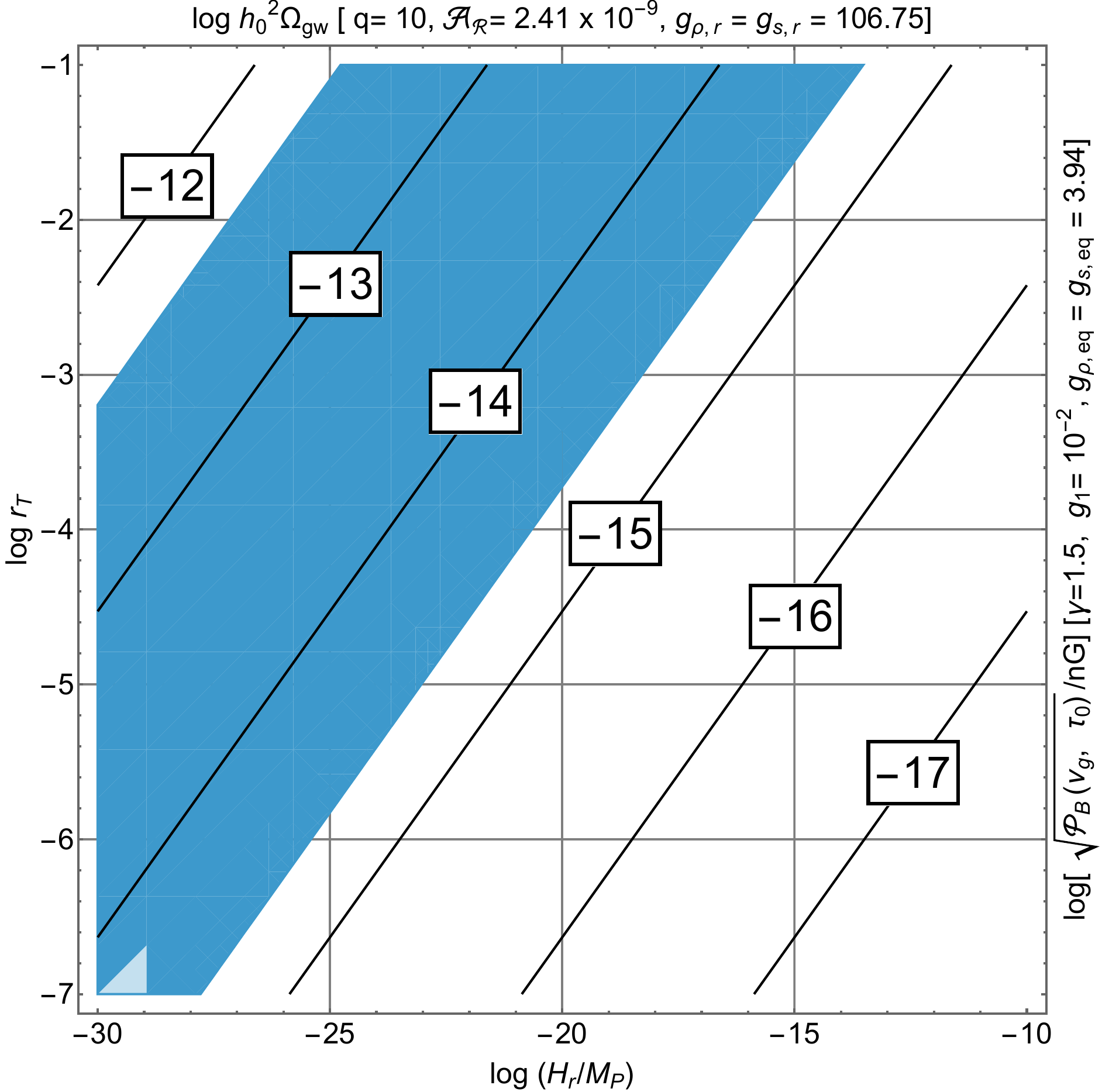}
\end{center}
\caption{\label{FIGURE8} 
The parameter space is now illustrated in the plane $(H_{r}/M_{P}, \, r_{T})$. Common 
logarithms are employed on both axes. The ranges of $h_{0}^2 \Omega_{gw}(\nu_{max},\tau_{0})$ associated with the shaded 
area correspond to the ones of Fig. \ref{FIGURE7} (i.e. between $10^{-6}$ and $10^{-11}$). The labels appearing on the 
various contours indicate the common logarithm of  $\sqrt{{\mathcal P}_{B}(\nu_{g},\tau_{0})}/\mathrm{nG}$.}
\end{figure}

\begin{figure}
\begin{center}
\includegraphics[width=7cm,height=7cm]{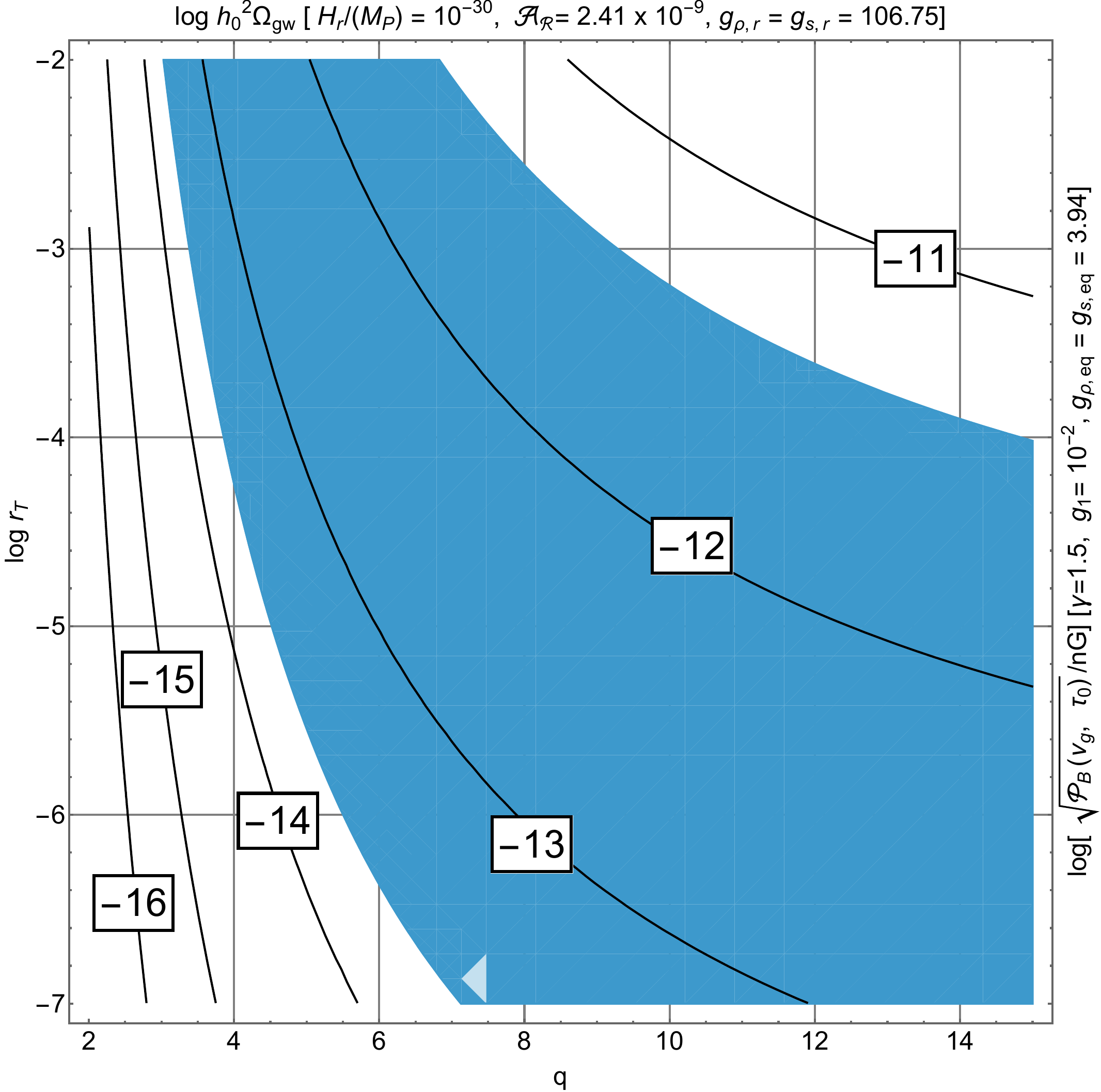}
\end{center}
\caption{\label{FIGURE9} The parameter space is examined in the plane 
$(q,\, r_{T})$. After setting the value of $H_{r}/M_{P}$ to $10^{-30}$, in the shaded region $h_{0}^2 \Omega_{gw}(\nu_{max},\tau_{0})$ ranges between $10^{-11}$ and $10^{-6}$. The labels of the contours suggest that the common logarithm 
of $\sqrt{{\mathcal P}_{B}(\nu_{g},\tau_{0})}/\mathrm{nG}$ ranges between $-16$ and $-11$. Thus the requirements of Eq. (\ref{REQ1}) are only satisfied in the case of an efficient dynamo action. Note the region of small $r_{T}$ (i.e. invisible gravitons) corresponds, as expected, to 
$q \gg 1$ (i.e. expansion rates much slower than radiation).}
\end{figure}

It is then useful to fix $\xi$ and investigate the plane $(\gamma, \delta)$; this 
analysis is illustrated in Fig. \ref{FIGURE6}. The shaded 
region (where the signal of relic gravitons is potentially large) is not affected by
$\gamma$ that does not enter the spectral energy density of the 
relic gravitons. In Fig. \ref{FIGURE4} the parameter space has been illustrated in the $(\log{\xi}, \delta)$ plane
but a similar analysis can be presented also in terms of $q$; we remind that the coherent 
oscillations of a potential may lead to an effective $q$-dependence of the expansion rate 
and this possibility is illustrated in Fig. \ref{FIGURE7}.
In Fig. \ref{FIGURE7}, for the sake of 
illustration,  the shaded area corresponds 
to the region where $h_{0}^2 \Omega_{gw}(\nu_{max},\tau_{0})$ ranges between $10^{-11}$ and $10^{-6}$ (on purpose this requirement is slightly different in comparison of the one employed in Fig. \ref{FIGURE4}).  While $r_{T}$ has been previously set to $0.03$ (which is 
close to the current observational limit), the variation of $r_{T}$ is specifically investigated in Figs. \ref{FIGURE8} and \ref{FIGURE9}. In particular 
in Fig. \ref{FIGURE8} the value of $q$ is  fixed (i.e. $q\to 10$) and, 
 as usual, we consider the situation where $H_{r} > 10^{-44}\, M_{P}$ and the protogalactic scales 
 reenters during the radiation stage. To investigate the explicit variation of $r_{T}$ we must 
 trade $\xi= H_{r}/H_{1}$ for $H_{r}/M_{P}$. Indeed since $H_{1} \simeq H_{\nu}$ 
 (and $H_{\nu}/M_{P} = \sqrt{\pi r_{T} {\mathcal A}_{{\mathcal R}}}/4$) the variable $\xi$ is implicitly affected by $r_{T}$ whose dependence must be excluded by considering $H_{r}/M_{P}$ rather than $H_{r}/H_{1}$.
From Fig. \ref{FIGURE8} it also follows that a drastic reduction of $r_{T}$ (well below the current 
 observational limits) does not reduce the high-frequency signal and is compatible 
 with the magnetogenesis constraints in their relaxed version.

To complete the discussion in Fig. \ref{FIGURE9} the value of $H_{r}/M_{P}$ 
is fixed and the variation of $r_{T}$ is examined together with 
the dependence upon $q$. As expected it appears that the region of large $q$ (corresponding 
to a rate much smaller than the one of radiation) is compatible with a significant reduction of $r_{T}$ while the high-frequency signal and the magnetogenesis requirements are preserved. 
Overall a signal coming from the relic gravitons in the ultra-high 
frequency range is compatible with a large magnetic field at the protogalactic scale and 
with a very small value of $r_{T}$ in the aHz domain. This situation 
can be dubbed by saying that invisible gravitons and successful large-scale 
magnetogenesis are not incompatible; both possibilities may lead to a large spike 
in $h_{0}^2 \Omega_{gw}(\nu,\tau_{0})$ between few GHz and the THz.
The amplitude of the spike can even be $10$ or $11$ orders 
of magnitude larger than the signal of the concordance paradigm where 
$h_{0}^2\Omega_{gw}(\nu,\tau_{0})$ is typically ${\mathcal O}(10^{-17})$ (or smaller).
Also in Fig. \ref{FIGURE9} when the signal of the relic gravitons is maximized the magnetogenesis requirements are only partially compatible with the limit deduced in Eq. (\ref{REQ1}). However, in the presence of an efficient dynamo action, the 
two classes of constraints are compatible since, in the shaded region,  $\sqrt{{\mathcal P}_{B}(\nu_{g},\tau_{0})}/\mathrm{nG} \geq 10^{-16}$.  Indeed, the allowed region of Fig. \ref{FIGURE9}  corresponds to $r_{T}\ll {\mathcal O}(10^{-2})$ (i.e. invisible gravitons in the aHz region) and large values of $q$ (i.e. an evolution slower than radiation in the decelerated phase 
preceding the radiation epoch). 

\subsubsection{Double decelerated stage of expansion}
If an initial decelerated stage expanding faster than radiation\footnote{When a double stage 
of decelerated expansion takes place before the radiation 
dominance, the two phases are characterized by the rates $\delta_{1}$ and $\delta_{2}$ (see Sec. \ref{sec4} and discussion therein).} (i.e. $\delta _{1} > 1$) is followed by a phase with rate slower than radiation (i.e. $\delta_{2} < 1$) $h_{0}^2 \Omega_{gw}(\nu,\tau_{0})$ develops a 
maximum at intermediate frequencies. The most interesting physical situation coincides 
with the possibility that this maximum falls exactly in the audio band.  For this succession of rates the high-frequency spectral index is negative (i.e. $m_{T}^{(1)} < 0$)  and  $h_{0}^2 \Omega_{gw}(\nu,\tau_{0})$ decreases for $\nu>  \nu_{2}$. Given that the intermediate spectral index is instead positive 
(i.e. $m_{T}^{(2)} > 0$),  $h_{0}^2 \Omega_{gw}(\nu,\tau_{0})$ increases for $\nu<  \nu_{2}$. The typical frequency of the intermediate 
maximum is therefore of the order of $\nu_{2}$. Since the most constrained intermediate range 
falls in the audio band it makes sense to consider the situation where $\nu_{2} = \nu_{au} = {\mathcal O}(100) \mathrm{Hz}$.
\begin{figure}
\begin{center}
\includegraphics[width=7cm,height=7cm]{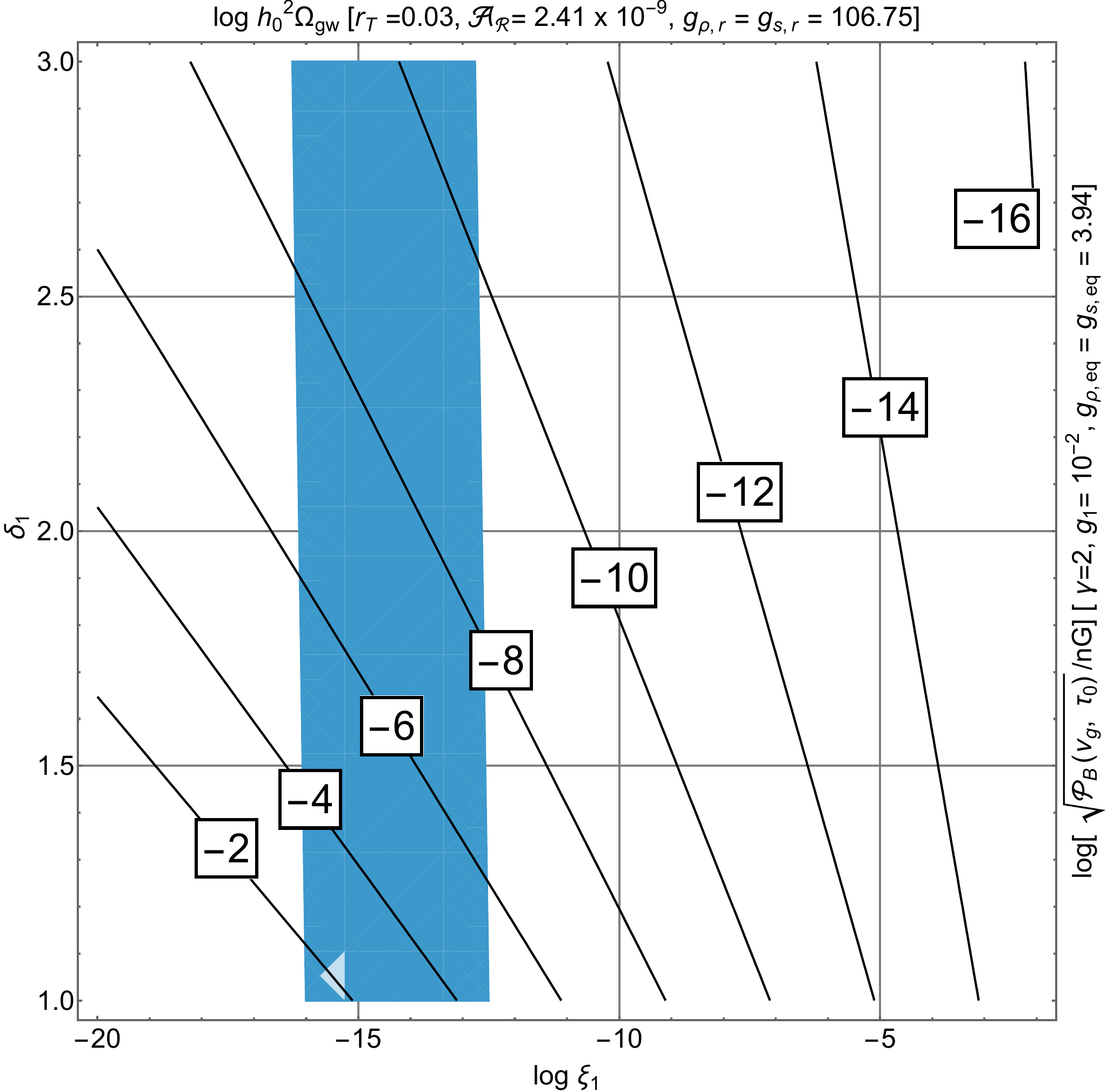}
\end{center}
\caption{\label{FIGURE10} The plane 
$(\xi_{1},\, \delta_{1})$ is analyzed when a double decelerated stage precedes radiation dominance.  The various parameters have been fixed to their fiducial values as illustrated in the plot and the dashed region corresponds to the constraints coming from the audio band (i.e. $10^{-16} < h_{0}^2 \Omega_{gw}(\nu_{2}, \tau_{0}) < {\mathcal O}(10^{-9})$). We also require  that $h_{0}^2 \Omega_{gw}(\nu_{max},\tau_{0}) < {\mathcal O}(10^{-6})$; as before the labels on the various contours denote the common logarithms of $\sqrt{{\mathcal P}_{B}(\nu_{g},\tau_{0})}/\mathrm{nG}$. }
\end{figure}

In Fig. \ref{FIGURE10} the shaded region corresponds to the requirement that 
$10^{-16} < h_{0}^2 \Omega_{gw}(\nu_{2}, \tau_{0}) < {\mathcal O}(10^{-9})$; in this range the 
upper bound comes from the direct constraints in the audio band while the lower
bound only represents a very optimistic reference value describing  
the claimed sensitivities in the frequency domain of $0.1$ kHz.
The shaded slice of Fig. \ref{FIGURE10} complies with the magnetogenesis 
requirements in their most demanding form (i.e. ${\mathcal P}_{B}(\nu_{g}, \tau_{0}) \geq 10^{-11} \, \mathrm{nG}$) and it is also consistent with a maximum of $h_{0}^2 \Omega_{gw}(\nu, \tau_{0})$ for $\nu= {\mathcal O}(\nu_{au})$. 
\begin{figure} 
\begin{center}
\includegraphics[width=7cm,height=7cm]{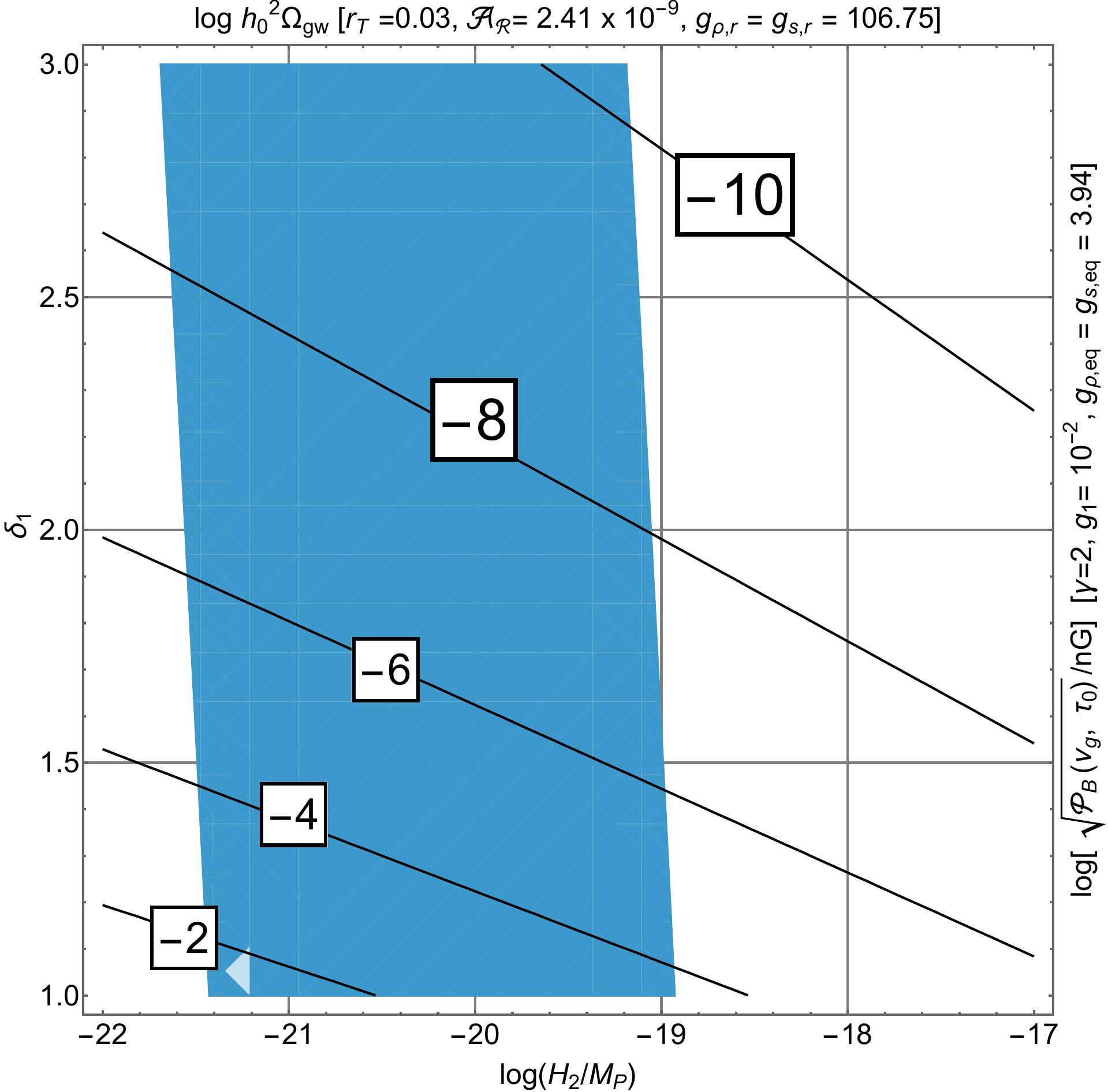}
\end{center}
\caption{\label{FIGURE11} The parameter space of Fig. \ref{FIGURE10} is now examined in the plane $(H_{2}/M_{P},\, \delta_{1})$. Common logarithms are employed on the horizontal axis. The shaded region corresponds to the range 
$10^{-16} < h_{0}^2 \Omega_{gw}(\nu_{2}, \tau_{0}) < {\mathcal O}(10^{-9})$ implying the compatibility of the local maximum 
with the current constraints coming from wide-band detectors. As in the case of Fig. \ref{FIGURE10} the common logarithm of $\sqrt{{\mathcal P}_{B}(\nu_{g},\tau_{0})}/\mathrm{nG}$ is always larger than $-11$. This means that a maximum in the spectral energy density of the relic gravitons is compatible with the condition (\ref{REQ1}) imposed by a successful magnetogenesis scenario.}
\end{figure}
The typical values of the physical power spectra $\sqrt{{\mathcal P}_{B}(\nu_{g},\tau_{0})}$ range between ${\mathcal O}(10^{-4})$ nG and ${\mathcal O}(10^{-10})$ nG for $\gamma = 2$.
We now recall that $\xi_{1} = H_{2}/H_{1}$ and since $H_{1}/M_{P} = \sqrt{\pi {\mathcal A}_{\mathcal R} r_{T}}/4$ we can always trade $\xi_{1}$ for $H_{2}/M_{P}$.  Because the condition $\nu_{2} = {\mathcal O}(\nu_{au})$ 
implicitly imposes a relation between $\xi_{2}$, $\xi_{1}$ and $\overline{\nu}_{max}$, when $\nu_{2} = {\mathcal O}(\nu_{au})$ the dependence of $h_{0}^2 \Omega_{gw}(\nu,\tau_{0})$ upon $\delta_{2}$ and $\xi_{2}$ can be effectively eliminated. 
From the technical viewpoint the relation  $\nu_{2} = {\mathcal O}(\nu_{au})$ implies $\xi_{2}^{\alpha(\delta_{2})} = \xi_{1}^{-1/2}\, (\nu_{au}/\overline{\nu}_{max})$. Therefore the requirement that $\nu_{2} = \nu_{au} = {\mathcal O}(100)\,\,\mathrm{Hz}$ 
simplifies the phenomenological discussion: instead of dealing with two scales (i.e. $\xi_{1}$ and $\xi_{2}$) and two rates (i.e. 
$\delta_{1}$ and $\delta_{2}$) the dependence upon $\xi_{2}$ and $\delta_{2}$ can be eliminated. 

Let us consider, as an example, $\nu_{max}$; this quantity depends, in principle, 
on $\xi_{1}$, $\xi_{2}$, $\delta_{1}$ and $\delta_{2}$. 
However because $\xi_{2}^{\alpha(\delta_{2})} = \xi_{1}^{-1/2}\, (\nu_{au}/\overline{\nu}_{max})$ the expression of $\nu_{max}$ becomes 
\begin{eqnarray}
 \nu_{max} = \xi_{1}^{-1/(\delta_{1}+1)} \nu_{au},\quad \xi_{1} < 1, 
\label{DPH1}
\end{eqnarray}
and only depends upon $\xi_{1}$. Consistently with the whole construction it always happens that $\nu_{max} > \nu_{au}$; this is because $\delta_{1}> 1$ and $\xi_{1} <1$ in Eq. (\ref{DPH1}).  
Thus although the spectral energy 
density of the relic gravitons evaluated at $\nu_{max}$ formally depends upon $\xi_{1}$ and $\xi_{2}$, the relevant constraints can be directly expressed in the plane $(\xi_{1}, \, \delta_{1})$. Thus, in case $\nu_{2} = \nu_{au} = {\mathcal O}(100)\,\,\mathrm{Hz}$
we have 
\begin{equation}
h_{0}^2 \Omega_{gw}(\nu_{max}, \tau_{0}) = h_{0}^2 \overline{\Omega}_{gw} \xi_{1}^{-4/(\delta_{1} +1)} b^4(\nu_{au}),
\label{DPH2}
\end{equation}
and $b(\nu_{au})=\nu_{au}/\overline{\nu}_{max}$. Similarly when $\nu \to \nu_{2} = \nu_{au}$ the spectral 
energy density becomes 
\begin{equation}
h_{0}^2 \Omega_{gw}(\nu_{2}, \tau_{0}) = h_{0}^2 \overline{\Omega}_{gw} \xi_{1}^{-n(r_{T},\delta_{1})} b^4(\nu_{au}),
\label{DPH3}
\end{equation}
where the spectral index depends upon $r_{T}$ and $\delta_{1}$ and it is now given by:
\begin{equation}
n(r_{T},\delta_{1}) = \frac{3 - (16-3 r_{T})/(16 - r_{T}) + 2 \delta_{1}}{\delta_{1} +1}.
\label{DPH4}
\end{equation}

With the same strategy leading to Eqs. (\ref{DPH2}) and (\ref{DPH3})--(\ref{DPH4}) we can also 
express the (physical) magnetic power spectrum in the case $\nu_{2} = {\mathcal O}(\nu_{au})$. 
Because the exact expression is a bit lengthy we prefer to focus on the scaling associated 
with the relevant parameters $\xi_{1}$ and $\delta_{1}$, namely 
\begin{eqnarray}
&&{\mathcal P}_{B}(\nu_{g}, \tau_{0}) = {\mathcal O}(10^{-16}) \xi_{1}^{(n_{B}-4)/(\delta_{1}+1)} 
\nonumber\\
&&\times b^4(\nu_{au})(\nu_{g}/\nu_{au})^{n_{B}},
\label{DPH5}
\end{eqnarray}
where $n_{B} = 3 - |2 \gamma -1|$.
With this logic in Fig. \ref{FIGURE11} the parameter space is illustrated 
in the plane $(H_{2}/M_{P}, \, \delta_{1})$. We stress that both in Figs. \ref{FIGURE10} and \ref{FIGURE11} the attention has been limited  
to the region $\delta_{1} > 1$ since only on this case the spectral energy density exhibits a true maximum in the audio band; this choice is consistent with a decreasing $h_{0}^2 \Omega_{gw}(\nu,\tau_{0})$ for $\nu_{au}< \nu < \nu_{max}$. For the same reason 
$\delta_{2} < 1$ since the spectral energy density must increase\footnote{We remind that, approximately, the spectral index can be written as $m_{T}^{(i)} = 2 - 2 \delta_{i} + {\mathcal O}(r_{T})$. 
Thus $m_{T}^{(1)} < 0$ for $\delta_{1} > 1$ (decreasing spectral energy density) and $m_{T}^{(2)} > 0$ for $\delta_{2} <1$ (increasing 
spectral energy density). Since $m_{T}^{(1)} < 0$ when $\delta_{1} > 1$ the spectral energy density decreases 
around $\nu_{max}$ an this explains why, in this case, the ultra-high frequency constraints 
stipulating that $h_{0}^2 \Omega_{gw}(\nu_{max}, \tau_{0}) < {\mathcal O}(10^{-6})$ are automatically satisfied.} for $\nu< \nu_{au}$. 
The trend of Fig. \ref{FIGURE10} is then confirmed by Fig. \ref{FIGURE11} where the region allowed by the constraints on the relic gravitons also exhibits a magnetic power spectrum compatible with the conditions of Eq. (\ref{REQ1}).  

We may then fix $H_{2}$ and consider the situation where $r_{T}$ is progressively reduced 
well below the current observational limit. 
\begin{figure} 
\begin{center}
\includegraphics[width=7cm,height=7cm]{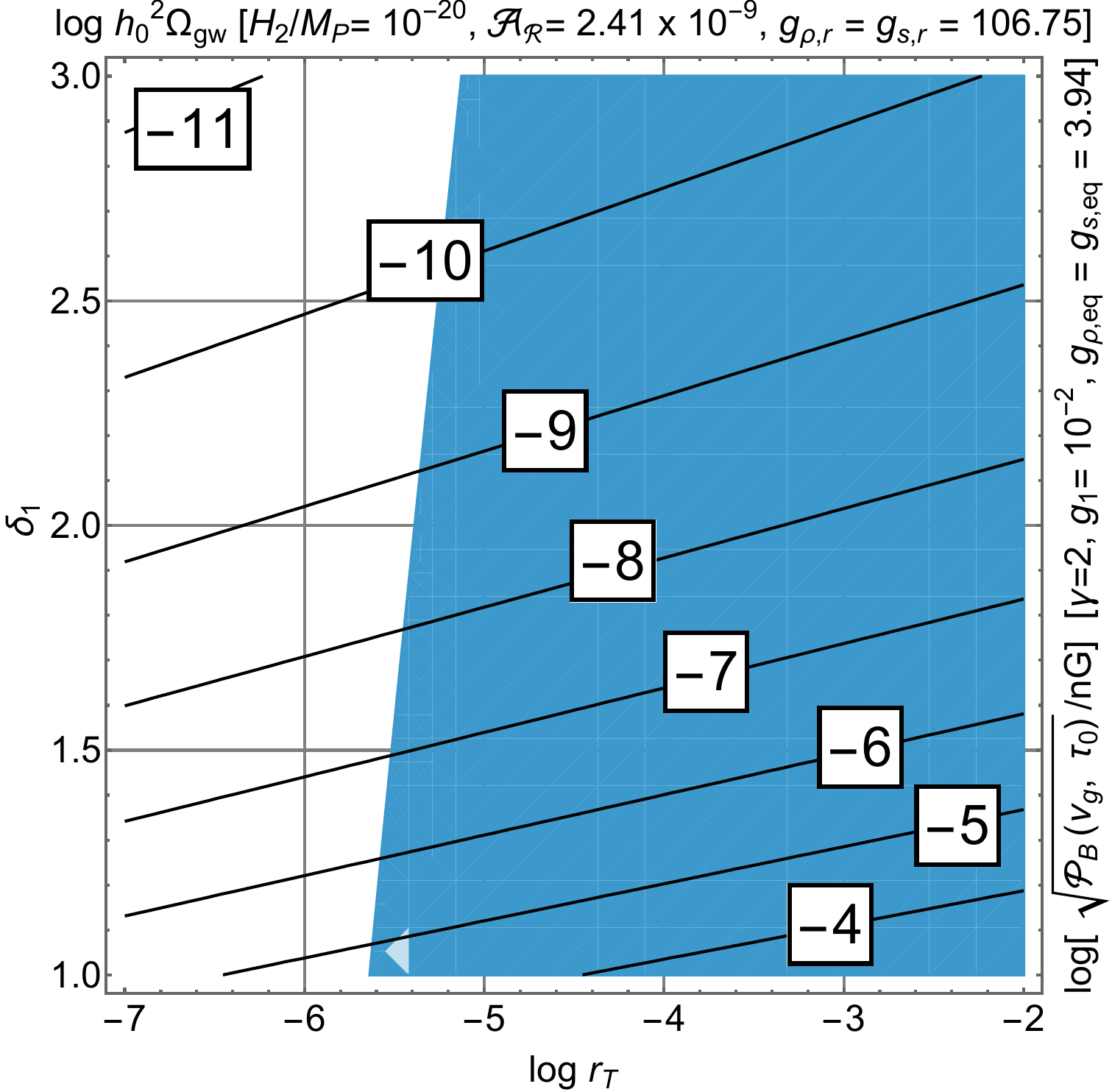}
\end{center}
\caption{\label{FIGURE12} The parameter space is examined in the plane 
$(r_{T}, \delta_{1})$. Common logarithms are employed on the horizontal axis. The value of $H_{2}/M_{P}$ has been fixed to $10^{-20}$. In the shaded region $10^{-16} < h_{0}^2 \Omega_{gw}(\nu_{2}, \tau_{0})< 10^{-9}$. According to the labels appearing in the various contours  $\sqrt{{\mathcal P}_{B}(\nu_{g},\tau_{0})}/\mathrm{nG} > 10^{-11}$ so that the requirements of Eq. (\ref{REQ1}) are satisfied together 
with the presence of a maximum in the relic graviton spectrum for $\nu = {\mathcal O}(\nu_{au})$.}
\end{figure}
\begin{figure} 
\begin{center}
\includegraphics[width=7cm,height=7cm]{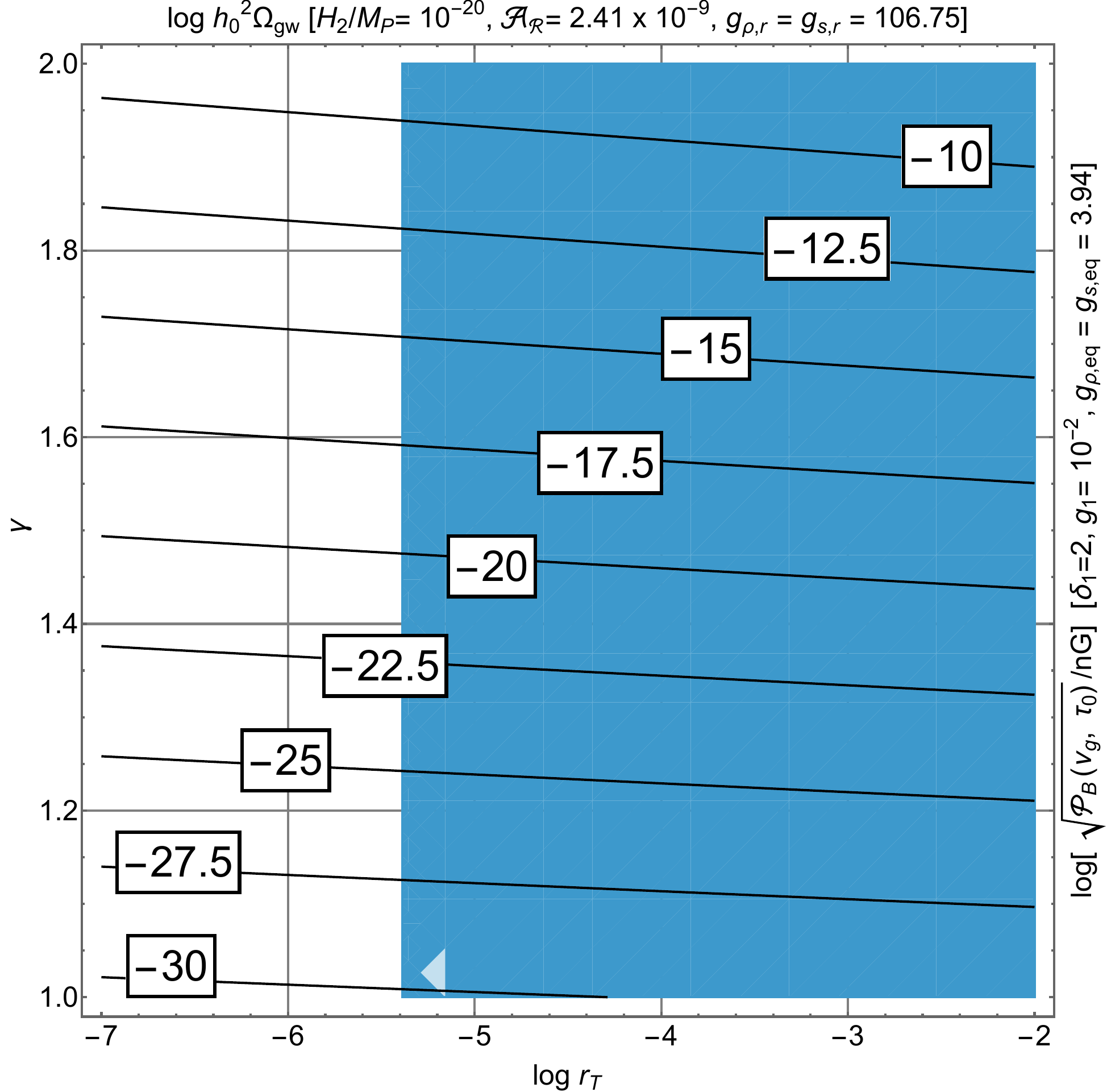}
\end{center}
\caption{\label{FIGURE13} The parameter space is further illustrated in the plane $(\log{r_{T}}, \, \gamma)$. The values of $\delta_{1}$ and $H_{2}/M_{P}$ have been fixed as $\delta_{1} \to 2$ and $H_{2}/M_{P}= {\mathcal O}(10^{-20})$. Common logarithms 
are employed on the horizontal axis.}
\end{figure}
This exercise is illustrated in Fig. \ref{FIGURE12} where $H_{2} = 10^{-20} \, M_{P}$ and the parameter space is studied in the plane $(r_{T}, \, \delta_{1})$.  Finally, in  Fig. \ref{FIGURE13} we have chosen $\delta_{1} \to 2$
and scrutinized  the plane $(\log{r_{T}}, \, \gamma)$. Both in Fig. \ref{FIGURE12} 
and \ref{FIGURE13} the lower limit of Eq. (\ref{REQ1}) is automatically enforced. The interesting 
feature exhibited by Figs. \ref{FIGURE12} and \ref{FIGURE13} is that the allowed values of $r_{T}$ fall in the range ${\mathcal O}(10^{-6}) < r_{T} \leq 0.03$. 

All in all the relic gravitons can be invisible 
in the aHz range (i.e. $r_{T} \ll 0.03$) even if  there is a lower bound on the tensor to scalar ratio.
This is in contrast with the case of a single decelerated stage preceding the radiation epoch (see e.g. Fig. \ref{FIGURE9} and discussion 
therein). While in the second part of this subsection the case of a local maximum in the audio band has been specifically studied, it is also possible to discuss, with the same approach, the situation where $\nu_{2} > \nu_{au}$. In particular 
an interesting example suggests that the intermediate maximum may occur for $\nu_{2} = {\mathcal O}(0.1) \mathrm{MHz}$. However since $\nu_{2} \gg \nu_{au}$ it would not make sense to enforce the limits coming from ground based detectors. 
In this frequency region the signal can be comparatively larger [i.e. $h_{0}^2 \Omega_{gw}(\nu_{2}, \tau_{0}) = {\mathcal O}(10^{-6})$] and the magnetogenesis constraints of Eq. (\ref{REQ1}) satisfied. This is why although the present logic has been to focus on the most constrained framework (i.e. a maximum for $\nu= {\mathcal O}(\nu_{au})$), it is not excluded that in other cases the concurrent constraints coming from the relic gravitons and from large-scale magnetogenesis will be equally satisfied.

\renewcommand{\theequation}{6.\arabic{equation}}
\setcounter{equation}{0}
\section{Concluding Considerations}
\label{sec6}
Prior to the synthesis of light nuclei the expansion rate of the 
Universe cannot be directly assessed and the only hope for an observational test relies on the detection of the diffuse backgrounds of relic gravitational radiation. In a nutshell this is the rationale behind the possibility that the ultra-low frequency gravitons are completely invisible in the aHz domain even if their spectral energy density in critical units could exceed the signal of the concordance paradigm both in the audio band and in the high frequency range (i.e. between the MHz and the THz). Since the postinflationary timeline also influences the evolution of other quantum fields amplified during an accelerated stage of expansion it is intersting to analyze the concurrent constraints arising from different kinds of phenomena. In particular a stage of increasing gauge coupling amplifies the quantum fluctuations of the gauge modes during inflation and, after the coupling flattens out, the late-time hypermagnetic power spectra during the decelerated stage are determined by the hyperelectric fields at the end of inflation. A quasi-flat hyperelectric spectrum (with blue tilt) amplified during inflation leads then to a nearly scale-invariant hypermagnetic spectrum prior to matter radiation equality, i.e. when the protogalactic wavelength effective horizon.  After electroweak symmetry breaking the hypercharge field projects on the electromagnetic fields and the result of the amplification gets further reduced. However the presence of a postinflationary phase slower than radiation automatically increases the physical gauge spectra. The decelerated timeline can then be concurrently constrained by requiring that {\em (i)} the relic gravitons are invisible in the aHz domain, {\em (ii)} the large-scale magnetic fields are significant at the scale of the protogalactic collapse and {\em (iii)} $h_{0}^2 \Omega_{gw}(\nu,\tau_{0})$ exceeds the signal of the concordance paradigm both in the high frequency domain and in the audio band. 

For the sake of concreteness the attention has been focussed on the possibility that gravitons are  invisible at low frequencies while their high-frequency effects are more prominent 
and would imply that ${\mathcal O}(10^{-10})< h_{0}^2 \Omega_{gw}(\nu, \tau_{0})< {\mathcal O}(10^{-6})$ for $ 0.1 \mathrm{kHz} < \nu < \mathrm{THz}$.  Two complementary situations have been analyzed here involve, respectively, the presence 
of a spike in the ultra-high frequency region (i.e. between the MHz and the THz)  and 
a maximum in the audio band. A large signal in the MHz or THz domains
is associated with a single decelerated stage expanding slower than radiation:
in this case the tensor to scalar ratio can be much smaller than the current 
observational value (i.e. $r_{T} \ll 0.03$) while the physical power 
spectra of the magnetic fields correspond to ${\mathcal O}(10^{-16})\,  \mathrm{nG} < \sqrt{{\mathcal P}_{B}(\nu_{g}, \tau_{0})} < {\mathcal O}(10^{-11})\, \mathrm{nG}$ over the typical scale of the gravitational collapse of the protogalaxy. If the postinflationary expansion rate prior to radiation dominance consists of two successive stages the spectral energy density at intermediate frequencies develops a maximum in the audio band where 
the direct constraints determined by the wide-band detectors can be directly exploited. These 
limits imply,  broadly speaking, that $h_{0}^2 \Omega_{gw}(\nu,\tau_{0}) < {\mathcal O}(10^{-9})$ for 
$\nu = {\mathcal O}(\nu_{au})$ where $\nu_{au}$ approximately ranges between few 
Hz and the kHz. A maximum in the audio band is compatible 
with a comparatively larger magnetic field ${\mathcal O}(10^{-11})\, \, \mathrm{nG} < \sqrt{{\mathcal P}_{B}(\nu_{g}, \tau_{0})} < {\mathcal O}(10^{-2})\, \mathrm{nG}$. 

All in all the evolution of the hypercharge fields is correlated with the spectra of the relic gravitons since both phenomena depend on the modifications of the postinflationary timeline prior to the nucleosynthesis epoch. The requirement of invisible gravitons in ultra-low frequency domain is then compatible with a spectral energy density that drastically exceeds the signal of the concordance paradigm at higher frequencies and, in this situation, the magnetogenesis constraints are satisfied at a different level of accuracy. The potential detection of relic gravitons in complementary ranges of comoving frequencies (e.g. either in the audio band or in the THz domain) determines the magnetic power spectra at the scale of the protogalactic collapse and vice-versa.

\section*{Acknowledgements}
 It is  a pleasure to thank A. Gentil-Beccot,  A. Kohls,  L. Pieper, S. Rohr and J. Vigen of the CERN Scientific Information Service for their kind assistance.

\end{document}